\documentclass[preprint2]{aastex}

\usepackage{natbib}
\usepackage{multirow}
\usepackage{lscape}
\usepackage{fancyheadings}

\newcommand{\Msun}{\ifmmode{\mathcal{M}_\odot}\else{$\mathcal{M}_\odot$}\fi}
\newcommand{\Lsun}{\ifmmode{\mathrm{L}_\odot}\else{L$_\odot$}\fi}
\newcommand{\kms}{$\mathrm {km s}^{-1}$}
\newcommand{\grado}{$^\circ$}

\slugcomment{Accepted for publication in ApJ}
\shortauthors{Marino et al.}
\shorttitle{IFS and multi-wavelength imaging of NGC~5668: an unusual flattening in metallicity gradient}

\begin{document}

\title{Integral Field Spectroscopy and multi-wavelength imaging of the nearby spiral galaxy NGC~5668,\altaffilmark{1}\,: {\bf an unusual flattening in metallicity gradient}.}

\author{R.A.\,Marino\altaffilmark{2,}\altaffilmark{3}, A.\, Gil de Paz\altaffilmark{2}, A.\,Castillo-Morales\altaffilmark{2}, J.C.\,Mu\~{n}oz-Mateos\altaffilmark{4}, S.F.\,S\'{a}nchez\altaffilmark{3}, P.G.\, P\'{e}rez-Gonz\'{a}lez\altaffilmark{2,}\altaffilmark{7}, J.\, Gallego\altaffilmark{2}, J.\, Zamorano\altaffilmark{2}, A.\, Alonso-Herrero\altaffilmark{5} and S.\, Boissier\altaffilmark{6}}

\altaffiltext{1}{Based on observations collected at the German-Spanish Astronomical Center, Calar Alto, jointly operated by the Max-Planck-Institut f\"{u}r Astronomie Heidelberg and the Instituto de Astrof\'{i}sica de Andaluc\'{i}a (CSIC).}
\altaffiltext{2}{CEI Campus Moncloa, UCM-UPM, Departamento de Astrof\'{i}sica y CC$.$ de la Atm\'{o}sfera, Facultad de CC$.$ F\'{i}sicas, Universidad Complutense de Madrid, Avda.\,Complutense s/n, 28040 Madrid, Spain} \email{ramarino@fis.ucm.es}
\altaffiltext{3}{Centro Astron\'{o}mico Hispano Alem\'{a}n, Calar Alto, (CSIC-MPG), C/Jes\'{u}s Durb\'{a}n Rem\'{o}n 2-2, E-04004 Almeria, Spain}
\altaffiltext{4}{National Radio Astronomy Observatory, 520 Edgemont Road, Charlottesville, VA 22903-2475}
\altaffiltext{5}{Instituto de Fisica de Cantabria, CSIC-UC, Avenida de los Castros s/n, 39005 Santander, Spain}
\altaffiltext{6}{Laboratoire d\' Astrophysique de Marseille, OAMP, Universit\'{e} Aix-Marseille \& CNRS UMR 6110, 38 rue Fr\'{e}d\'{e}ric Joliot-Curie, 13388 Marseille cedex 13, France}
\altaffiltext{7}{Associate Astronomer at Steward Observatory, University of Arizona, AZ, USA}

\begin{abstract}
We present the analysis of the full bi-dimensional optical spectral cube of the nearby spiral galaxy NGC~5668, observed with the PPAK IFU at the Calar Alto observatory 3.5m telescope. We make use of broad-band imaging to provide further constraints on the evolutionary history of the galaxy. This dataset will allow us to improve our understanding of the mechanisms that drive the evolution of disks. We investigated the properties of 62 H~II regions and concentric rings in NGC~5668 and derived maps in ionized-gas attenuation and chemical (oxygen) abundances. We find that, while inwards of r\,$\sim\,36''\,\sim$\,4.4kpc\,$\sim$\,0.36\,$(\frac {D_{25}}{2})$ the derived O/H ratio follows the radial gradient typical of spiral galaxies, the abundance gradient beyond r$\sim36''$ flattens out. The analysis of the multi-wavelength surface brightness profiles of NGC~5668 is performed by fitting these profiles with those predicted by chemo-spectrophotometric evolutionary models of galaxy disks. From this, we infer a spin and circular velocity of $\lambda$=0.053 and v$_{c}$=167\,km\,s$^{-1}$, respectively. The metallicity gradient and rotation curve predicted by this best-fitting galaxy model nicely match the values derived from the IFU observations, especially within r\,$\sim36\arcsec$. The same is true for the colors despite of some small offsets and a reddening in the bluest colors beyond that radius. On the other hand, deviations of some of these properties in the outer disk indicate that a secondary mechanism, possibly gas transfer induced by the presence of a young bar, must have played a role in shaping the recent chemical and star formation histories of NGC~5668.
\end{abstract}
\date{Accepted for publication in ApJ. }



\section{Introduction}

\begin{figure*}
\begin{center}
\resizebox{0.49\hsize}{!}{\includegraphics{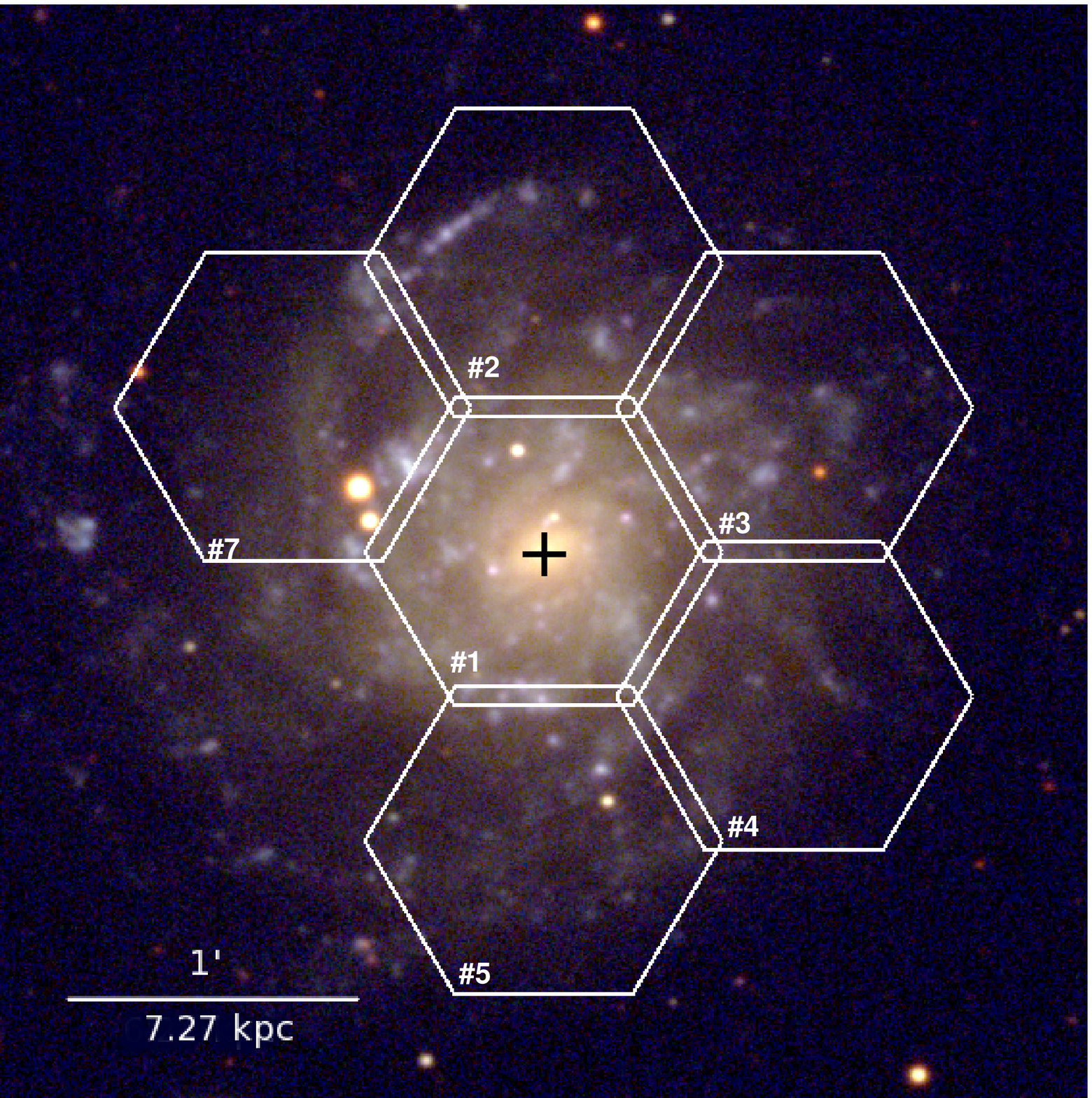}}
\resizebox{0.49\hsize}{!}{\includegraphics{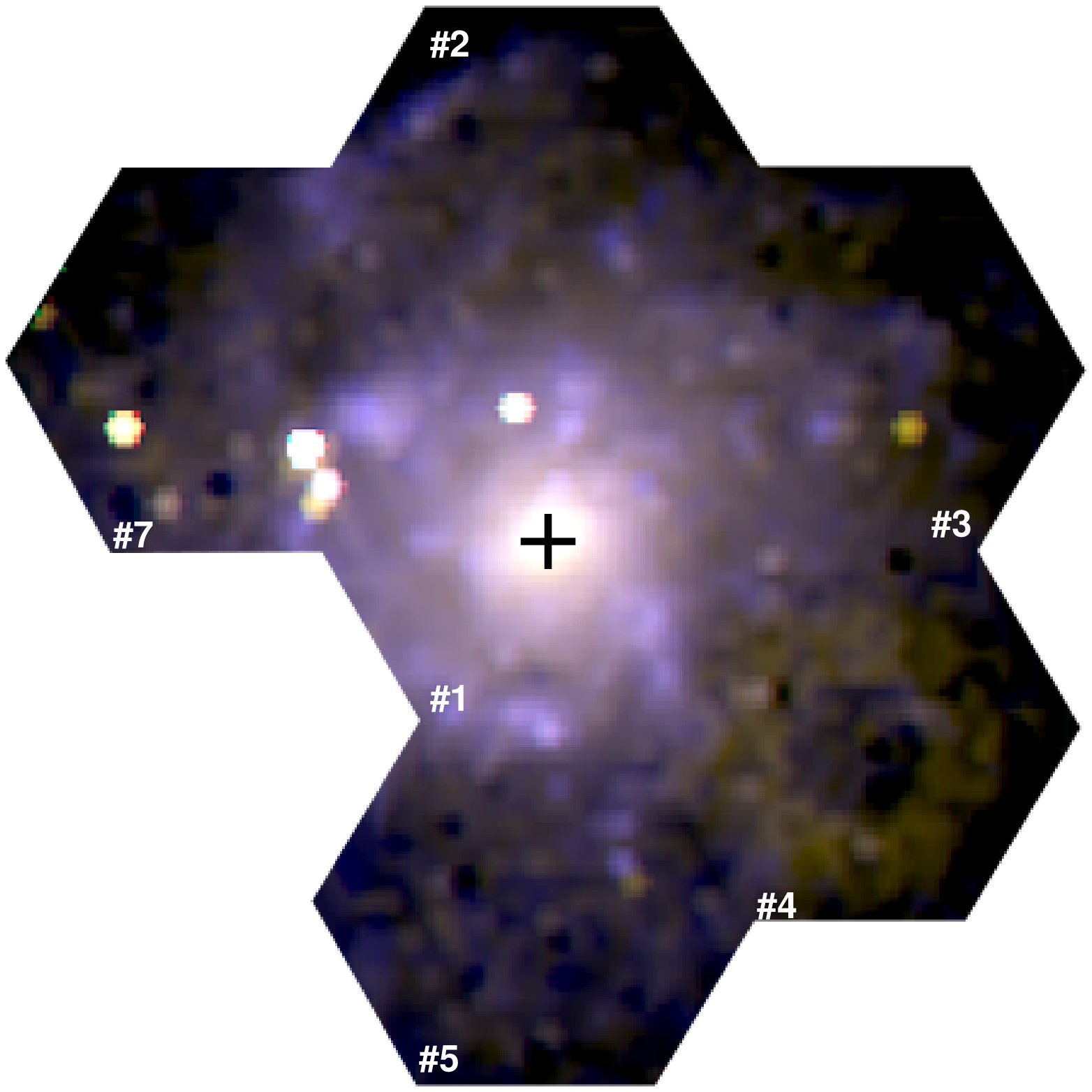}}
\caption[Fig.1]{\footnotesize
	{\bf Left:} SDSS optical {\it ugr} image (from 4458 to 7706 \AA) of NGC~5668 centered at RA(2000) 14$^{h}$33$^{m}$24.3$^{s}$ and Dec(2000) +04$^{\circ}$33$^{\prime}$24.3$^{\prime\prime}$. North is up and East is to the left. Plate scale is 0.396 arcsec/pix. {\bf Right:} Synthetic false-color image obtained from the PPAK data cube and the response curves of the SDSS {\it u}, {\it g} and {\it r} filters. Note that the luminous spot in pointing 7 is caused by cross-talk effects from a nearby (on the CCD) field star. Plate scale is 1 arcsec/pix in this case. Center and orientation are the same as in the left panel.\label{SDSS}}
\end{center}
\end{figure*}

The evolution of galactic disks is one of the most important and yet not fully understood topics in extragalactic astronomy. Despite significant progress in the recent past regarding our understanding of the history of both the thick and thin disks, important questions remain unanswered: How old are the disks seen in the spiral galaxies today? How did they chemically evolve? Are they growing inside-out, as proposed to explain the color and metallicity gradients of our own Milky Way? Do they have an edge? How efficient is stellar radial diffusion? \\
Until recently, the study of the properties of spiral disks has been limited to broad-band imaging data and/or long-slit spectroscopy, which has severely limited the reach of previous works. Radial color profiles have been widely used to probe the inside-out scenario of disk evolution (\citealt{bell, mac, pohlen, mun07}). However, radial variations of metallicity and dust extinction conspire to create color gradients similar to those due to changes in stellar age alone. Analysis of color-magnitude diagrams allow for more robust measurements of stellar ages as a function of radius, but these kind of studies are currently limited to very nearby galaxies where individual stars can be resolved (\citealt{djong}; \citealt{barker}; \citealt{gogarten}). Long-slit spectra overcomes the problems of broad-band photometry, but at the expense of focusing along a pre-defined spatial axis \citep{yoadalc}.
Our effort is committed to add another dimension to the study of nearby spiral galaxies thanks to the use of wide-field integral-field spectrometers (IFS) combined with images covering a wide wavelength range from the ultraviolet to the infrared. This paper presents the analysis of the IFU data taken for NGC~5668 within a more ambitious project aimed to map a sample of a dozen nearby galaxies included in the HGRS (Herschel Galaxy Reference Survey;  \cite{boselli}) using mosaics obtained with two of the largest IFU available to date: PPAK (Pmas fiber PAcK) IFU of PMAS (Potsdam Multi-Aperture Spectrophotometer) at the CAHA 3.5-m and VIMOS (VIsible MultiObject Spectrograph) at ESO-VLT.

The evolution of galaxy disks is a complex process as many are the mechanisms that might alter their photometric, chemical, and kinematical properties: minor mergers, formation of bars or rings, density waves, stellar diffusion, gas in-fall, and, in the case of galaxies in clusters, also ram pressure striping and galaxy harassment. Despite this rather discouraging scenario, a picture for the formation of disks has emerged in recent years that attempts to explain most of their observational properties: the inside-out formation scenario. Although this scenario has been around for a long time now, it has been only recently that inside-out growth of galaxy disks has been applied and quantified beyond the MW. Both the extinction-corrected color gradients in nearby galaxies, \citet{mun07}, and weak dependence of the mass-size relation with redshift (\citealt{truj04}, \citealt{barden}, \citealt{truj06}) support an inside-out scenario for the evolution of disks. However, there are several observational results that conflict or, at least, cannot be explained by this otherwise elegant scenario. In this regard, many nearby spiral disks show that, while the color and metallicity gradients of the inner disk are compatible with the inside-out scenario, beyond the SF-threshold radius \citep{martkenn} the gradients flatten and even reverse, by getting redder toward the outer regions (\citealt{bakos}; \citealt{azzollini}; \citealt{vlajic}). In the specific case of the chemical abundances, the models based on the standard inside-out scenario of galaxy disk formation predict a relatively quick self enrichment with oxygen abundances 1/10 the Solar value after only 1 Gyr of evolution \citep{boipra}. These models also predict an almost universal negative metallicity gradient once this is normalized to the galaxy optical size d$[O/H]$ of $\sim-$0.8 dex/R$_{25}$; \citep{praboi}.
Regarding this, the few observational abundance measurements obtained to date in the outer edges of disks show a flattening or even an increase in the abundance toward the outermost parts of the disk. This is also true in the case of the extended UV disks recently discovered by the GALEX satellite (\citealt{gil05}; \citealt{gil07}; \citealt{thilker}), which show oxygen abundances that are rarely below 1/10 (Solar value). In this regard, \cite{bresolin09a} found that the normalized radial metallicity gradient in M83 changes from $-$0.030 dex kpc$^{-1}$ in the inner disk to $-$0.005 dex kpc$^{-1}$ in the outer parts, see also \cite{bresolin12}. We also refer the reader to \cite{yoac10}~ for another recent example in this same regard. Similar results were obtained on the metallicity gradient of the outer disk of NGC~300 from single-star CMD analysis \citep{vlajic}.\\ 

In this work we present the pilot study of the full bi-dimensional spectral cube of the nearby spiral galaxy NGC~5668, (Fig$.$~\ref{SDSS}), obtained with the PPAK Integral Field Unit at the Calar Alto (CAHA) observatory 3.5 m telescope. Despite the relatively modest collecting area of the CAHA 3.5-m telescope, the broad spectral coverage (from 3700 to 7000 \AA) and adequate spectral resolution ($R = 500$) of PPAK with the V300 grating make this instrument one of the best tools for studying stellar populations, dust content, and physical conditions of the gas (temperature, density, chemical abundances) in spatially-resolved galaxies.

\section{The nearby spiral galaxy NGC~5668} 

NGC~5668 is a nearly face-on late-type spiral galaxy classified as an Sc(s)II-III type according to \cite{sandage} and as an SA(s)d by \cite{devac}. There is a weak bar or oval inner structure $12''$ in size visible on the optical image, which leads to a small shoulder in the surface brightness profile published by \cite{schu94}. The outer disk (beyond $R = 100''$) is slightly asymmetric and more extended towards the North. For this work we adopted a distance of $\sim$ 25 Mpc, $(m-M) = 31.99$ mag, assuming a cosmology with $H_0$ = 73\,km\,s$^{-1}$Mpc$^{-1}$, $\Omega_{\mathrm{matter}}$ = 0.27, $\Omega_{\mathrm{vacuum}}$ = 0.73, for a recession velocity of 1813 km\,s$^{-1}$ corrected to the reference frame defined by the 3K Microwave Background Radiation \citep{fix}\footnote{Source: NASA$/$IPAC Extragalactic Database (\url{http://nedwww.ipac.caltech.edu/})}. A broad-band optical image of this galaxy from the Sloan Digital Sky Survey \citep{york} is shown in Fig$.$~\ref{SDSS}.
The galaxy presents an inclination of 18$^{o}$ and a total magnitude $B = 12.13 \pm 0.03$ mag \citep{schu96}. The total mass within the optical disk was estimated to be 5.7$\times$10$^{10}$ M$_{\odot}$ \citep{schu96}\footnote{They adopted a distance of 30 Mpc consistent with a cosmology with $H_0$ = 50\,km\,s$^{-1}$Mpc$^{-1}$.}. 
NGC~5668 has been found to host a number of HVC’s (High Velocity Clouds) by \citet{schu96} using the Arecibo telescope. These authors detected high velocity wings in the line shape, which were attributed to HVC’s in the galaxy. High velocity clouds and High-Residual-Velocity-Regions (HRVR; residual velocity field and the shell/chimney candidates, regions which have a systematic deviation from rotational velocity) were found also in the ionized gas and interpreted as regions with vertical motions related to ongoing star forming processes in the disc on the basis of Fabry-Perot H$\alpha$ observations \citep{jimenez}. \cite{schu96} also computed $L_{FIR} = (5.8 \pm 1.2)\times 10^{9} \Lsun$ and $L_{H{\alpha}} = 1.0 \pm 0.3 \times 10^{8} \Lsun$. \\
Such high FIR and H$\alpha$ luminosities (and SFRs therefore) would naturally result in a high Supernova rate in this galaxy. Indeed, NGC~5668 is known as a \textquotedblleft SNe factory\textquotedblright\, due to the discovery of multiple supernova explosions in recent epoch, namely SN2004G \citep{nakano}, SN1952G, SN1954B \citep{boffi}. NGC~5668 has also been recently observed by a number of instruments and facilities, including SAURON at the WHT, SDSS, Spitzer and as a part of the Medium-deep Imaging Survey of GALEX. This data-set in combination with the PPAK mosaic obtained as part of this work should allow a very detailed analysis of the evolution- and chemical-enrichment history of NGC~5668. 

\section{Observations and Data Reduction}

\subsection{IFS Observations}

We have observed the nearly face-on spiral galaxy NGC~5668 with the PPAK IFU of PMAS at the Calar Alto (CAHA, Spain) observatory 3.5 m telescope \citep{kelz}. The observations were carried out on June 22-24, 2007. We used the PPAK mode that yields a total field-of view (FoV) of $74''\times 65''$ (hexagonal packed) for each pointing. We covered a total area of roughly $2 \times 3$\,arcmin$^{2}$ with a mosaic of 6 PPAK pointings (see Fig$.$~\ref{SDSS}). We used the V300 grating covering a wavelength range of 3700\,-\,7100\,\AA, (10\,\AA\,FWHM, corresponding to $\sigma\sim$\,300 km\,s$^{-1}$ at $\mathrm{H}\beta$). With this spectral configuration we covered all the optical strong emission lines in which we are interested. A total of 112 individual images were taken during the observing run. These images include 19 frames on-source for the six pointings obtained in NGC~5668 (see Table~\ref{obs} for a summary of the science observations), bias frames (22 in total), sky-flats (19), focus images (6), HgCdHe arcs (14), tungsten-lamp flats (14), observations of the spectro-photometric standard star Hz44 (3), dome-flats (6).
The spatial sampling is determined by a hexagonal array of 331 densely packed optical fibers for the object (science-fibers), 36 fibers for the sky and 15 calibration fibers. Each fiber has a diameter of 2.68$^{\prime\prime}$ and a pitch 3.458$^{\prime\prime}$ with a high filling factor in one single pointing (65\%). For a more exhaustive description of the instrument see \cite{kelz}.

\subsection{IFS data reduction}

The reduction procedure applied to NGC~5668 follows the techniques described in \cite{sanchez06} using R3D. This is a software package developed specifically for the reduction of fiber-based IFS data, which has been extensively used for the reduction of PMAS data, in combination with IRAF\footnote{IRAF is distributed by the National Optical Astronomy Observatories, which are operated by the Association of Universities for Research in Astronomy, Inc., under cooperative agreement with the National Science Foundation.} packages and Euro3D software \citep{sanchez04}.
Six pointings were observed for NCG~5668 leading to total covered area of $2\times3$\,arcmin$^{2}$. Each pointing results from the combination of three/four images, each one with a exposure time of 1000 seconds. Our combined raw mosaic includes a total of 2292 spectra (1982+310 science+internal calibration spectra) covering the spectral range 3700\,-\,7131\,\AA. The pre-reduction of the IFS data consists of all standard corrections applied to the CCD that are common to the reduction of any CCD-based data:
\begin{itemize}
\item {\it Bias subtraction:} A master bias frame was created by averaging all the bias frames observed during the night and subtracted from the science frames.
\item {\it Cosmic-ray rejection:} We median combined different exposures (at least 3) of the same pointing. This procedure is able to remove 95\% of all cosmic-ray events from each pointing.
\end{itemize}
After these corrections, the IFS data reduction consists of: 
\begin{itemize}
\item {\it Spectra extraction:} We identified the position of the spectra on the detector for each pixel along the dispersion axis. We then co-added the flux within an aperture of a 5 pixels around the trace of the spectra and we extracted the flux corresponding to the different spectra at each pixel along the dispersion axis. Finally the 1D extracted spectra are stored in a row-staked-spectra file (RSS).
\item {\it Wavelength calibration:} We corrected for all distortions and determined the wavelength solution by using an arc calibration lamp exposure. Calibration data cubes were taken at the beginning and at the end of the night by illuminating the instrument with a HgCdHe lamp. The V300 wavelength calibration was performed using 14 lines in the considered spectral range. The $rms$ of the best-fitting polynomial (order 4) was 0.45\,\AA \,(i.e.~FWHM/20).
\item{\it Fiber-to-fiber transmission:} We corrected the differences in fiber-to-fiber transmission by using twilight sky exposures from which a Fiber-Flat frame was computed.
\item{\it Sky emission subtraction:} For each pointing, a sky image of 300 seconds (shifted $+5'$ in Dec) was taken. We median combined the different sky images to create a sky frame which was then subtracted from the science data frame.
\item{\it Relative flux calibration:} We relative-flux calibrated our data using observations of the spectrophotometric standard star Hz44 ($\alpha$(2000)= 13h 23m 35.37s, $\delta$(2000)= +36d 08' 00.0'', \citealt{oke}) obtained during the night with the same instrumental setup as for the object. We applied all the previous steps to the calibration star frame and extracted the standard star spectrum from the brightest 2.7\arcsec-wide PPAK fiber centered on the star. Finally, we derived the ratio between counts per second and flux. In this way we obtained the instrumental response which is applied to the science frames in order to flux-calibrate them. Note that although this procedure also provides an absolute calibration for the data this is uncertain due to potential flux loses during the observation of the standard star and variations in the transparency throughout the night.   
\item{\it Differential Atmospheric Refraction (DAR) correction:} We calculated a total theoretical DAR value using a Calar Alto public tool obtaining a value of 1$"$ (the mean value of airmass used is 1.3). In this case the differential atmospheric refraction can be considered negligible compared with the fiber size (2.7$"$ in diameter). Therefore, we considered the DAR correction is not required for this data cube. In any case we never obtained any flux measurements from regions with less than 5 adjacent fibers in size.
\item{\it Locating the spectra in the sky:} A position table that relates each spectrum to a certain fiber gives the location of the spectra in the sky.
\item{\it Mosaic reconstruction:} We built a single RSS file for the whole mosaic by adding pointings previously normalized to a reference ('master') pointing. The 'master' pointing is that having the best sky subtraction and most optimal observing conditions regardless of the geometric position of the pointing in the mosaic. In our case this 'master' pointing was number 5 (offsets (0,-60), see Fig$.$~\ref{SDSS}). This procedure makes use of the overlapping fibers between different pointings (11 fibers in total) to re-scale each new added pointing using the flux ratio in the continuum region 5869-5893\,\AA\ between the two pointings (the newly added and the 'master' one).
\item{\it Absolute Flux Re-Calibration:} We also computed an absolute flux re-calibration based on SDSS photometry. Only two of the five filters of SDSS ( {\it g} and {\it r}, $\lambda_{eff}=4694$\, and 6178\,\AA, respectively) are used because these are the ones for which their pass-bands are fully covered by our spectra. We convolved the whole datacube of NGC~5668 with the SDSS g- and r-filter pass-bands to derive the absolute spectrophotometry. Firstly, we measured the AB magnitude within the footprint of the entire PPAK mosaic and in 7 concectric annuli for both PPAK and SDSS images. Then, these magnitudes were converted to fluxes using the prescription in the SDSS documentation\footnote{\url{http://www.sdss3.org/dr8/algorithms/fluxcal.php}} in order to calculate the flux ratio in both bands. The resulting average scaling factor found is 0.9 for the two data pairs. This result implies a factor 0.04 in magnitudes.  Note that the error in SDSS photometry alone is estimated to be 2\% mag. In our case we have an uncertain of 5\% in fluxes so the total absolute error associated to line fluxes is 8\%. Obviously, relative measurements are more precise. All results showed in the following analysis (and in Table 3) are recalibrated based on these SDSS photometry measurements.
\end{itemize}

\subsection{Imaging observations}
\begin{figure*}
\begin{center}
\resizebox{0.488\hsize}{!}{\includegraphics{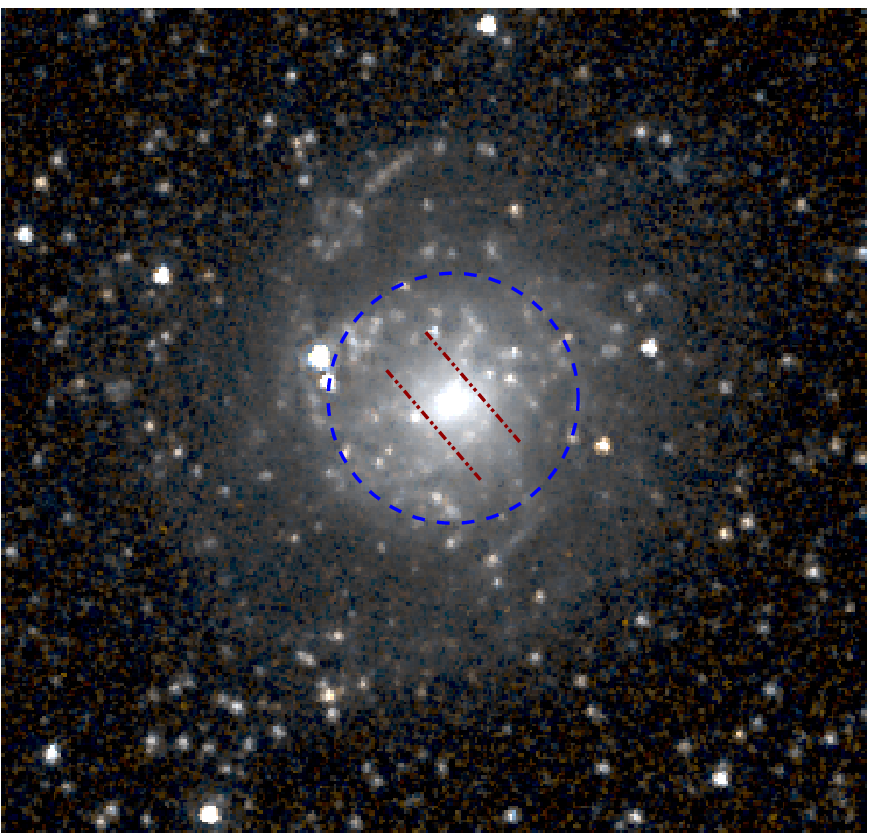}}
\resizebox{0.49\hsize}{!}{\includegraphics{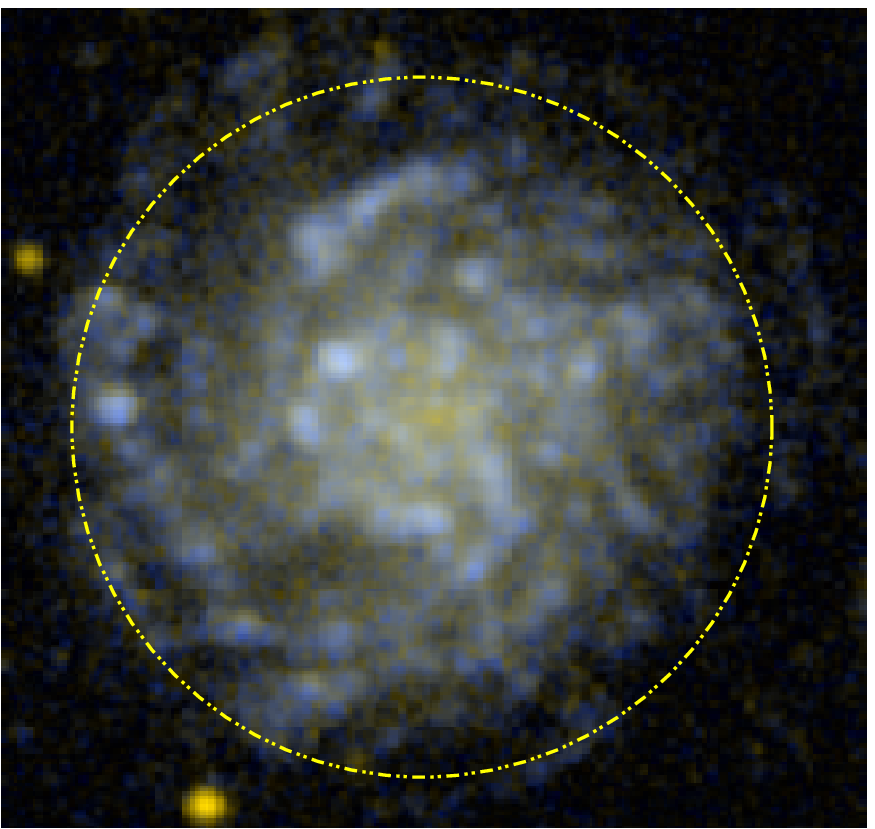}}
\caption[Fig.2]{\footnotesize {\bf Left:} IRAC color image of NGC5668, using the 3.6$\micron$ and 4.5$\micron$ images. Since these bands primarily probe old stars, no significant color variations are found. The red dashed lines show the width of the weak bar visible in the optical images ($\sim$12" in size) and the blues dashed circle corresponds to the break radius of the metallicity gradient (r$\sim$36"). {\bf Right:} GALEX color image, using the FUV and NUV bands as the blue and red channels, respectively, and a linear combination of both in the green one. The diameter of the yellow dashed circle is the optical one quoted in the RC3 catalog, at 25.5 mag/arcsec$^2$ in the B band ($\sim$3.3'). Both frames are 250" in size. North is up and East is to the left. \label{IracGalex}}
\end{center}
\end{figure*}

In order to complement the IFS data of NGC~5668 presented here, we have also compiled multi-wavelength broadband images acquired with different facilities: UV imaging data from GALEX, optical data from SDSS and near-IR data from Spitzer (see Fig$.$~\ref{IracGalex}).
The GALEX telescope \citep{martin05} observed NGC~5668 as part of its Medium-deep Imaging Survey. Images at both the FUV ($\lambda_{eff}=151.6$\,nm) and NUV ($\lambda_{eff}=226.7$\,nm) bands are available for this galaxy. The final pixel scale is 1.5\arcsec/pixel, and the PSF has a FWHM of\,$\sim$\,5-6\arcsec\, in both bands \citep{morris}. At the distance of NGC~5668, this corresponds to a physical scale of $\sim$\,700\,pc. The uncertainty of the zero-point, which is calibrated against white dwarfs, is estimated to be less than 0.15\,mag \citep{gil07}.

Optical images in the $ugriz$ bands were retrieved from the SDSS \citep{york} Data Release 7 archive \citep{aba}. The two different scans on which NGC~5668 lies were flux-matched and mosaicked together using a custom-built task in IRAF. We applied the photometric calibration specified in the SDSS DR7 Flux Calibration Guide\footnote{\url{http://www.sdss.org/dr7/algorithms/fluxcal.html}}, relying on the calibration factors of the scan used as reference when creating the mosaic. According to the information provided by the SDSS project, the photometric relative errors are of the order of 2\%-3\%. The plate-scale is 0.396\arcsec\, per pixel with a PSF FWHM of 1.4\arcsec\, (170\,pc at the distance of the galaxy).

Near-IR images at 3.6\,$\micron$ and 4.5\,$\micron$ were taken by Spitzer \citep{werner} using the IRAC instrument \citep{fazio}. The corresponding Post Basic Calibrated Data (PBCD) were downloaded from the Spitzer archive\footnote{Spitzer Proposal ID 69: A Mid-IR Hubble Atlas of Galaxies, PI: G. Fazio}. These images are already flux-calibrated and delivered in units of MJy\,sr$^{-1}$. The pixel scale is 0.6\arcsec\, per pixel, and the PSF has a FWHM of 1.7\arcsec\, in both channels, probing a spatial extent of 200\,pc \citep{reach}. The photometric error ($\sim$2\%) is dominated by the aperture corrections that need to be applied to account for the scattered light on the array. This total uncertainty is estimated to be around 10\%\footnote{\url{ http://ssc.spitzer.caltech.edu/irac/calib/extcal/}} in this case.

\section{Analysis}

\subsection{Spectroscopy}
\label{spec}

Many important topics in astrophysics involve the physics of ionized gases and the interpretation of their emission line spectra. Powerful constraints on theories of galactic chemical evolution and on the star formation histories of galaxies can be derived from the accurate determination of chemical abundances either in individual star forming regions or distributed across galaxies or even in galaxies as a whole. In this sense chemical abundance are a fossil record of its star formation history. In addition, the distribution of H~II regions is an excellent tracer of recent massive star formation in spiral galaxies. The previous evolutionary history of the gas, and the local ionization, density and temperature are important features to understand the physical conditions prevailing in the emitting regions where they are emitted. 

For all these reasons, our first aim is to identify H~II regions in our target, or at least H~II complexes (see Fig$.$~\ref{regions})\footnote{Hereafter we use the terms 'H~II regions' and 'H~II complexes' indistinctly, even although individual H~II regions are not spatially resolved at the physical resolution achieved in our IFS data.}. We extract from the PPAK data cube the emission line fluxes corresponding to each of these regions. Furthermore, we also compute 2D maps of the flux of strong emission lines, such as [O\,{\textsc{ii}}]\,$\lambda\lambda$\,3726,3729\AA\AA \footnote{The [O\,{\textsc{ii}}]\,$\lambda\lambda$\,3726,3729\AA\AA\, doublet is spectrally unresolved in our data, therefore and hereafter we will refer to it as [O\,{\textsc{ii}}]\,$\lambda$\,3727\AA.}; [O\,{\textsc{iii}}]\,$\lambda\lambda$\,4959,5007\AA\AA; $\mathrm{H}\beta$; [N\,{\textsc{ii}}]\,$\lambda$\,6548\AA; $\mathrm{H}\alpha$; [N\,{\textsc{ii}}]\,$\lambda$\,6583\AA\, and [S\,{\textsc{ii}}]\,$\lambda\lambda$\,6717,6731\AA\AA. 

\subsubsection{Region selection}

\begin{figure*}
\centering
\includegraphics[width=0.85\textwidth]{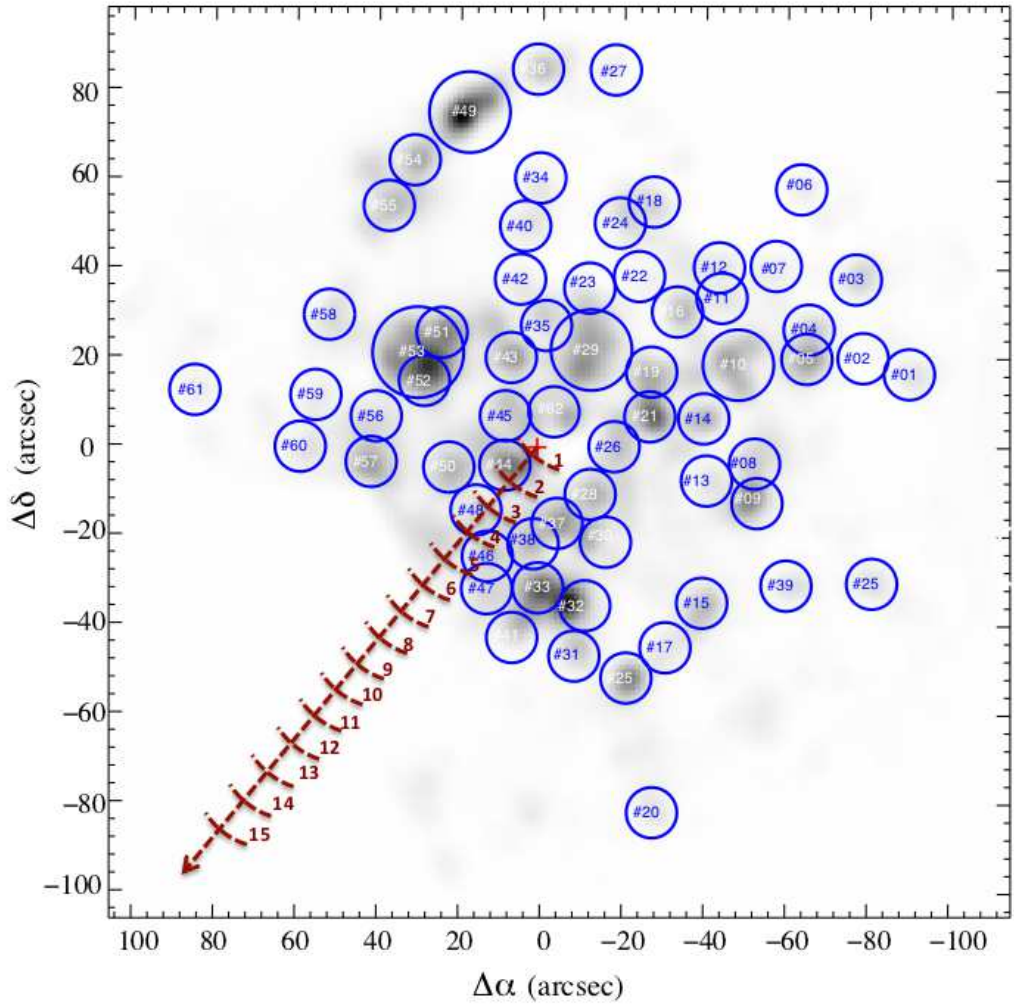}
\caption[Fig.3]{\footnotesize PPAK continuum-subtracted $\mathrm{H}\alpha$ emission map. The H~II regions selected in NGC~5668 are shown as blue circles see section 4.1.1 for details. The red dashed lines indicate the radii of the rings and the red cross represents the central fiber of the mosaic. North is up and East is to the left.\label{regions}}
\end{figure*}

In spiral galaxies, H~II regions are strongly concentrated along the spiral arms and in the galactic plane and are the best objects for tracing the structure of spiral arms in galaxies. We are interested in the study of the optical emission lines from H~II regions because they represent the primary mean of performing gas-phase diagnostics in galaxies (\citealt{pagel86}, \citealt{oster}). These data were analyzed using IRAF and E3D \citep{sanchez04} tasks and some custom-made IDL (Interactive Data Language) scripts.
In order to select the H~II regions of NGC~5668 we computed a synthetic $\mathrm{H}\alpha$ image (the brightest emission line in the optical spectrum of H~II regions under most physical conditions) from the data cube using E3D. The spectral window for the $\mathrm{H}\alpha$ line was defined as a narrow wavelength range from 6589\,\AA\, to 6613\,\AA\,. Two bands (6457-6553\,\AA\, and 6667-6730\,\AA) close to $\mathrm{H}\alpha$ line were also extracted for continuum subtraction. We normalized the \textquotedblleft continuum\textquotedblright\, maps to match the width of the $\mathrm{H}\alpha$ intensity map. From the synthetic $\mathrm{H}\alpha$ image we can then identify H~II regions. We visually selected 62 H~II complexes in the galaxy from this image (see Fig$.$~\ref{regions}). The apertures were defined so they would approximately reach the same surface brightness level in $\mathrm{H}\alpha$. We also imposed that these apertures would be large enough to include at least 5 PPAK fibers (in order to minimize atmospheric-refraction effects) and that each region would be isolated from bright neighbors. As a result of the conditions imposed the majority of regions are the sum of 5 PPAK fibers, except in the case of regions number 17, 37 and 62, which are the superposition of one bright and several, marginally-resolved faint H~II regions, and region 58 which extends 15\arcsec\, in the SE-NW direction approximately following the spiral pattern of the galaxy northern arm. See Fig$.$~\ref{regions} for the positions and sizes of the regions of interest selected for this study. We extracted the spectrum coming from each region by averaging the signal for each fiber belonging to a particular region. 

In many spiral galaxies of early and intermediate Hubble type (Sa-Sc), active star formation is organized in a ring-like structure that often contains a large fraction of the entire star formation activity of the galaxy. To investigate this and the ubiquitous radial variation of the physical properties in spiral galaxies, we also selected 18 concentric annuli, centered on the peak of the optical continuum emission. A width of about $5''$ proved to be a good compromise in term of the trade-off between spatial resolution and depth. The outermost ring analyzed is located at $R_{\mathrm{last}} = 95''$.

In a first step we analyze the data to verify the quality of the spectra extracted in all regions: we identify and eliminate bad fibers (e.g. those placed on bright field stars or affected by cross-talk), we interpolate over the spectrum at the position of the brightest sky lines to clean out our spectra. We trim the wavelength range  in each spectrum to avoid the poor instrumental efficiency at the very edges. This leads to a useful wavelength range of 3702-6997\,\AA.

\subsubsection{Emission-line fluxes}
The radiation emitted by each element of volume in a region depends upon the abundance of the elements, determined by the previous evolutionary history of the galaxy, and the local ionization, density and temperature. The most prominent spectral features are emission lines, many of these are collisionally-excited lines. In the V300 grating setup the main emission lines observed are:
[O\,{\textsc{ii}}]\,$\lambda$\,3727\AA\, doublet; [O\,{\textsc{iii}}]\,$\lambda\lambda$\,4959,5007\AA\AA; $\mathrm{H}\beta$; [N\,{\textsc{ii}}]\,$\lambda$\,6548\AA; $\mathrm{H}\alpha$; [N\,{\textsc{ii}}]\,$\lambda$\,6583\AA\, and [S\,{\textsc{ii}}]\,$\lambda\lambda$\,6717,6731\AA\AA.  

We identified two different types of regions according to the intensity of the continuum: (1) regions where the continuum was bright enough so some absorption features were clearly visible and, therefore, the spectrum of the underlying stellar population could be fit using stellar populations synthesis models and (2) regions with fainter and/or noisier continua where no spectral absorption features could be identified.\footnote{The limit in the continuum intensity between the two types of spectra was set to $1 \times 10^{-16}$\,erg\,sec$^{-1}$\,\AA$^{-1}$\,. For those spectra where the continuum was fainter (in either $\mathrm{H}\beta$ or $\mathrm{H}\alpha$) than this number no attempt to carry out a full-spectral fitting to the continuum was attempted.}

We refer to section ~\ref{stellar} for a description of the procedure followed to fit the underlying continuum in the case of the former regions; here we discuss the procedure followed to derive accurate emission line fluxes in regions with negligible or noisy continua. Initially, we have identified 73 H~II regions in our synthetic $\mathrm{H}\alpha$ image but we excluded some of these regions because of their low SNR or because they were later identified as bad columns or field stars.

We first masked out the main emission lines and used a third-order polynomial function to fit the continuum. We then selected two spectral ranges excluding emission lines in three different wavelength ranges [O\,{\textsc{ii}}]\,$\lambda$\,3727\AA; $\mathrm{H}\beta$; $\mathrm{H}\alpha$ regions and measured the continuum level in each of these ranges. In a second step we averaged the two continua and subtracted the result from each emission line spectra in order to obtain a {\it decontaminated} flux.
We analyzed the spectra using the IRAF package {\sc onedspec} and {\sc stsdas} packages. For well isolated emission lines, such as [O\,{\textsc{ii}}]\,$\lambda$\,3727\AA, $\mathrm{H}\delta$, $\mathrm{H}\gamma$\footnote{Note that we detect these lines only in 23 HII Regions, so we not use $\mathrm{H}\delta$ and $\mathrm{H}\gamma$ to derive physical properties.}, [O\,{\textsc{iii}}]\,$\lambda$\,5007\AA, we fit the emission line profile to a single gaussian function. For partially-blended emission lines, such as the [N\,{\textsc{ii}}]\,$\lambda$\,6548\AA; $\mathrm{H}\alpha$; [N\,{\textsc{ii}}]\,$\lambda$\,6583\AA\, triplet, and [S\,{\textsc{ii}}]\,$\lambda\lambda$\,6717,6731\AA\AA\, doublet, we used the task {\sc ngaussfit} that fits simultaneously multiple gaussian functions. We calculated the typical signal-to-noise ratio (SNR) of our data. In the case of [O\,{\textsc{ii}}]\,$\lambda$\,3727\AA\, the SNR ranges between 3.4 and 23. In the case of $\mathrm{H}\beta$ we obtain similar SNR values, ranging between 2.5 and 23. For the part of our spectra where $\mathrm{H}\alpha$ emission line is located the SNR values varies between 6 to 67.

Before analyzing the data in terms of chemical abundances or star formation rates, the observed emission-line fluxes must be corrected for various effects. For example, the presence of dust between the zone of emission and the observer or the possible underlying stellar absorption in the case of the hydrogen recombination lines can alter considerably the intrinsic emission line fluxes. We correct for underlying stellar absorption in the hydrogen Balmer lines using the values obtained from the analysis of the ring spectra (see section ~\ref{stellar} for fitting details). These best-fitting equivalent widths in absorption show very little variation with radius ranging between -1.9 and -2.3\,\AA\ in H$\beta$ and -1.7 and -1.8\,\AA\ in H$\alpha$. We apply an average correction of -2 and -1.7\,\AA, respectively for $\mathrm{H}\beta$ and $\mathrm{H}\alpha$, to the spectra of the 58 H~II regions where the continuum emission was too faint for carrying out a full spectral fitting. The emission fluxes were then corrected for reddening using the $\mathrm{H}\alpha$/$\mathrm{H}\beta$ Balmer decrements after adopting an intrinsic ratio of 2.86 \citep{oster}. Additionally, we verified that in the case of regions of high equivalent width in emission (typically no less than 5\,\AA\, in $\mathrm{H}\beta$) the corrected $\mathrm{H}\beta$/$\mathrm{H}\gamma$ ratios were consistent with the predictions for case-B recombination value at a typical $T_e$ of $\sim$10$^{4}$\,K. In Table 3 we present the results obtained from the analysis of H~II regions and concentric annuli in NGC~5668.

\begin{figure*}[]
\begin{center}
\resizebox{0.49\hsize}{!}{\includegraphics{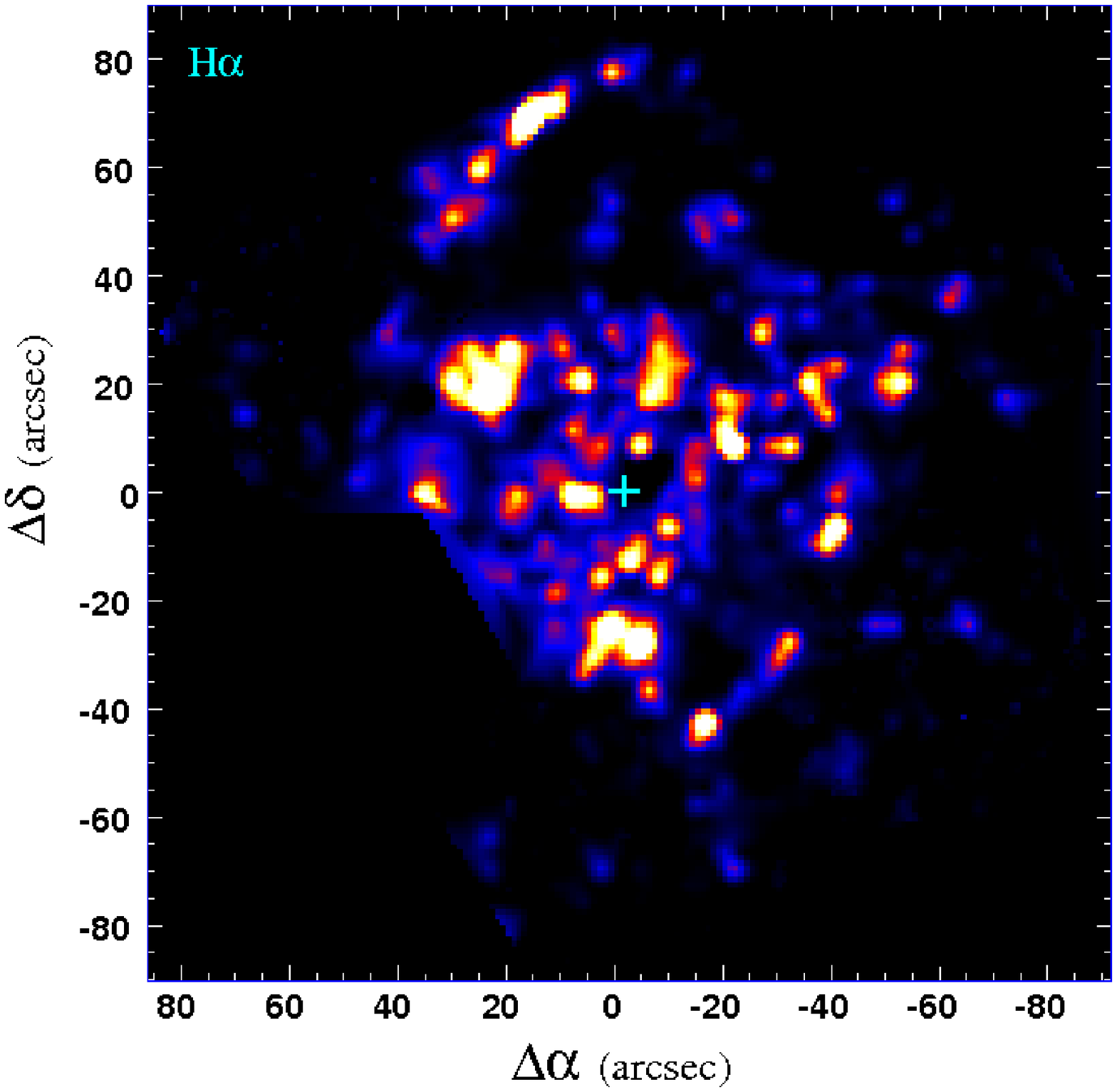}}
\resizebox{0.49\hsize}{!}{\includegraphics{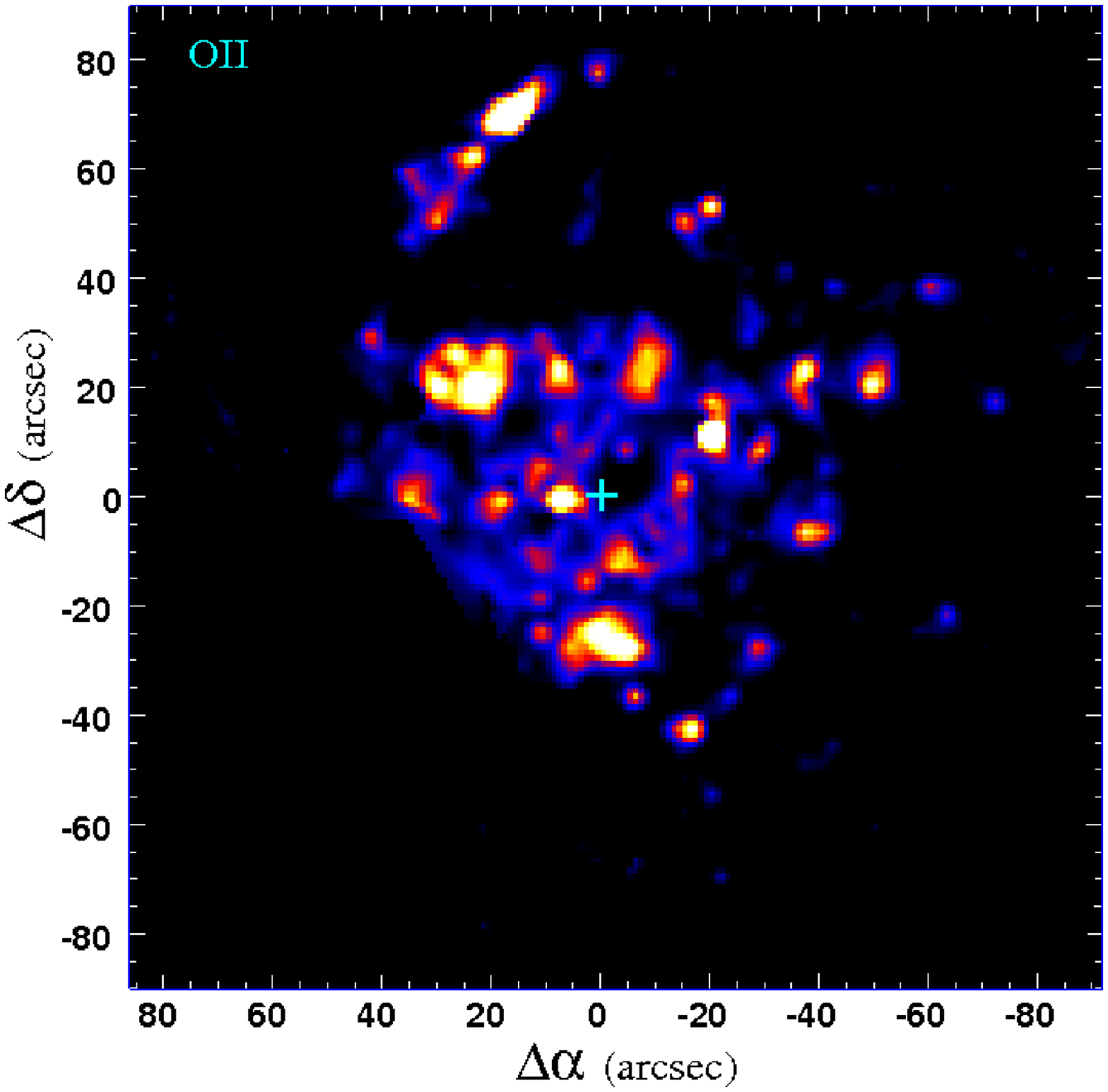}}\\
\resizebox{0.49\hsize}{!}{\includegraphics{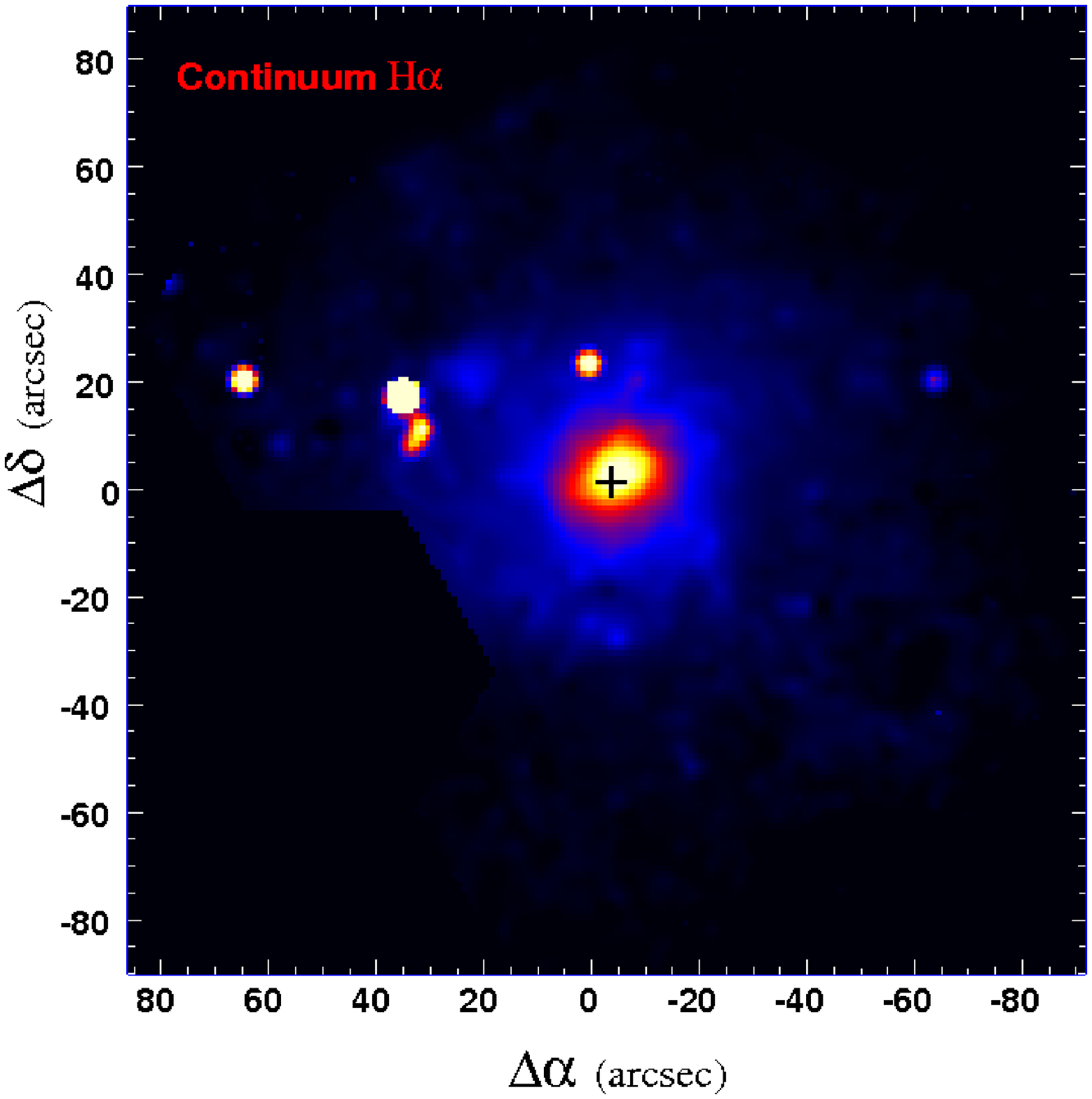}}
\resizebox{0.49\hsize}{!}{\includegraphics{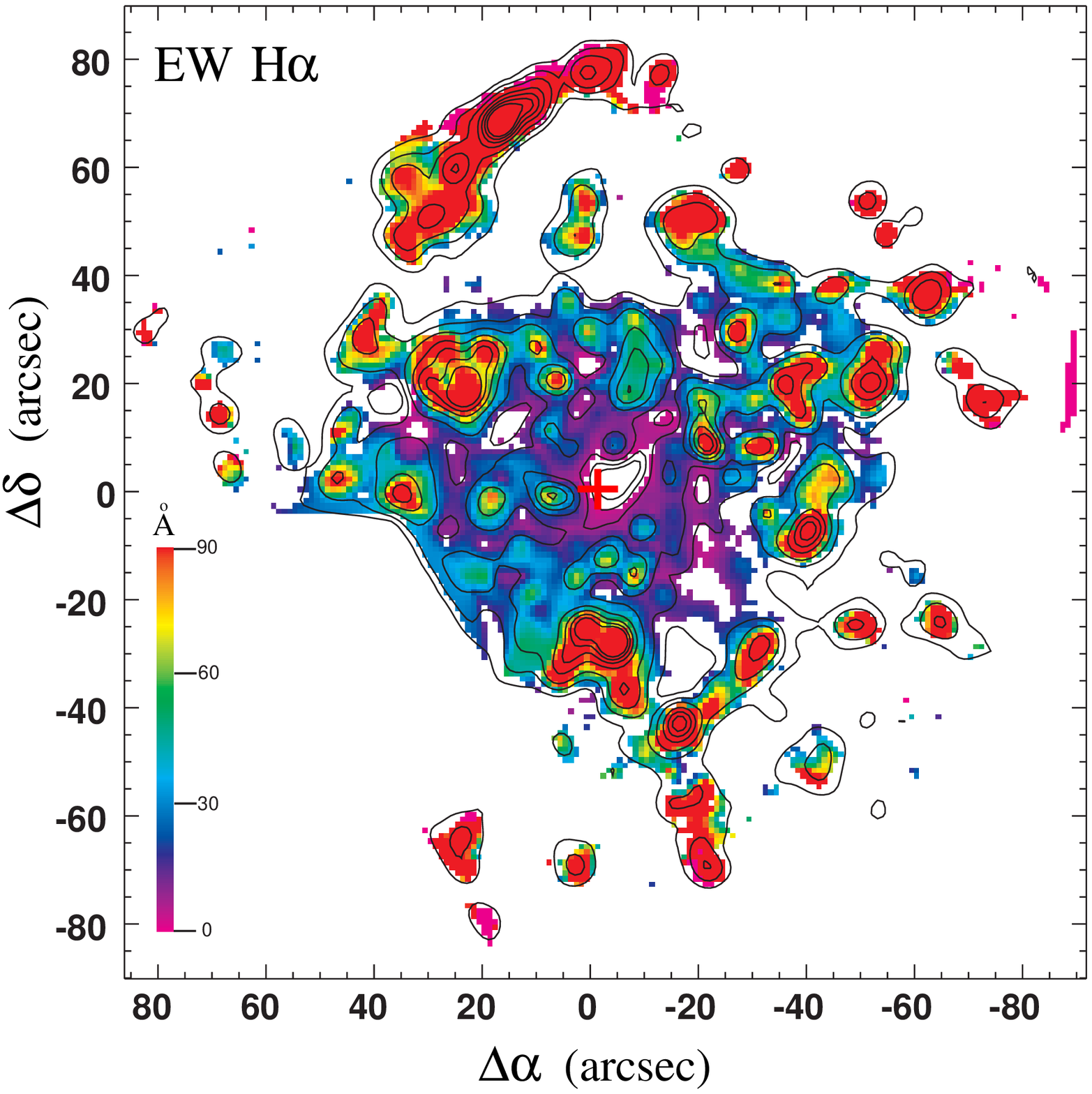}}
\caption[Fig.5]{\footnotesize
NGC~5668 synthetic maps. {\bf Top:} Emission lines maps at $\mathrm{H}\alpha$ (left) and [O\,{\textsc{ii}}]\,$\lambda$\,3727\AA\, doublet (right). {\bf Bottom:} $\mathrm{H}\alpha$ continuum (left) and EW (right); we calculated these maps assuming an average absorption equivalent width in H$_{\beta}$ of -2\,\AA. $\mathrm{H}\alpha$ emission line fluxes in this plot are represented by the isocontours. The mosaic center is marked with a cross. North is up and East is to the left in all cases.\label{maps}}
\end{center}
\end{figure*}

Following a similar analysis on a pixel-by-pixel basis, we also generate maps of ionized-gas extinction, radial velocity, emission line fluxes and stellar absorption equivalent width. In Fig$.$~\ref{maps} we present the maps of the $\mathrm{H}\alpha$ line flux; [O\,{\textsc{ii}}]\,$\lambda$\,3727\AA\, doublet flux; $\mathrm{H}\alpha$ continuum intensity and equivalent width (EW). Note that while the EW(H$\alpha$) in emission gets locally higher as we move towards the outer parts of the disk, this is mainly due to the decrease in the intensity of the adjacent underlying ($R$-band) continuum. Indeed, when we compute the azymuthally-averaged EW(H$\alpha$) from the ring spectra this changes less dramatically, with values ranging from 20 to 50\,\AA\ in all cases except for the very outer rings (see Table~\ref{ewrings}). The very high EW(H$\alpha$) values measured in the rings beyond $\sim$\,70\,\arcsec\ are likely due to strong H$\alpha$ emission associated to the bright H~II complex located in the galaxy Northern spiral arm. Note that in the very outer disk, where the total SFR is low, stochasticity in the number and luminosity of H~II regions might lead to significant fluctuations in the EW(H$\alpha$) compared with values averaged over timescales of a few hundred Myr or even a rotation period.

\subsubsection{Stellar-absorption correction}
\label{stellar}

\begin{figure*}
\begin{center}
\includegraphics[scale=0.9]{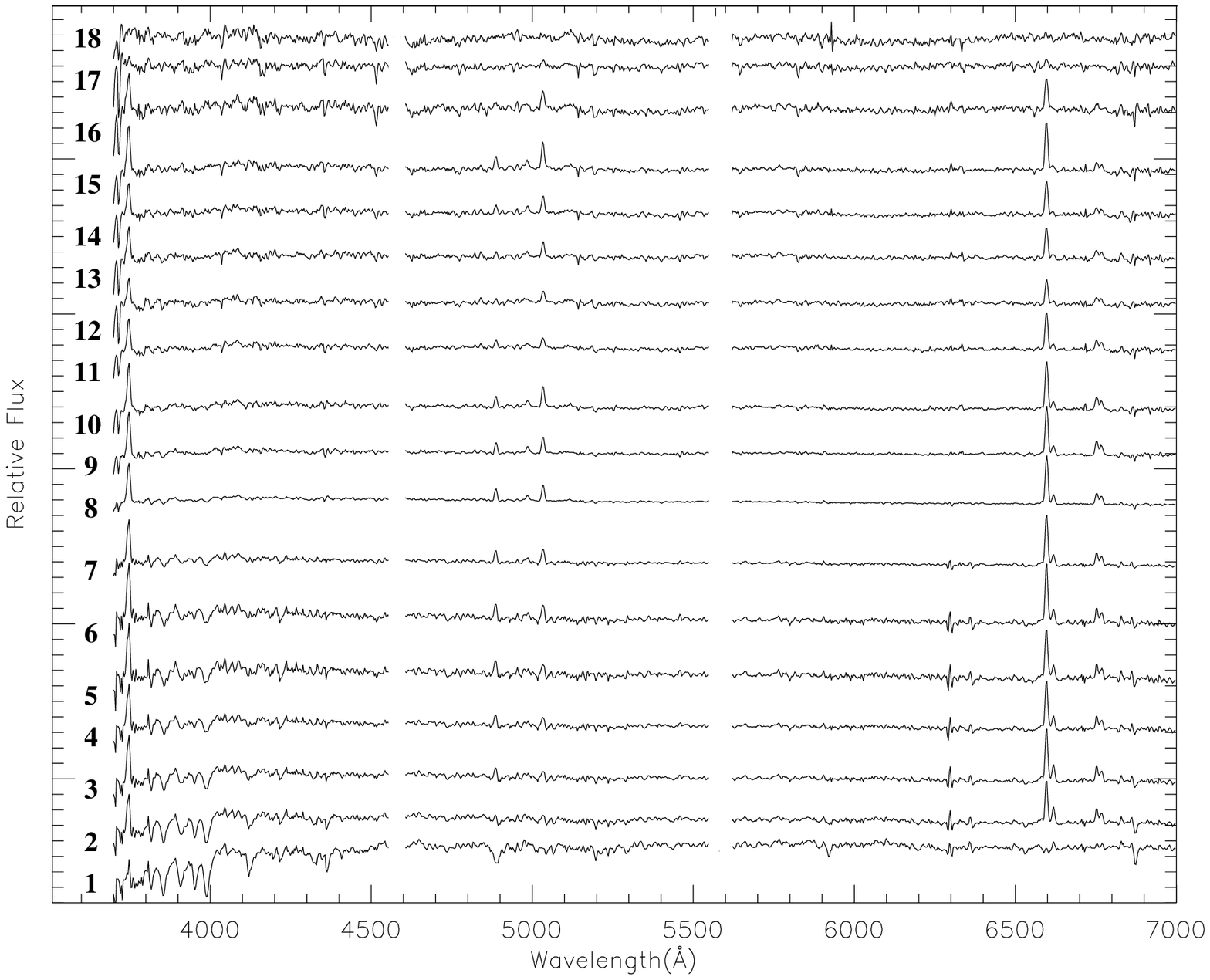}
	\caption[Fig.4]{\footnotesize PPAK spectra extracted of the rings selected in NGC~5668. From bottom to top we show the spectra of rings with increasing radius. The spectra are obtained summing all fibers within each ring. Two bright sky-lines residual ranges are masked. These spectra are plotted in linear scale and shifted in the y-axis for a better visualization. \label{rings}}
\end{center}
\end{figure*}

As discussed above, for the rings and some H~II regions the continuum is sufficiently strong that allows us to clearly identify absorption features in their spectra. In this way we can fit the absorption and emission components in the spectra simultaneously. In this case we made use of a own IDL script that finds the best-fitting stellar population synthesis model among the ones in libraries by \cite{brucha} (hereafter BC03) and \cite{pat06} (also known as MILES). This procedure allowed us to measure intensities for the emission lines already corrected for stellar continuum absorption (nebular continuum emission is negligible at the EW($\mathrm{H}\alpha$) measured anywhere in this object).
Both libraries yield good fits to our spectra, yet the analysis presented here is based on the results obtained using the BC03 library since this provides a good coverage of physical parameters and good spectral resolution and has been more extensively tested than MILES.  We explore the whole range in age and metallicity in the BC03 library for simple stellar populations (SSPs) with a spectral resolution of 3 \AA\, across the whole wavelength range from 3200 \AA\, to 9500 \AA\,. These templates have 11 values in age (5, 25, 100, 290, 640, 900 Myr, and 1.4, 2.5, 11, 13, 17 Gyr) and for each of them six different metallicity values: z=0.0001, 0.0004, 0.004, 0.008, 0.02, 0.05. 
Our fitting program is able to find the model that best fits the underlying stellar population (stellar continuum) in this BC03 stellar library by minimizing the residuals between the model and observed spectra for a set of input parameters. 
The program finds the best fitting parameters for the recession velocity, velocity dispersion, normalization factor and stellar-continuum attenuation in the V band for each region. 

In order to attenuate the stellar continuum as a function of wavelength we adopt a dependence of the optical depth with wavelength of the form $\lambda^{-0.7}$, which, according to the results of \cite{charlot}, is a good approximation to the extinction law in star forming galaxies at these wavelengths. This analysis provides us with the best-fitting SSP model for each spectrum along with best-fitting values for the recession (tropocentric) velocity, velocity dispersion, normalization factor, and continuum dust attenuation. 

Note that the age and metallicities derived are in general highly uncertain but, on the other hand, the H$\alpha$ and H$\beta$ equivalent widths in absorption are much more precise (see \citealt{marmol} for a detailed description on the feasibility in the derivation of the properties of the underlying stellar population). The residual spectrum obtained after the subtraction of the best-fitting SSP model is then used for the subsequent analysis of the emission-line fluxes. 

We calculate the errors in all line fluxes from the rms in the spectra (in the region adjacent to each emission feature) after the best-fitting stellar synthesis model has been subtracted. We assume that these rms measurements were due mainly to photon noise and to sky subtraction. These errors were scaled to those spectra for which the fainter continuum emission prevented carring out a full spectral fitting. Finally, we used the errors in the continuum to calculate the errors associated to the line fluxes \footnote{The errors are calculated from the residual fit-models of the underlying stellar population; we are assuming that these residuals are primarily due to photon noise and sky subtraction. We note that we have pointings with a slightly different spatial resolution and several combinations of exposure time and number of individual exposures; for this reason the continuum level emission and the associated noise was derived for each pointing separately. The sky subtraction to the spectra of NGC 5668 was done, see the data reduction steps in \cite{sanchez06}, by median average the 331 sky-spectra. The resulting spectrum was then subtracted from the cube of NGC 5668. The analysis of these sky-subtracted spectra showed that the error in the continuum level emission increases linearly with the number of fibers (but not to the square root). This indicates that for apertures including a large number of fibers the total error is dominated by the sky-subtraction error.}.

\subsection{Integrated spectrum}

\begin{figure*}
\begin{center}
\includegraphics[scale=0.9]{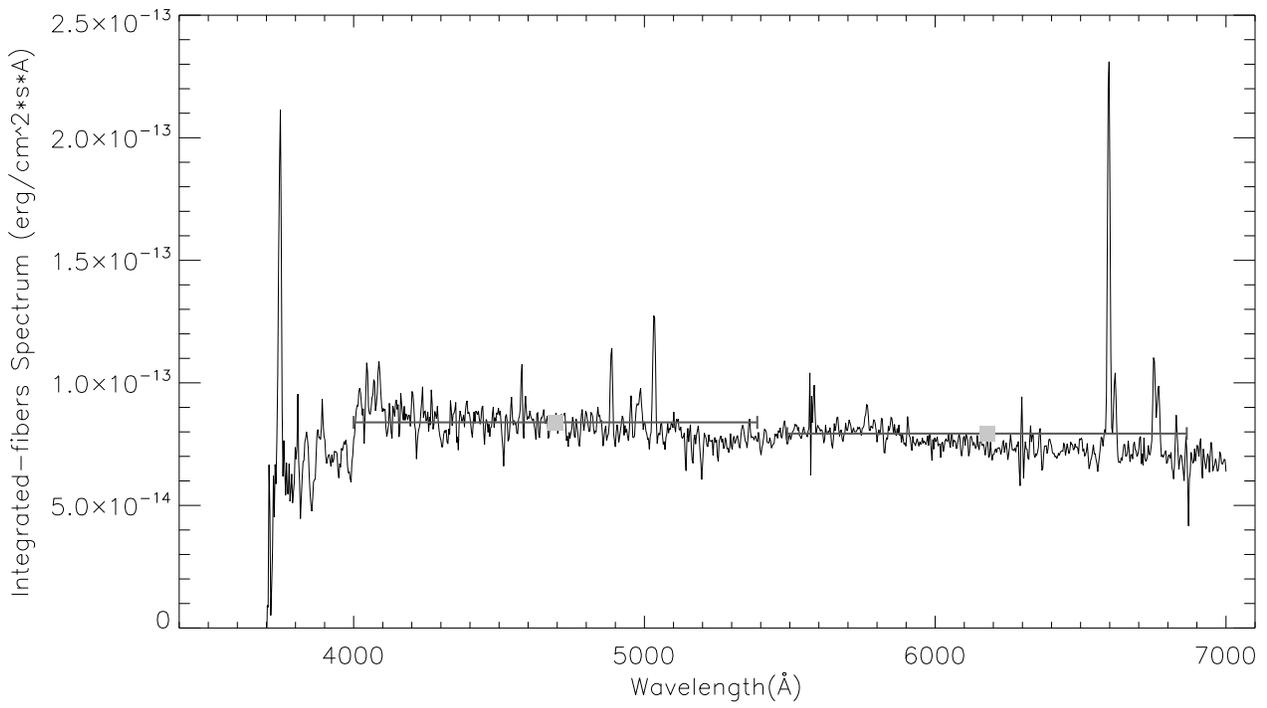}
	\caption[Fig.5]{\footnotesize The integrated spectrum of NGC~5668 is shown as a black line. The SDSS  {\it g'} and {\it r'} band photometry data are shown as grey squares. Horizontal error bars represent the FWHM of each filter.
\label{integrated}}
\end{center}
\end{figure*}

Some advantages of using IFUs are that all the spectra are obtained simultaneously and that we can use the IFU as a large-aperture spectrograph. 
From our datacube we add up the spectra of all the fibers to create an integrated spectrum of NGC~5668 (see Fig$.$~\ref{integrated}).
In this figure we also compare this spectrum with the corresponding fluxes derived from the Sloan Digital Sky Survey photometry. Note that the {\it g'} and {\it r'} bands overlap completely with our spectral range. Since the spectroscopy data were not taken under photometric conditions and given that some flux losses are expected as part of the PPAK observations of our spectrophometric standard star, a 8\% difference was found between the two datasets. This offset, which is wavelength independent, has been already applied to the spectrum shown in Fig$.$~\ref{integrated} although was not applied to our spectral datacube and the line fluxes measured hereafter. Similar wavelength-independent offsets are also found in other works based on PPAK spectroscopic observations \citep{fabian}. In any case, this does not have any impact on the physical properties derived (dust attenuation, oxygen abundance, electron density) as these are determined by the ratio of emission-line fluxes alone. In Table 3 (last row), we present the emission-line flux ratios found for the integrated fibers spectrum of NGC~5668.

\subsection{Imaging}
\label{imag}
Prior to the measurement of the surface brightness profiles in NGC~5668, foreground stars, background galaxies and cosmetic artifacts had to be masked out from our multi-wavelength imaging data. Radial profiles were obtained using the IRAF task {\sc ellipse}. We measured the mean intensity within concentric circular annuli centered on the galaxy's nucleus, using a radial increment of 6\arcsec\, between adjacent annuli. Photometric errors were derived as explained in \cite{gilmad} and \cite{mun09b}. The uncertainty of the mean intensity at a given radius is computed as the quadratic sum of two terms: the poissonian noise of the galaxy's light and the error in the sky level. The latter being the result of the combination of both pixel-to-pixel variations an large-scale background errors.

\section{Results} 


\subsection{Attenuation}
More than 80 years have passed since Trumpler's discovery of color excesses provided the first definitive proof of the existence of interstellar dust \citep{trumpler}. However the nature of interstellar dust still remains unclear and how dust both reddens and attenuates the light from the stars is one of the least understood of the physical phenomena which take place in galaxies.

The dust content of galaxies is a critical issue given its important impact on the observational properties of galaxies (mainly in the optical and UV). The extinction in a galaxy depends first on the amount of dust and its composition, and also on the distribution of dust relative to the light sources \citep{calz01}. For instance, some authors favor a foreground screen dust geometry model (\citealt{calz94}, \citealt{calz96}), while others propose hybrid models with the dust partially distributed in a foreground screen and partially concentrated in the star-forming regions \citep{charlot}.
In this subsection we discuss the methods used to estimate dust attenuation. We calculate the attenuation of the stellar continuum from the UV data available on NGC~5668 and these estimates are then compared with the ionized-gas attenuation derived from the Balmer decrement in both individual H~II regions and concentric annuli.

\subsubsection{UV-continuum attenuation}
\label{UVsec}
In this case we estimate the attenuation of the stellar continuum using UV data alone. Because the UV radiation is preferentially emitted by young stars ($\sim$100\,Myr) and because the dust is most efficient in attenuating UV light, rest-frame UV observations can lead to incomplete and/or biased reconstructions of the recent star formation activity and star formation history of galaxies where dust absorption is expected to be significant, unless proper corrections are applied.
Radiative transfer models suggest that the total-IR (TIR, 3-1100$\mu$m) to UV luminosity ratio method (i.e. \citealt{buat92}; \citealt{xu}; \citealt{meurer95}; \citealt{meurer99}) is the most reliable estimator of the dust attenuation in star-forming galaxies because it is almost completely independent of the extinction mechanisms (i.e. dust/star geometry, extinction law, see \citealt{xu}; \citealt{meurer99}; \citealt{gordon}; \citealt{witt}).
However, this would require having both far-infrared data at the same resolution as the UV data. 
In the case of galaxies at the distance of NGC~5668 this would provide very little spatial information on the radial variation of the dust attenuation even if data from state-of-the-art infrared facilities such as Herschel would be available. 
Even though IRAC and MIPS images for NGC~5668 are available in the Spitzer archive, so theoretically, we can obtain profiles in 8, 24, 70 and 160 $\mu$m but the poor resolution at 160 $\mu$m (38\arcsec) would yield a total-IR profile with just a couple of data-points.
Under certain circumstances, the UV attenuation can be indirectly estimated using the slope of the UV spectrum (denoted as $\beta$), in the sense that redder UV colors are indicative of larger attenuations. This so-called IRX-$\beta$ relation has been widely used in starburst galaxies, where most of the UV light comes from newly-born stars (\citealt{calz94}; \citealt{heckman}; \citealt{meurer99}). In normal star-forming spirals, more evolved stars can also contribute to the UV flux, shifting the IRX-$\beta$ relation to redder UV colors and increasing the overall scatter (\citealt{bell02}; \citealt{buat05}; \citealt{seibert}; \citealt{cortese06}; \citealt{gil07}; \citealt{dale07}). The use of radial profiles instead of integrated measurements seems to reduce the global scatter (\citealt{boissier}; \citealt{mun09a}).

Here we apply the IRX-$\beta$ relation provided by \cite{mun09a} to FUV-NUV measurements obatined from the GALEX images of NGC~5668. This relation was specifically calibrated for normal nearby spirals, so it should be applicable to NGC~5668 as well.
However, before applying this technique to subregions, we checked this assumption using the total galaxy IR luminosity of NGC~5668 from the IRAS fluxes quoted by \cite{wang} leading to a total IR flux of $\log(F_{\mathrm{TIR}})=13.64 \times10^{-23}$ (erg\,s$^{-1}$\,cm$^{-2}$). This value combined with the FUV flux obtained from our GALEX FUV image leads to $\log(\mathrm{TIR}/\mathrm{FUV})\sim0.18$. On the other hand, for a global UV color of FUV-NUV=0.27, the IRX-$\beta$ yields $\log(\mathrm{TIR}/\mathrm{FUV})\sim0.39$. This is somewhat larger than the observed value, but still lies within the 1$\sigma$ scatter around the IRX-$\beta$ relation of \cite{mun09b}.
Thus, using the radial profiles described in Section~\ref{imag}, we determined the radial variation of the FUV-NUV color in steps of 6\arcsec. The UV color was then used to obtain profiles of TIR/FUV and TIR/NUV ratios using the prescriptions of \cite{mun09b}. These values were then transformed into dust attenuation in the FUV (A$_{FUV}$) and NUV (A$_{NUV}$) by means of the recipes by \cite{cortese08}, which take into account the IR excess associated to the extra dust heating due to the light from old stars. The resulting attenuation profiles are shown in Table~\ref{UVtable}. Note that the absolute uncertainty in the TIR/FUV ratio from the FUV-NUV is larger than the relative change in such ratio across the galaxy, see \cite{mun09b}.

Now that we have derived the extinction in the UV, we can calculate the extinction in other bands using an attenuation curve. Such reddening corrections have normally been undertaken using a number of extinction curves, including those of \cite{seaton} (in the UV), \cite{cardelli}, \cite{savage}, \cite{ardeberg} and \cite{fitz}. These \textquotedblleft standard\textquotedblright \,curves are often used interchangeably, on the understanding that they should give broadly similar results. The parameterized curve by Cardelli, Clayton, \& Mathis (1989) law is commonly used to fit the extinction data both for diffuse and dense interstellar medium.
In our case, the continuum attenuation is obtained using at different wavelengths two different extinction laws, \cite{cardelli} and \cite{calz01}. We based our calculation on the NUV data, adopting the relation with the color excess E(B-V) given by \cite{cardelli}, E(B-V)= A$_{NUV}$/8.0 and for \cite{calz94}, E(B-V)= A$_{NUV}$/8.22 \citep{gil07}. We then adopt the average values of R$_{V, Cardelli}$ = 3.1 and R$_{V, Calzetti}$ = 4.05. For sake of clarity Fig.7 plots only the results obtained in the case of the Calzetti extinction law as the two curves are consistent within the errors (but offset by -0.09 magnitudes in the case of the Cardelli law).

\subsubsection{Ionized-gas emission-lines attenuation}

We now made use of the Balmer decrement in individual H~II regions and concentric annuli for determining the amount of dust extinction in the ionized-gas emission along the line of sight. The relative ratios of the Balmer lines of hydrogen are often used as extinction indicators due to the fact that they are observationally convenient (being in the optical band), strong and their intrinsic relative flux ratios are fairly well determined from atomic theory. Under Case B conditions with temperatures $\sim$10,000 K and electron densities $\sim$100 cm$^{−3}$, the theoretical $\mathrm{H}\alpha$/$\mathrm{H}\beta$ ratio should be close to 2.86 \citep{oster}. Since extinction is more severe in $\mathrm{H}\beta$ than at $\mathrm{H}\alpha$ wavelengths, the net effect is to increase the observed Balmer-line ratio. As we discussed above, the results are dependent of the dust model of choise. Here we assume that the dust is located in a uniform screen between us and the gas and that the extinction law in the optical is similar to the Galactic one, as parametrized by Cardelli et al$.$ (1989).

\subsubsection{Optical-continuum attenuation}
\begin{figure*}
\resizebox{0.60\hsize}{!}{\includegraphics[scale=0.73]{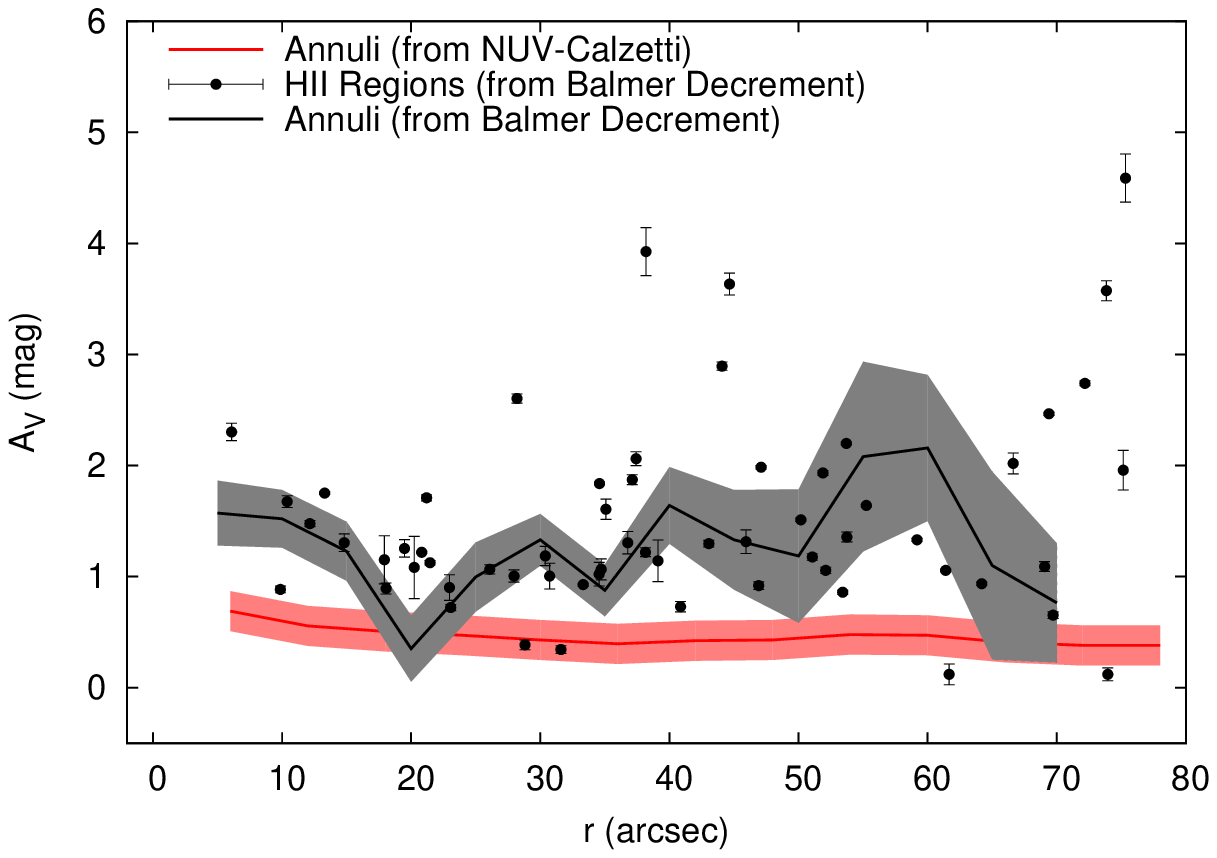}}
\resizebox{0.41\hsize}{!}{\includegraphics{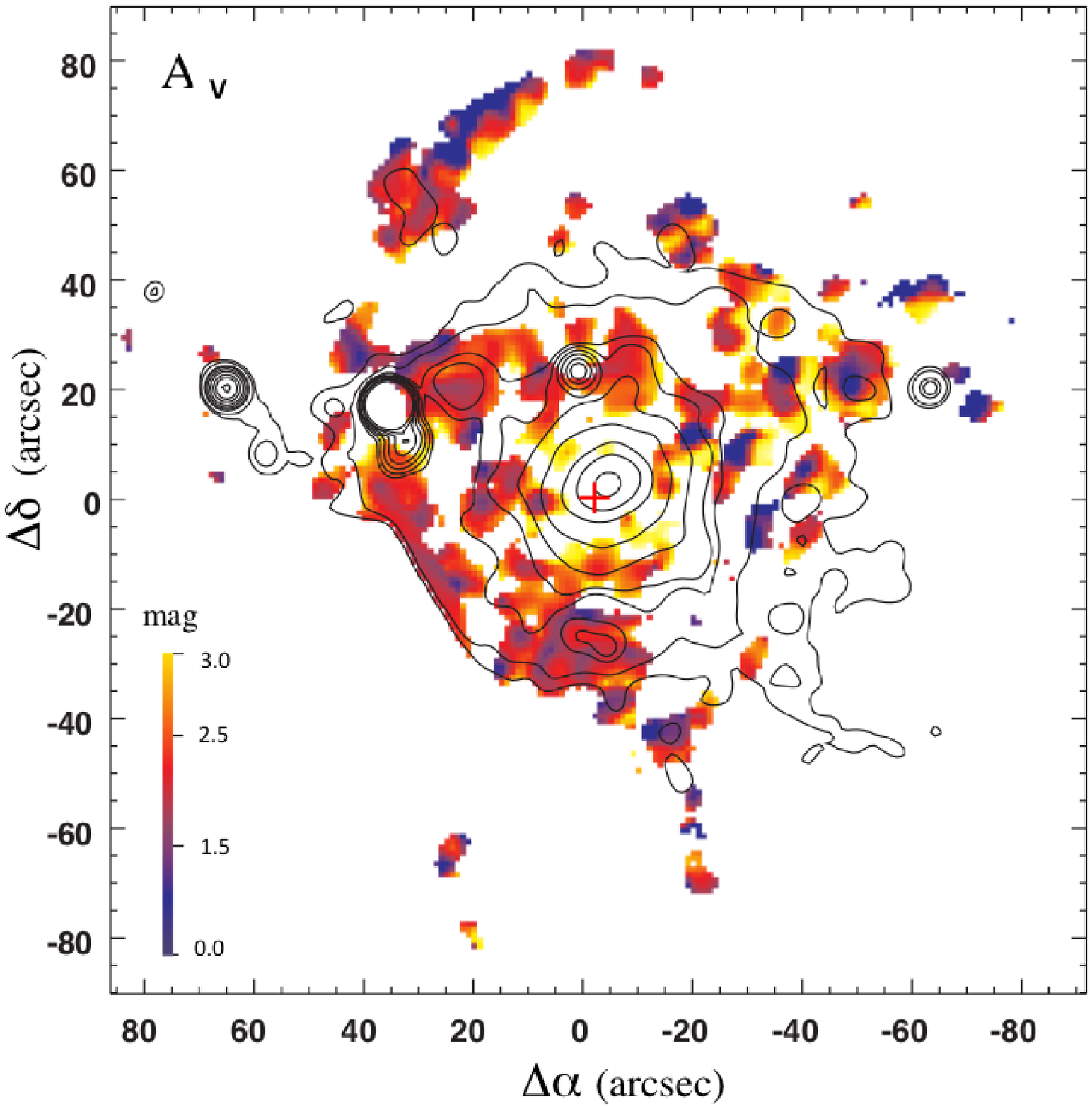}}
	\caption[Fig.6]{\footnotesize
	{\bf Left}: Representation of the radial variation of dust attenuation. Filled circles represent the values obtained from individual H~II regions based on the Balmer decrement measurements. The black line represents Balmer decrement data for the concentric annuli. The red line is the attenuation (and $\pm 1 \sigma$ error band) calculated from the slope of the UV continuum (or, equivalently, the FUV-NUV color) adopting a Calzetti extinction law. {\bf Right}: Dust extinction map (MW + internal) of the ionized gas in NGC~5668 computed from the $\mathrm{H}\alpha$/$\mathrm{H}\beta$ ratio. The H$\alpha$ continuum values are represented as isocontours. North is up and East is to the left. \label{atten}}
\end{figure*}

The last method that we have used for the extinction is based in our IDL fitting-program for the underlying stellar population (see~\ref{spec}). From the output values for each concentric annuli we find a mean value of A$_{V}$$\sim$1.7 mag. This value has a very large uncertainty due to the limited wavelength range of our spectroscopic observations which leads to strong age-metallicity-extinction-SFH degeneracies (see \citealt{gilmad02}).
The results are summarized in Fig$.$~\ref{atten}. In general, the ionized-gas dust attenuation has a mean value of $\sim$1.5 magnitude, which is in agreement to what is found in other spiral disks \citep{gil07}. We find that the gas attenuation is larger than the continuum attenuation by $<$E(B-V)$_{continuum}$ $>$ =0.36 $\times$ $<$ E(B-V)$_{ionized-gas}$ $>$ when the Calzetti attenuation law is adopted; note that ionized-gas color excess (derived from the Balmer decrement) is corrected for a Milky Way foreground reddening, $E(B-V)=0.037$. This is somewhat expected given that, as shown in \cite{calz00} and \cite{stasinska}, in nearby spiral galaxies there is evidence that stars, gas and dust are typically decoupled (see e.g$.$ \citealt{maiz}) so the attenuation inferred from the $\mathrm{H}\alpha$/$\mathrm{H}\beta$ ratio is typically higher than that inferred from the spectral continuum of the same wavelength (e.g., \citealt{calz94}; \citealt{mayya}; \citealt{calz97}). 
\cite{pogg99} and \cite{pogg00} explain such results as due to selective-dust-extinction effects where a large fraction, but not all, of the dust in galaxies is associated with star formation regions, which absorb a significant fraction of the light emitted by the young stars.

\subsection{Electron density}
The electron density, $N_{e}$, is one of the key physical parameters that characterize gaseous nebulae. Most of the density estimates found in the literature are based on measurements of a sensitive emission-line ratio. In the presence of internal variations of electron density, however, these single line-ratio measurements may not be representative of all ionizing zones. The measurements are based on the fact that a relation exists between the collisional de-excitation of atoms and electron density, \citep{oster}. If an ion emits similar amounts of energy from different energy levels at nearby wavelengths, then the ratio of the emission line intensities can be used to obtain an estimate of the electron density and there are two line ratios that can be used: [O\,{\textsc{ii}}]\,$\lambda\lambda$\,3726,3729\AA\AA\, or [S\,{\textsc{ii}}]\,$\lambda\lambda$\,6717,6731\AA\AA. In our case, we make use of the [S\,{\textsc{ii}}]\,$\lambda\lambda$\,6717,6731\AA\AA\, line ratio and the electron density of the region responsible for the [S\,{\textsc{ii}}] emission can be determined as:

\begin{equation}
R([S\,{\textsc{ii}}])=\frac{I([S\,{\textsc{ii}}]\,\lambda\,6717)}{I([S\,{\textsc{ii}}]\,\lambda\,6731)}\simeq1.49\left(\frac{1+3.77x}{1+12.8x}\right)
\end{equation}

\noindent where $x$, the density parameter, is defined as $x=10^{-4}\cdot n_{e}\cdot t^{-1/2}$. Solving this equation for the density parameter we can obtain $N_{e}$ by assuming $T_{e}$ = $T_{[S\,{\textsc{ii}}]}$/10$^{4}$, where $T_{[S\,{\textsc{ii}}]}$ is the electron temperature of the region responsible for the [S\,{\textsc{ii}}] emission (\cite{mccall}). Thus, in order to properly estimate this electron density a previous knowledge on the electron temperature of the region responsible for the [S\,{\textsc{ii}}] emission lines is required. Assuming the calibration of T[NII] as a function of the $R_{23}$ line ratio (see equation~\ref{r23}) given by \cite{thur} and an average difference of 3000 K between T[N\,{\textsc{ii}}] and T[S\,{\textsc{ii}}] (with the former being lower; \cite{garnett}) we estimated $T_{[S\,{\textsc{ii}}]}$ in the equation above. Note that although this estimate for the temperature is not very precise (as it does not rely on the use of temperature-sensitive line ratios) is accurate enough for correcting the densities obtained from the [S\,{\textsc{ii}}]\,$\lambda\lambda$\,6717,6731\AA\AA\, line ratio for temperature effects. The large uncertainties associated to the determination of T[S\,{\textsc{ii}}] prevent us from extracting further conclusions from the density measurement derived. Despite that, we find a mean value for $N_{e}$ of 190 cm$^{-3}$ for the H~II regions, in agreement with the typical value densities in H~II regions (N$_{e}\sim$\,10$^{2}$\,cm$^{-3}$, \citealt{oster}).

\begin{figure*}
	\centering
	\includegraphics[scale=1.4]{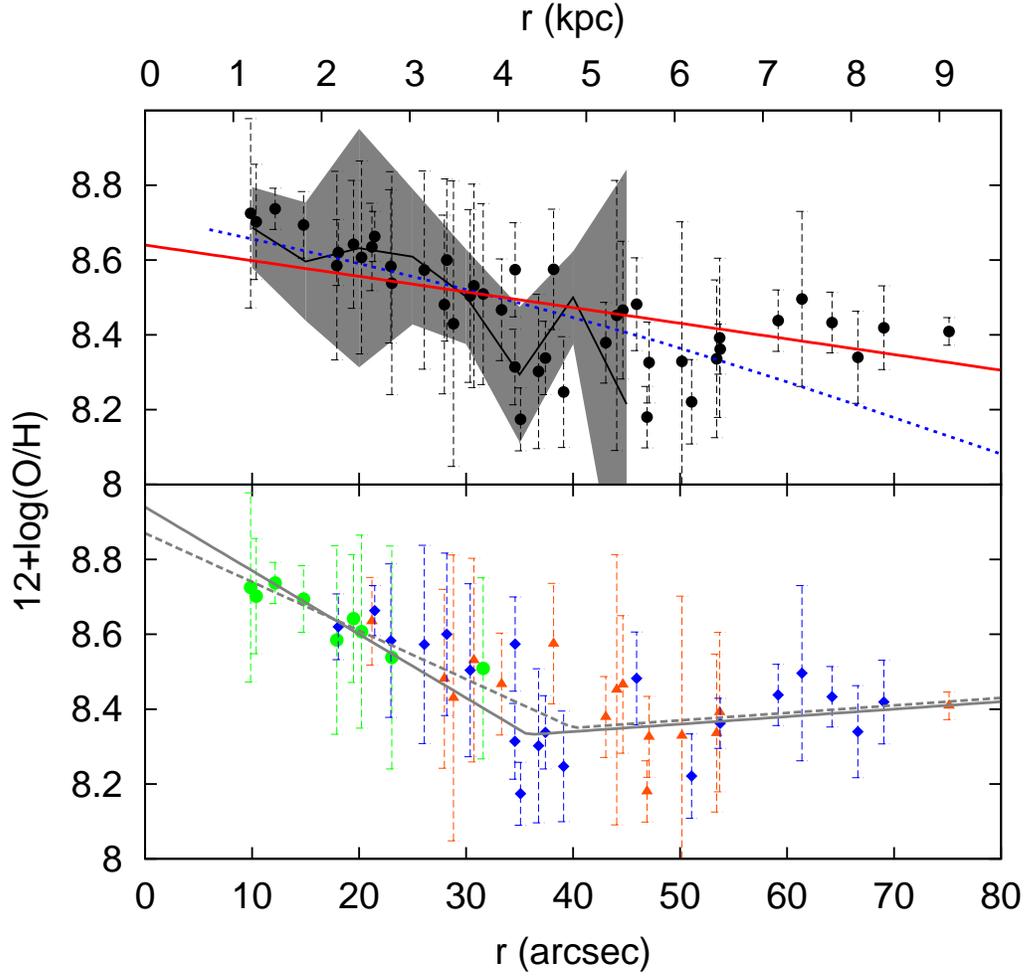}
	\caption[Fig.7]{\footnotesize
Radial abundance gradient in NGC~5668, where filled-symbols correspond to individual H~II regions while the grey-shaded area represents the values obtained for the concentric annuli. Top panel: The red line symbolizes the linear regression fit to the H~II regions data. We obtained a single gradient of −0.035 $\pm$ 0.007 dex/kpc. The blue dashed line represents the gradient obtained from the models of Boissier \& Prantzos (1999, 2000), with an offset of −1 dex in the metallicities plotted (see text for details). Bottom panel: In this plot we draw two types of fits: the grey line reproduces a double fit weighted by the errors; the gradient has a value of −0.140 $\pm$ 0.016 for inner part and 0.002 $\pm$ 0.019 dex for outer part. The grey dashed line reproduces the unweighted double fit that yields gradients values of −0.1073 $\pm$ 0.0165 (inner) 0.002 $\pm$ 0.019 (outer). The color coding of the points represents the different line ratios used to calculate the metallicity for the central point of the ratios probability distribution (based on Kewley \& Dopita 2002 recipe): green points are calculated from [N\,{\textsc{ii}}]/[0\,{\textsc{ii}}], red triangles represent the values obtained with [N\,{\textsc{ii}}]/$\mathrm{H}\alpha$ and blue diamonds show $R_{23}$-based values. Note that the presence of three uncertain points (regions 13, 17, 59 have an error major of 0.3) has not influence on the fitting parameters values.
\label{met} }
\end{figure*}

\subsection{Chemical abundances}

H~II regions identify the sites of recent massive star formation in galaxies. The rapid evolution of these stars, ending in supernovae explosions, and the subsequent recycling of nucleosynthesis products into the interstellar medium, make of H~II regions essential probes the present-day chemical composition of star-forming galaxies across the Universe. The study of nebular abundances is therefore crucial for understanding the chemical evolution of galaxies. Obtaining a direct measurement of chemical abundances in H~II regions requires a good estimate of the electron temperature. Unfortunately, this implies detecting very faint auroral lines such as [O\,{\textsc{iii}}]\,$\lambda$\,4363\AA\ or [N\,{\textsc{ii}}]\,$\lambda$\,5755\AA, that are very faint at the abundance levels of most spiral disks. Thus, we must use instead calibrations based on the predictions of photo-ionization models or on empirical measurements for strong lines. Among the various strong-line methods, the $R_{23}$ indicator originally proposed by \cite{pagel79} stands out as arguably the most popular. 

\begin{equation}
R_{23}=\frac{f([O\sc{III}] \lambda\lambda 4959, 5007)+f([O\sc{II}] \lambda \lambda3726, 3729)}{f(H\beta)}
\label{r23}
\end{equation}

Like all strong-line diagnostics, $R_{23}$ as an abundance indicator has a statistical value, based on the fact that the hardness of the ionizing radiation correlates with metallicity. Numerous calibrations of $R_{23}$ in terms of the nebular chemical composition can be found in the literature.
We choose as best set of values for the oxygen abundance those given by the \cite{kewdop} recipe, which basically adopts a calibration of the [N\,{\textsc{ii}}]/[O\,{\textsc{ii}}] ratio for oxygen abundances above half solar and their own calibration of $R_{23}$ for abundances below that value\footnote{We used the value of $12 + log(O/H)_{\odot} = 8.93$ as in Kewley \& Dopita (2002).}. The $R_{23}$ abundance diagnostic depends strongly on the ionization parameter, for this reason we also calculated it following the indications given in Kewley and Dopita (2002). These authors used a combination of stellar population synthesis and photoionization models to develop a set of ionization parameter and abundance diagnostics based only on the use of strong optical emission lines. These techniques are applicable to all metallicities. In particular, for metallicities above half solar, the ratio [N\,{\textsc{ii}}]/[O\,{\textsc{ii}}] provides a very reliable diagnostic since it is ionization-parameter independent and does not have a local maximum. This ratio has not been used historically because of concerns about reddening corrections. However, the use of classical reddening curves is sufficient to allow this [N\,{\textsc{ii}}]/[O\,{\textsc{ii}}] diagnostic to be used with confidence as a reliable abundance indicator. Note that the calibration of the [N\,{\textsc{ii}}]/[O\,{\textsc{ii}}] line ratio as a diagnostic of $12 + log(O/H)$ relies on the dependence of the (N/O) abundance ratio on the oxygen abundance. Thus, no attempt has been made in this work to derive the Nitrogen abundance in this work. 

The iterative method used as part of this recipe allows both these parameters to be obtained without the need of using temperature-sensitive line ratios involving very faint emission lines, that are particularly elusive in the case of spiral disks. One drawback of using $R_{23}$ (and many other emission-line abundance diagnostics; see \citealt{perez}) is that it depends also on the ionization parameter {\itshape q} defined here as $q = S_{H0} / n$, where $S_{H0}$ is the ionizing photon flux through a unit area, and {\itshape n} is the local number density of hydrogen atoms. 
Some calibrations have attempted to take this into account (eg. \citealt{mcgaugh}), but others do not (eg. \citealt{zari}). Another difficulty in the use of $R_{23}$ and many other emission-line abundance diagnostics is that they are double valued in terms of the abundance ($12 + log(O/H)$). Thus, one of the main difficulties in its adoption is related to the necessity of locating which of two branches (upper and lower) a given H~II region belongs to, since $R_{23}$ is degenerate \citep{bresolin09a}. This is because at low abundance the intensity of the forbidden lines scales roughly with the chemical abundance while at high abundance the nebular cooling is dominated by the infrared fine structure lines and the electron temperature becomes too low to collisionally excite the optical forbidden lines. When only double-valued diagnostics are available, an iterative approach which explicitly solves for the ionization parameter, as the one used here that is based on the Kewley and Dopita logical flow diagram, helps to resolve the abundance ambiguities. In Table 7 we present the results obtained. The metallicity values vary between $12 + log(O/H) = 8.15$ and $12 + log(O/H) = 8.7$ (i.e. from approximately 1/3 the solar oxygen abundance to nearly the solar value). The lack of some regions in this table is due to the poor signal-to-noise or simply not detection of $\mathrm{H}\beta$ emission (12 regions).

Simulations of the line flux errors were carried out in order derive the errors in the oxygen abundances. These simulations assumed a gaussian probability distribution for each line flux and that these were not correlated between the different lines. The resulting errors, which are shown in Table 7, could then be considered as
upper limits to the actual line-flux errors (where a contribution of correlated errors is expected).

In order to quantify the abundance gradient, we have carried out three linear regressions to the data-points, the first one is a fit to all points weighted by their errors ($\chi^{2}_{red}$ = 1.45); the gradient has a value of $-0.035\, dex/kpc$, ($-0.0042\, dex/arcsec$). The second and the third fits are double fits to the data with a free parameter, the radius of break, but one is weighted by the errors (black solid line) and the other is unweighted (black dashed line; yield both $\chi^{2}_{red}$ = 0.54). 

The Kewley \& Dopita recipe allow us to make use of the most optimal abundance indicator among [N\,{\textsc{ii}}]/[O\,{\textsc{ii}}], [N\,{\textsc{ii}}]/$\mathrm{H}\alpha$ and $R_{23}$. We find that the three abundance diagnostics (showed in the bottom panel of Fig$.$~\ref{met} with different colors) are homogeneously mixed, Fig$.$~\ref{met}. Thus, the [N\,{\textsc{ii}}]/$\mathrm{H}\alpha$ and $R_{23}$-based abundance values are well mixed and can be derived at almost any galaxy radius, while the [N\,{\textsc{ii}}]/[O\,{\textsc{ii}}] indicator is lost in some H~II regions, mostly at large radii; this is, as pointed out in \cite{kewdop}, because nitrogen is predominantly a primary nucleosynthesis element in the range $12 + log(O/H) \le 8.6$ (see also Fig$.$3 in \cite{kewdop}) and the calibration of the (N/O) abundance ratio and the [N\,{\textsc{ii}}]/[O\,{\textsc{ii}}] line ratio is not sensitive to the the oxygen abundance under these circumstances.

The results are summarized in Table~\ref{tabmet} and Fig$.$~\ref{met} where we plot the three gradients as a function of de-projected galactocentric radius. We find that inwards r\,$\sim36''$ ($\sim$4.4 kpc) the O/H ratio follows a exponential profile with a slope of −0.140 $\pm$ 0.016 (dex/kpc) and $12 + log(O/H)_{r=0} \simeq 8.9$, similar to the normalized radial gradient found in other spiral disks. The outer abundance trend flattens out to an approximately constant value of $12 + log(O/H)_{r=0} \simeq 8.27$ (with a slight gradient of 0.002 $\pm$ 0.019 (dex/kpc)) and could even reverse (see Section~\ref{disc} for a discussion on the possible causes for such flattening). 
The abundance gradient derived, $-0.035 dex/kpc$ ($-0.0042 dex/arcsec$), is somewhat shallower than the one for the Milky Way ($-0.08 dex/kpc$; \citealt{boipra} and references therein). A similar trend was found from the analysis of the spectra of the concentric annuli. The analysis of these spectra was limited to the innermost nine annuli because beyond this ring both the continuum and line emission become too faint to derive reliable emission-line fluxes. In Fig$.$~\ref{met} we plot the change in abundance for all annuli with a solid black line (bottom panel) where the uncertainties associated are represented by the grey-shaded area. The blue-dashed line represents the metallicity profile predicted by the best-fitting model of Boissier \& Prantzos (1999, 2000; hereafter BP2000 models). This profile has been shifted by −1 dex to match the metallicity scale derived from our spectroscopic data. Note that this kind of zero-point offsets in metallicity are not unexpected due to the significant uncertainties in the yields used in the disk evolution models (see discussion in \citealt{mun11}); besides, even the empirical oxygen abundances can be subject to large systematic offsets \citep{moustakas}. The relative changes in metal abundances and therefore their radial profiles, on the other hand, are much more robust to these unknowns.

\subsection{Galaxy disk modeling}
In order to gain further insight into the evolution of NGC~5668, we have fitted its multi-wavelength surface brightness profiles with the BP2000 models. These models describe the chemical and spectro-photometric evolution of spiral disks as a function of only two variables: the dimensionless spin parameter, $\lambda$, and the circular velocity in the flat regime of the rotation curve, $V_\mathrm{C}$. Within these models, galactic disks are simulated as a set concentric rings that evolve independently one from each other (for simplicity, radial mass or energy flows are not considered). The gas infall rate at each radius decreases exponentially with time, with a time-scale that depends on both the total mass of the galaxy and the local mass surface density at that radius. Once in the disk, gas is transformed into stars following a Kennicutt-Schmidt law multiplied by a dynamical term, which accounts for the periodic passage of spiral density waves. The mass distribution of each new generation of stars follows a \cite{kroupa} initial mass function (IMF). The finite lifetimes of stars of different masses is taken into account; when they die, they inject metals into the ISM, thus affecting the metallicity of subsequent stars. The local metallicity at the time of formation is taken into account when determining the lifetimes, yields, evolutionary tracks and spectra of each generation of stars. The model was first calibrated against several observables in the Milky Way (see Boissier \& Prantzos 1999) and then extended to other galaxies with different values of $\lambda$ and $V_\mathrm{C}$ (Boissier \& Prantzos 2000), using several scaling laws derived from the $\Lambda$-Cold Dark Matter framework of galaxy formation \citep{mo}.

For each pair of values of $\lambda$ and $V_\mathrm{C}$, the model outputs radial profiles at different wavelengths, which can be then compared to the actual profiles of our galaxy. In order to probe the spatial location of stars of different ages $-$and therefore better constrain the model predictions$-$ we measured surface brightness profiles at all GALEX FUV and NUV bands, the $ugriz$ bands from SDSS and the 3.6 and 4.5\,$\micron$ Spitzer bands. The radial variation of internal extinction in the UV was estimated indirectly from the FUV$-$NUV color profiles, as explained in section~\ref{UVsec}. The UV extinction was then extrapolated to the optical and near-IR bands, assuming a Milky Way extinction curve, convolved with a sandwich model to account for the relative geometry of stars and dust (see \citealt{mun11} for details).

\begin{figure*}
\begin{center}
	\includegraphics[scale=0.7]{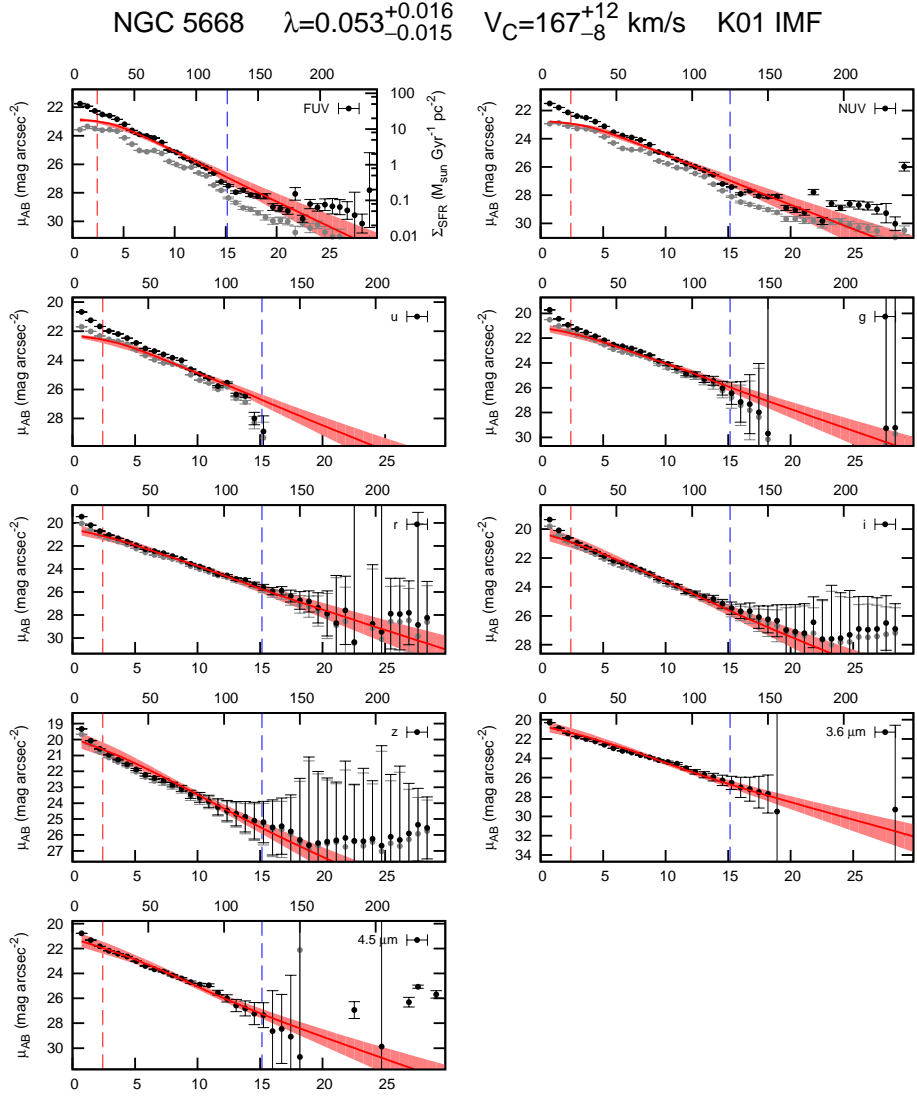}
	\caption[Fig.8]{\footnotesize Best-fitting model of NGC~5668 using the Kroupa (2001) IMF. The gray profiles are only corrected for foreground Milky Way extinction, whereas the black ones are also corrected for internal attenuation. The latter profiles are the ones used to constrain the disk-evolution models in the radial range spanned between the red and blue vertical lines (in order to exclude the bulge and the low S/N outer parts). The best-fitting model is shown as a red solid line, and the red shaded band comprises all models with $\chi^2 \leq 2\chi^2_{\mathrm{min}}$. The radius along the semimajor axis is shown both in kpc (bottom $x$ axis) and arcseconds (top $x$ axis).\label{perf}}
\end{center}
\end{figure*}

\begin{figure*}
\begin{center}
	\includegraphics[scale=0.7]{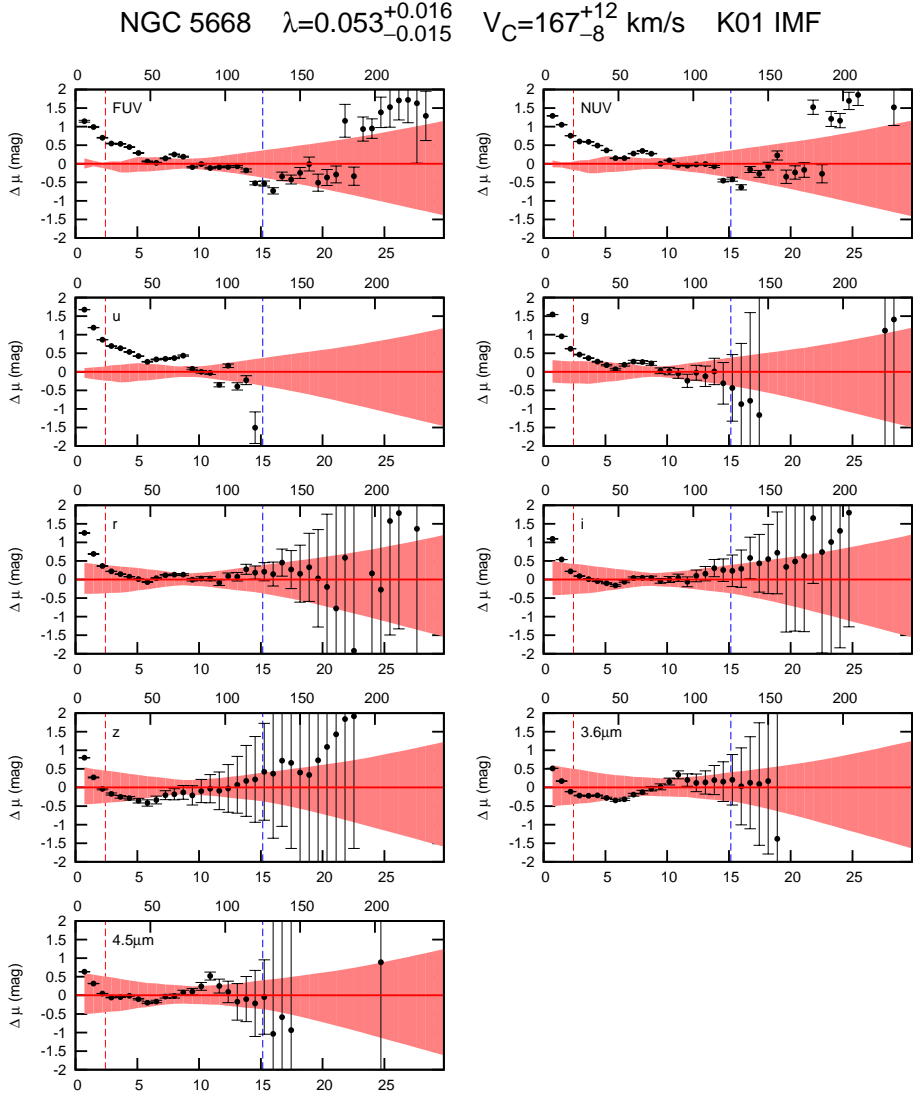}
	\caption[Fig.8]{\footnotesize Residuals of the best-fitting BP2000 model of NGC~5668 using the Kroupa (2001) IMF. The red line represents the best-fitting model and the red shaded area represents all models with $\chi^2 \leq 2\chi^2_{\mathrm{min}}$. The black points are the residuals of the black profile corrected for internal and foreground Milky Way attenuation. The radius along the semimajor axis is shown both in kpc (bottom $x$ axis) and arcseconds (top $x$ axis).\label{res}}
\end{center}
\end{figure*}

\begin{figure*}
\begin{center}

	\includegraphics[scale=0.68]{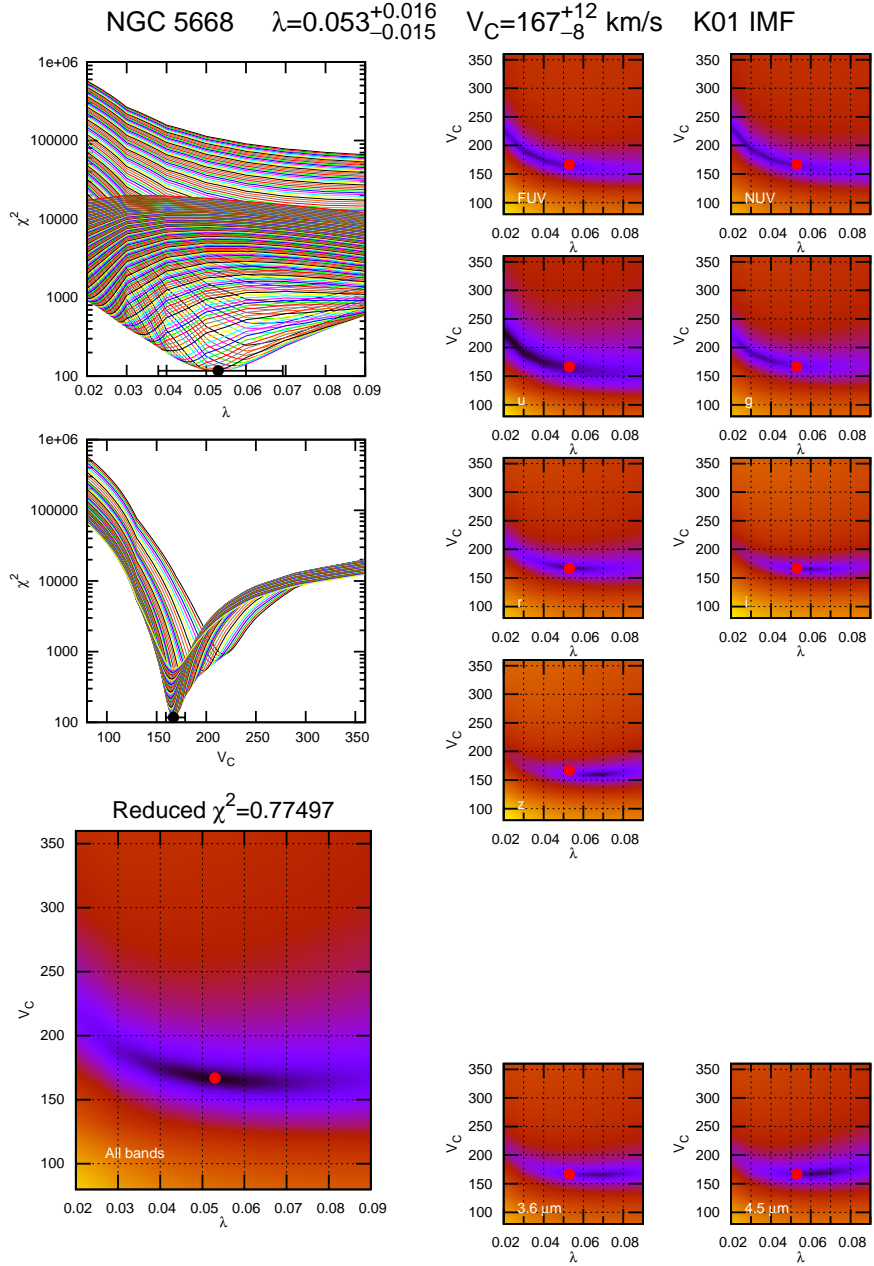}
	\caption[Fig.9]{\footnotesize Two-dimensional $\chi^2$ distribution of the disk-evolution model fiiting procedure. In the top-left panel we plot $\chi^2$ as function of the spin ($\lambda$) for different circular velocities ($V_\mathrm{C}$). The best-fitting value and its estimated uncertainty are marked with a circle with error bars. The middle-left plot is analogous and shows $\chi^2$ as a function of $V_\mathrm{C}$ for different spins. These are projections of the surface shown in the bottom-left panel, where the red dot marks the best model. While we constrain the model by fitting all bands at the same time, we keep track of the individual $\chi^2$ distributions at each wavelength. These are shown in the small panels to the right (the red point has been replicated in all of them for clarity). The missing small panels would correspond to J, H, and K 2MASS images that we decided to exclude from this analysis as they are not deep enough to provide additional information. For more details see \cite{mun11}. \label{chi}}
\end{center}
\end{figure*}

We compared these extinction-free profiles with the model predictions for a grid of values of $\lambda$ and $V_\mathrm{C}$, and used a $\chi^2$ minimization algorithm to find the best fitting values. The results of the fit are shown in Fig$.$~\ref{perf} and in Fig$.$~\ref{res} we present the residuals. Each panel corresponds to a different wavelength. The gray profiles are only corrected for foreground Milky Way extinction, whereas the black ones are corrected for internal extinction as well. The red and blue dashed lines bracket the region of the profile used for the fit. On one hand, starlight inside $r\simeq20$\arcsec\ is dominated by emission coming from the bulge and the oval. On the other hand, beyond $r\simeq125$\arcsec\, the signal-to-noise ratio becomes very low, and contamination from background sources could be an issue. The model that best reproduces simultaneously all multi-wavelength profiles is shown with a red line. The shaded band encompasses all models whose $\chi^2$ is less than twice the $\chi^2$ of the best model ($\chi^2_{\mathrm{min}}\simeq$0.8)\footnote{Note that the scatter of the data-points around the model is not entirely due to noise, but also to real features of the galaxy such a smooth model cannot reproduce. This precludes applying the classical statistical formula to translate the $\chi^2$ distribution into confidence intervals. The criterion of using $\chi^2 \leq 2\chi^2_{\mathrm{min}}$ is just orientative, based on the results shown in Fig$.$~\ref{perf}.\,\,Note also that we are not using the $\chi^2$ distributions in a strict statistical way, since many of the required mathematical conditions (gaussianity of errors, etc) are not met (see \cite{mun11} for a more extensive discussion).}. The resulting $\chi^2$ of the fitting model is obtained after a two-stage fitting procedure because, as one would expect, this model does not reproduce the very small-scale variations of the surface brightness profiles of our disk. So in a first run of the fitting procedure we assume that the total uncertainty for each point is due to the quadratic sum of the zero-point and photometric errors, plus an extra uncertainty in the model predictions of 10\%. These values represent the initial guess for the intrinsic error of the model and in this stage the $\chi^2$ is $\sim$5. In the second step we calcute the {\it rms} of the best-fitting model with respect to the galaxy profiles and we pass these error profiles to the code as initial uncertainties to start the second run. In this way the new reduced $\chi^2$ values are close to unity.
In principle, when we fit a set of data-points with a given model, it is often implicitly assumed that deviations between the observed data-points and the model are due to the measured uncertainties of the former. However, in practice one also has to account for the fact that the models themselves are not a perfect representation of nature and have their own \textquotedblleft uncertainties\textquotedblright. The model yields smooth profiles that, by construction, cannot reproduce the fine structure of the actual profiles. In order to account for this, we performed for NGC~5668 this fit in a two-stage fashion, see also Sec$.$5 of \cite{mun11} for a detailed discussion.
When fitting the multi-wavelength profiles we should not ignore the internal degeneracy in the determination of the $\lambda$ and $V_\mathrm{C}$ values as these are not completely independent parameters. For this reason in Fig$.$~\ref{chi} we show the two-dimensional $\chi^2$ distribution obtained for NGC~5668 in the case of fitting all bands simultaneously and in the case of each band separately. The corresponding best-fitting values are $\lambda=0.053^{+0.016}_{-0.015}$ and $V_\mathrm{C}=167^{+12}_{-8}$\,km\,s$^{-1}$. Where the errors are those expected when one is interested in deriving each of the two quantities separately.

In general, all bands are succesfully reproduced by the BP2000 models but the fit is not equally good at all wavelengths. For example, we can appreciate in the top pannels of Fig$.$~\ref{res} that in the cases of FUV, NUV and u-band data the quality of the fit is not so good because these bands are more sensitive to the recent variations in the star formation history and on the recipe used to calculate the exctinction (or even limitations intrinsic to the models in order to reproduce the UV part of the spectrum at certain metallicities, as explained in Fig$.$9 of \cite{mun11}). The deviations in UV bands from the BP2000 models are compatible with the typical range of errors, as pointed out in \cite{mun11}.

\begin{figure*}
\begin{center}
	\includegraphics[scale=1.2]{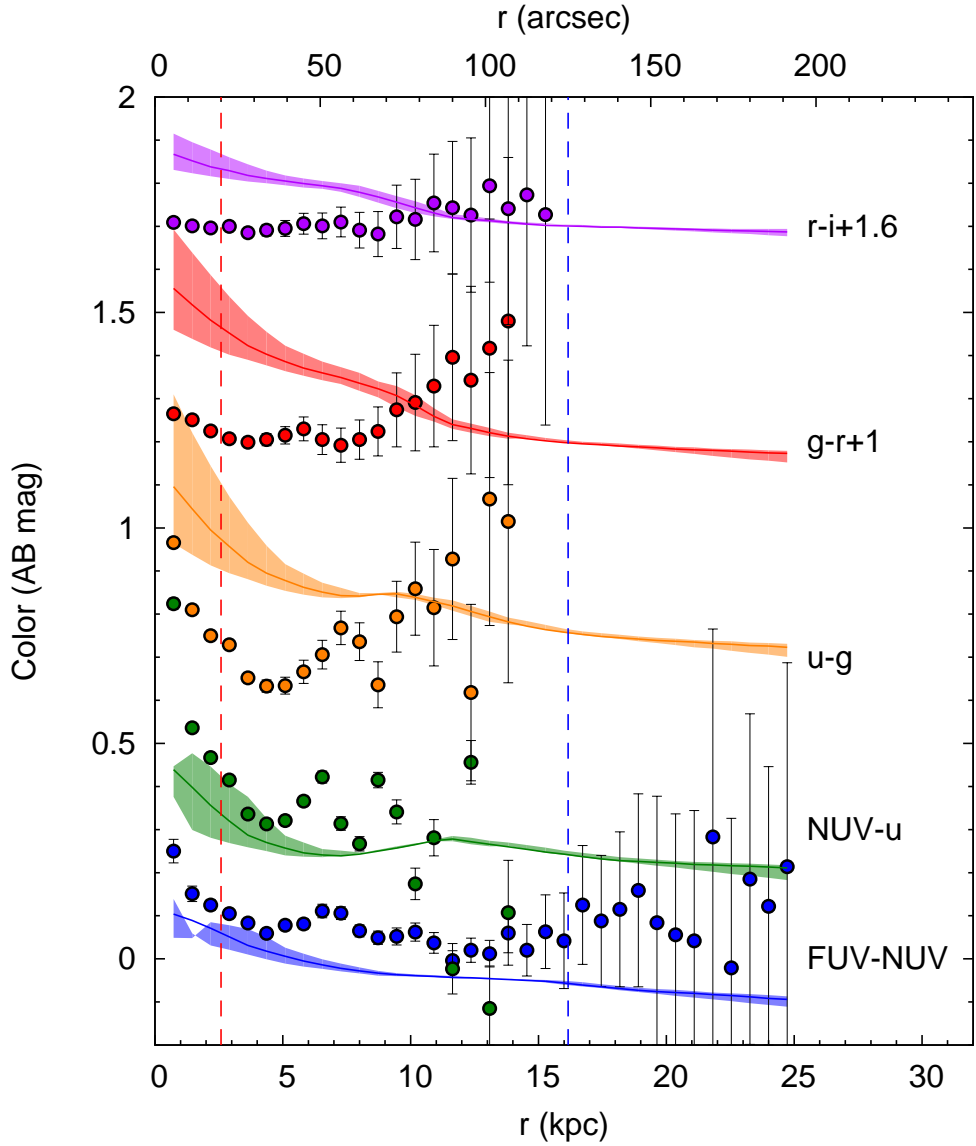}
	\caption[Fig.10]{\footnotesize Observed color profiles of NGC~5668 (points) corrected for internal extinction as in the black points of Figure~\ref{perf}. These observed colors are compared with the ones predicted by the best-fitting model to light profiles (lines and bands). The red and blue vertical dashed lines bracket the spatial range used during the fit. Note that since the fit is performed simultaneously at all radii and all wavelengths from the FUV to 4.5\,$\micron$, departures up to a few tenths of a magnitude are expected for a particular color and radius. \label{color}}
\end{center}
\end{figure*}

While the BP2000 models provide a good fit to the overall shape of the surface brightness profiles of NGC~5668 (see Fig$.$~\ref{perf}), the analysis of the deviations from these otherwise idealized models could give important clues on the details of the star formation history of this galaxy (e.g$.$ gas and/or stellar radial transfers). In this regard, Fig$.$~\ref{color} shows that while the best-fitting model predicts a systematic bluing in the colors towards the outer parts the disks, the measured colors flatten or even get redder beyond $\sim$30-40\arcsec. This radius interestingly coincides with the position where the slope of the metallicity gradient changes sign. We also find small offsets ($\leq$0.2\,mag) between the observed and predicted colors. This is partially due to uncertainties in the model predictions for some of these colors. 

Note that the model has to reproduce at the same time the shape of the observed surface brightness profiles in all bands simultaneously and in addition it has to reproduce the \textquotedblleft average\textquotedblright \, level in surface brightness. This kind of fitting introduces some limits in the range of the model parameters in a way that the colors of these models are located in a relatively narrow range as a consequence of the high degeneration within those model (surface brightness) profiles that match the data at all wavelengths simultaneously\footnote{ One could decide to fit only the color profiles instead. However, that would lead to solutions that might not fit at all the overall shape of the surface brightness profiles and that would be very sensitive to uncertainties in the model predictions regarding specific colors (associated to limitations in the yields, stellar libraries at non-solar abundances, uncertainties in the luminosity at near-infrared and UV wavelengths; see \cite{mara} and \cite{mun11}, respectively).}.

Small variations of a few tenths of a magnitude will not have a significant impact on the overall shape of the surface brightness profiles, but will reveal themselves more clearly in the color profiles (see \citealt{mun11}). 
On the other hand, the SFH of the different annuli in the model are tied one with each other (due to the analytical way in which we implement radial changes) but the actual SFH within the galaxy will exhibit more complex radial variations from one ring to the next one.

\begin{figure*}
\begin{center}
\includegraphics[scale=0.75]{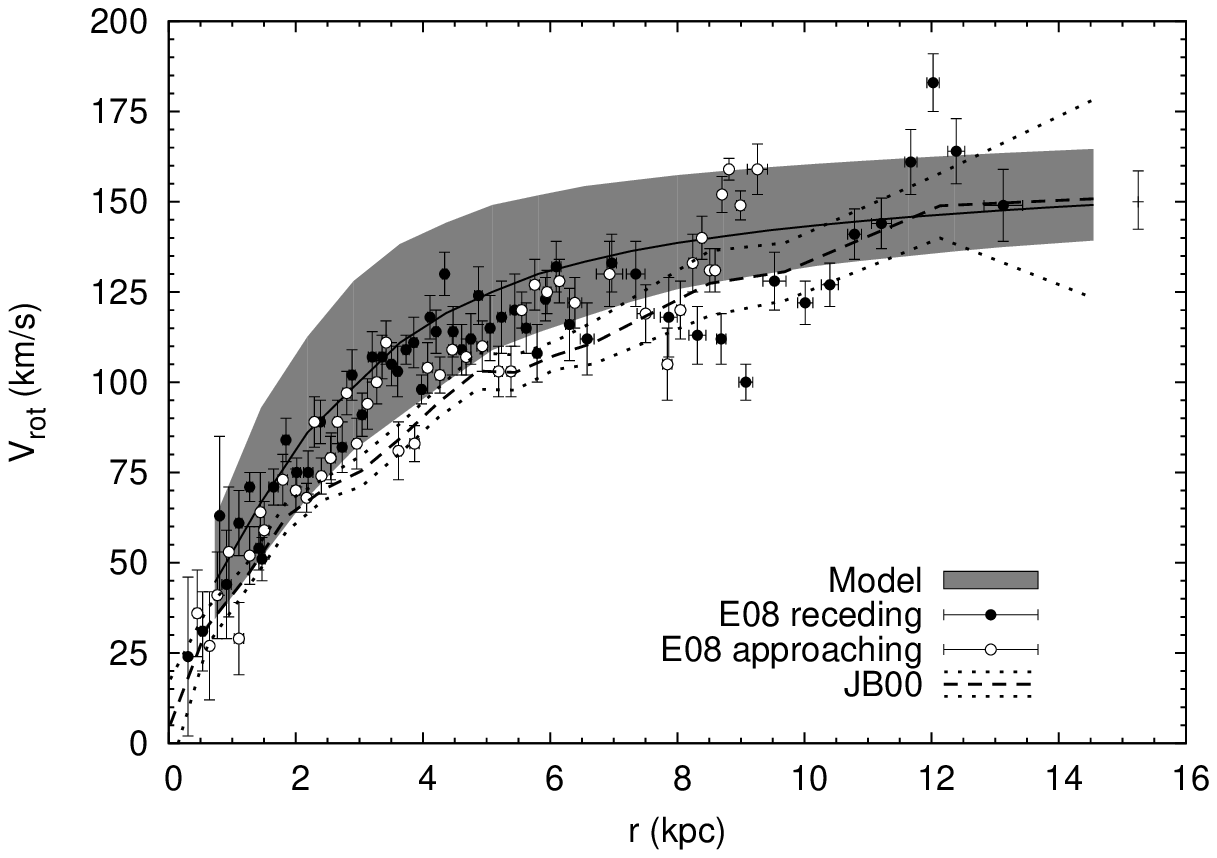}
\resizebox{0.40\hsize}{!}{\includegraphics{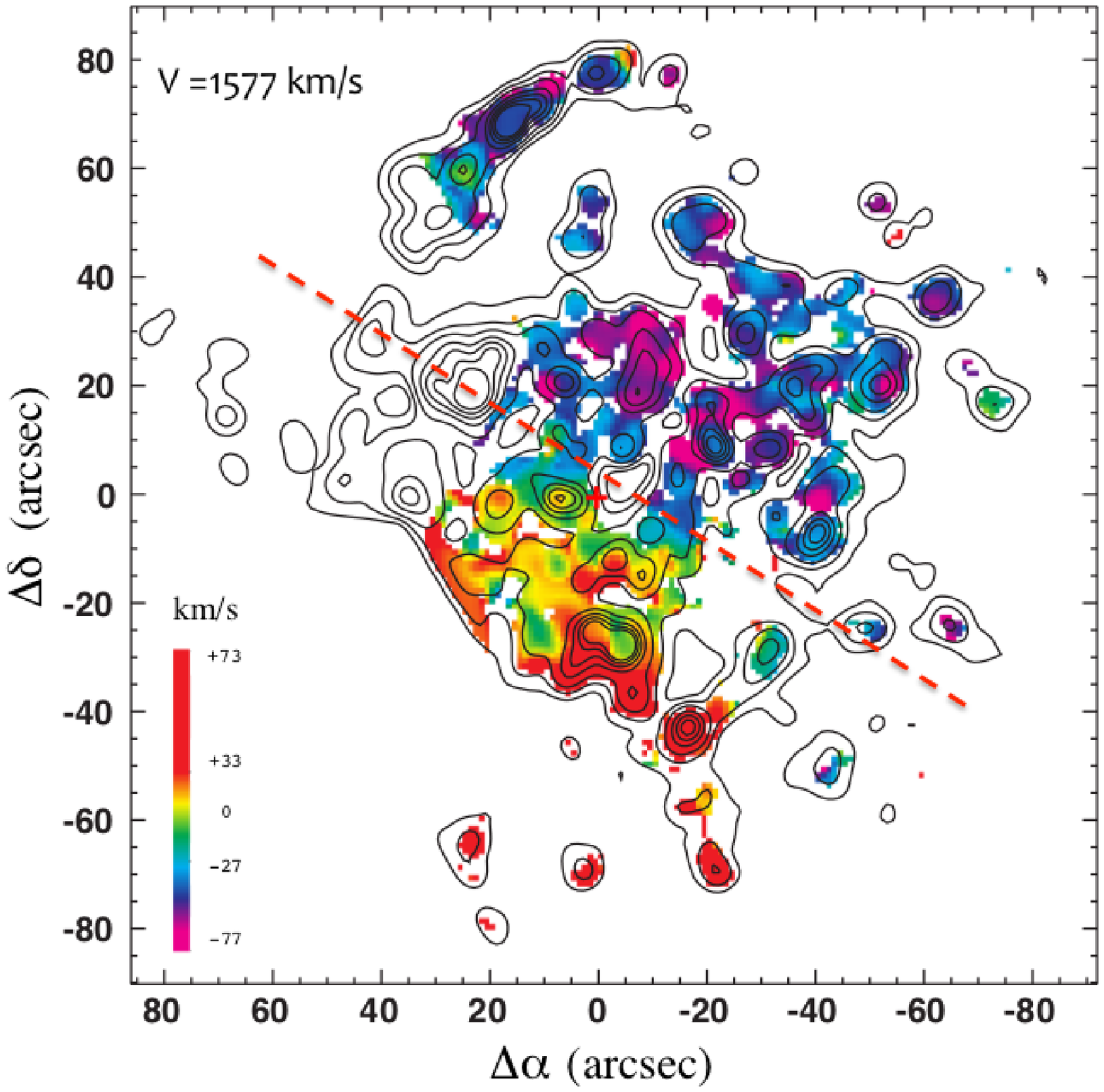}}
	\caption[Fig.10]{\footnotesize {\bf Left:} comparison between the observed and predicted rotation curve of NGC~5668. The black solid line shows the rotation curve corresponding to the model that best fits the multi-wavelength profiles, and the gray band shows the associated uncertainty. The empirical data come from the Fabry-Perot H$\alpha$ studies of Jimenez-Vicente \& Battaner (2000) and Epinat et al. (2008). The rotation curve of the former is plotted with a dashed line, the dotted ones showing the corresponding errors. The values of Epinat et al. (2008) are marked with solid points (receding side of the disk) and open ones (approaching side). Since NGC~5668 is almost face-on, small variations in the adopted inclination angle could rescale the empirical rotation curves. The vertical tickmark near the end of the curve shows the variations induced by changes of$\pm1^{\circ}$  in the inclination angle. {\bf Right:} $\mathrm{H}\alpha$ velocity map of the ionized gas derived from the H$\alpha$ emission line (H$\alpha$ is represented by isocontours). North is up and East is to the left. One pointing was masked out because the derived velocity values were not reliable.  \label{vel}}
\end{center}
\end{figure*}

\section{Discussion}
\label{disc}

The analysis presented above shows that most of the properties of the disk of NGC~5668 can be well reproduced by the chemical and spectro-photometric BP2000 models. This is true both for the global metallicity gradient as well for the shape of the galaxy surface brightness profiles. This is also true for the rotation curve (see Fig$.$~\ref{vel}). However, there are observables (chemical abundances and colors) and, especially, radial intervals among them that are not so well fitted by these simple models. The objective must then be to establish the mechanism(s) that drive the departures of these observables and the associated star-formation and chemical histories in NGC~5668 from those predicted by the $inside-out$ scenario of disk formation on which the BP2000 models are based on. In particular, we ought to explain the bimodality \citep{chiappini} of the abundance gradient (and color profiles) measured in NGC~5668: flatter in the outermost regions and steeper in the inner ones. 

\subsection{Bimodal distribution mechanisms}

The re-distribution of stars, gas and dust in galaxies can have important consequences in the metallicity distribution and, in particular, in deviations of the chemical and photometric properties of disks with respect to the predictions of static disk-evolution models. In this regard, radial changes (up$-$ or down$-$ bending) in the slope of the abundance gradient of nearby spiral galaxies \citep{vilest} are detected and are often related to gaseous and stellar radial mixing processes (\citealt{spitzer}; \citealt{barbanis}; \citealt{shaver}; \citealt{fuchs}; \citealt{sellbin}; \citealt{roskar08}; \citealt{haywood}; \citealt{bresolin09b}; \citealt{vlajic}). 

The existence of this kind of bimodal gradients are often proposed to have been induced by the presence of a bar-like potential (\citealt{diaz}; \citealt{vila}; \citealt{edmunds}; \citealt{zari}). Indeed, in the case of strong-barred galaxies a change from a shallow to a steep metallicity profile is commonly observed and explained in the context of radial mixing of the gas induced by the bar (\citealt{marroy}; \citealt{roywal}). On the other hand, bars can also lead to a steepening of the metallicity profile towards the inner disk if the gas transfer results in  significant star formation in situ (\citealt{friedli94}; \citealt{marroy}; \citealt{roy}; \citealt{roywal}). 

This latter behavior has been found in dynamical simulations of the formation of bars (\citealt{friedli94}; \citealt{friebenz}). In these simulations the presence of a steep-shallow break in the metallicity profile is the result of an intense chemical enrichment by star formation in the bar combined with the dilution effect of the outward flow beyond the break. According to these models the presence of such a break indicates that the bar has recently formed, i.e. in the last Gyr. Observationally, the metallicity distribution of a number of galaxies hosting young bars have been successfully reproduced by this or a similar scenario (\citealt{roywal}; \citealt{considere}). 

However, other mechanisms have been also proposed to explain bimodal metallicity distributions. Thus, in the context of the spiral density-wave theory, star formation is expected to be proportional to $\Omega-\Omega_{p}$ \citep{oort}. In \cite{jensen} this prediction was successfully tested against the metallicity gradients measured in a number of grand-design spirals. Should this scenario be valid one would expect to find a minimum in the chemical abundance at the position of the corotation radius (see also the recent works by \citealt{acha} and \citealt{scarano}), especially in the case of galaxies where the spiral structure is long-lived and quasi-stationary (McCall 1986; Acharova et al. 2005). This is to the fact that $\Omega-\Omega_{p}$ is null (by definition) at corotation and, therefore, star formation should less efficient resulting in a lesser degree of metal enrichment. In this scenario a maximum in the colors should be also observed at the approximate position of the corotation radius as the optical-near-infrared colors would be dominated by those of the underlying stellar population since the current day SFR is expected to be low.

Finally, a flattening of the outer color and metallicity profiles is believed to be also caused by the increasing contribution of old migrated stars to the stellar content of disks (\citealt{zari}; \citealt{binnselw}; \citealt{sellbin}; \citealt{roskar08} and more recently \citealt{minchev}; \citealt{roskar11}). In this case, as we move further out into the outer disk we find progressively older stars, which have been migrating for a longer period of time, leading to a positive age and color gradient in these regions. With regard to the metal abundance and according to the simulations of \cite{roskar08} radial migration leads to the mixing of the old stellar populations, which results in flatter gradients at early times and in the very outer regions of present-day disks as these regions are primarly populated by migrated stars. In the few observational and theoretical studies on the luminosity-weighted age profiles of the disks a similar flattening or even an up-bending is found (\citealt{bakos}; \citealt{pat09}; \citealt{yoac10}). While some authors favor stellar migration as the main driver of the shaping of these age profiles \citep{yoac10} some others indicate that this might due to a decrease in the star formation in the external parts of the disk with time caused by a reduction of the volume density of the gas in these regions \citep{pat09}. 

The different scenarios proposed above to explain a steep-shallow break in the metallicity gradient of galaxies, namely a young bar, reduced star formation at the corotation radius, stellar migration or evolution of the star-formation threshold with redshift have different imprints not only on the specific shape of the metallicity gradient but also in other properties such as the colors. In the case of NGC~5668 the change in the shape of the metallicity gradient takes place at a radius of $\sim$36\arcsec\,(4.4\,kpc or 2.8 disk-scale lengths), well within the region where in-situ star formation takes place in the disk, at an approximate surface brightness of $\sim$22\,mag\,arcsec$^{-2}$ (3.6\,$\micron$ Spitzer band). At this radius the contribution of migrated stars to either the colors, luminosity-weighted age, or chemical abundance is expected to be negligible compared with that from stars formed in-situ (see \citealt{bakos}, \citealt{vlajic}). Again, as this brakes takes place well within the star-forming disk of NGC~5668 changes in the color and metallicity gradients associated with a possible evolution of the star-formation threshold with redshift is highly unlikely. With respect to the possibility that the minimum in the metal abundance profile of NGC~5668 could be due to the presence of the corotation radius at this break two things can be said. Firstly, NGC~5668 is a rather flocculent spiral\footnote{The lack of strong spiral arms in NGC~5668 is evident even after the examination of the IRAC near-infrared images of the object (see \citealt{scarano} and references therein).}, where spiral density waves are expected to be weak or absent and, consequently, the effects of the spiral arms on the radial distribution of the star formation should be minimal (if any; \citealt{mccall2}). Secondly, the color profiles in $(u-g)$ and $(g-r)$ show a minimum at the position of the metallicity break, which is the opposite to what we would expect if that position corresponds to the corotation radius. 

\subsection{Bar formation in NGC~5668}

The only scenario that remains to be analyzed in detail is the possibility that the deviations of the metallicity and color profiles from those predicted by the best-fitting BP2000 model of NGC~5668 are due to the presence of a nascent bar and the effects associated to it. It is well known that one of the most important drivers in the evolution of galaxies are bars (or non axisymmetric central light distributions or ovals, \citealt{atha}). Bars exist in about two-thirs of disk galaxies \citep{sellwilk} and isolated galaxies are known to be able to develop a barred morphology spontaneously from internal instabilities caused by cooling processes (\citealt{miller}; \citealt{kalnajs}; \citealt{bintre}; \citealt{sellwilk}). These gravitational instabilities are commonly the result of enhanced gas accretion in the disk \citep{lindblad}. Some of the observational properties of NGC~5668 already reveal a potential active gas-accretion phase in this galaxy, such as the presence of both HVCs and HRVCs (\citealt{jimenez} and references therein) and the high total star formation rate in this object. 

In this regard, another relevant parameter is the HI content of NGC~5668 compared with objects of similar total mass and morphological type. In order to find out whether the HI content NGC~5668 is particularly high or low, we resort to the so-called HI deficiency parameter. This quantity, defined by \cite{haygio}, compares $-$in logarithmic scale$-$ the observed HI mass of a given galaxy and the typical HI mass of isolated galaxies of similar morphological types and linear sizes. The difference between both values is performed in such a way that positive HI deficiencies correspond to galaxies with less gas than similar field galaxies, and viceversa. Following the prescriptions of \cite{haygio}, we find that an Sd galaxy with the same optical diameter as NGC~5668 is expected to have $\log(M_\mathrm{HI}) = 9.66$. \cite{solanes96} extended the work of Haynes \& Giovanelli (1984) to a larger sample of galaxies. While they only include Sa-Sc galaxies in their sample, we can safely apply the fitting coefficients of Sc's to our Sd spiral \citep{solanes01}. By doing so we obtain a reference HI mass of $\log(M_\mathrm{HI}) = 9.64$, completely consistent with the previous estimation. According to \cite{schu96}, the actual HI mass of NGC~5668 is $\log(M_\mathrm{HI}) = 10$, which implies a negative HI deficiency of $\sim-0.35$. Considering that the usual 1$\sigma$ scatter of the HI mass for a given type and size is 0.2$-$0.3, dex (including all types of galaxies), we conclude that our galaxy is roughly 1-2$\sigma$ gas-richer than its {\it normal} counterparts. While this difference might not be significant it shows that if anything, 
NGC~5668 has a larger HI content than the average of the spiral galaxy population of its type.
\cite{schu94} found in NGC~5668 a weak bar or oval inner structure of $12''$ size visible in both optical and near-infrared images. \cite{atha} found that in spiral galaxies having oval structures or weak bars star formation can occur prolifically along them, especially in the central parts and at the two ends of the bar region. This would result in bluer colors in these regions. The presence of a local minimum in color at a galactocentric distance of $\sim$30-40 arcsec seems to favor this scenario for NGC~5668. Note, however, that the oval seen in its images is significantly smaller than the radius where the minimum in color and metallicity is found. One possible explanation for this difference might come from the fact that, while the light from the superposition of x$_{1}$ orbits that shape the oval is only clearly seen in the central $12"$, the instabilities associated to it could take place further out, where the superposition of these orbits is not yet significant enough for being detected via photometry only \citep{debatt}.

Another possible explanation for the existence of the oval distortion in NGC~5668 is the possible interaction of NGC~5668 with UGC9380
\cite{schu96}. This is a dwarf galaxy at a relative distance of 200 kpc to the southeast of NGC~5668, with a systemic velocity of 1690 km\,sec$^{-1}$ and a relative velocity of $\sim$108 km\,sec$^{-1}$. A recent or on-going tidal interaction between these two galaxies could have produced the high velocity features in NGC~5668 and triggered the formation of both the oval and the incipient bar, as reported by N-body simulations of minor mergers \cite{eliche06} and \cite{eliche11}.

The fact that these blue colors do not extend all the way from this galactocentric distance to the center of the galaxy is likely due to a combination of different effects. Firstly, the secular inside-out model for the evolution of the disk of NGC~5668 predicts a reddening of $\sim$0.2\,mag (0.15\,mag) in $u-g$ ($g-r$) from 36 to 6\arcsec\,(see Fig$.$~\ref{color}). Moreover, the reddening in the ($FUV-NUV$) color indicates a change of $\sim$0.09\,mag in the $B-V$ color purely due to dust attenuation [$E(B-V)$] within this same radial range. Finally, the light contribution of the bulge in the center, despite being small, can also lead to relatively red colors. These three effects are superpimposed on the change in colors induced by star formation in the bar region and whose effects on the colors appear to be noticeable only at both its ends.

Regarding the age of this young bar, in the case of the barred galaxy NGC~3359, \cite{marroy} calculated the age of bar by using the equation of turbulent transport in a shear flow \citep{roykun}. Below we repeat this exercise for the case of NGC~5668. If we consider a cartesian coordinate system centered in the galaxy center and we take x$_{2}$ as the radial direction of the local circular orbit (we do not take in account perpendicular effects), $l\sim$300\,pc as the mean free path of clouds between collisions \citep{roberts}, we can then calculate the time for the gas to diffuse in a length $\Delta x_{2}$ in the radial direction as:

\begin{equation}
\tau_{x_{2}}=\frac{\Delta {x_{2}}^{2} }{v l}
\end{equation}

In our case $\Delta x_{2}$ = 4.4 kpc, wich coincides with $r_{BREAK}$ is the radius at which we detect the flattening of the metallicity gradient and the minimum in the color profiles. From the radial-velocity map kindly provided by J. Jim\'enez-Vicente (priv$.$ comm$.$; see also \citealt{jimenez}) we determine the maximum of the non-circular motions to be $v$=10 \kms. This yields $\tau_{x_{2}}\sim\,10^{9}$ yr, which can be considered as a upper limit to the bar age since the radial transport induced by a bar is a stationary flow. The lower limit is given by simple division of $\Delta x_{2}$/$v$. The best estimate for the age of the young bar in NGC~5668 is then $10^{8} \le \tau_{x_{2}} \le 10^{9}$\,yr.
\\
Summarizing, while the overall observational properties of NGC~5668 are well fitted by the inside-out scenario of disk formation, the deviations in the color and metallicity profiles are best interpreted in the context of the presence of a nascent bar where significant in-situ star formation is (or have been recently) taking place. The formation of the bar is believed to be due to instabilities in the gas-rich inner disk of NGC~5668 possibly helped in the interaction with a companion. This scenario is compatible with the relatively large HI content and star formation rate of NGC~5668 and with the presence of HVCs and HRVCs in its velocity field, evidence of significant non-rotational gas motions in the disk.

\section{Conclusions}
\label{conc}

In this paper we have carried out an extensive and detailed study of the chemical and photometric properties of the nearby spiral galaxy NGC~5668. This detailed study has been possible thanks to the combined use of Integral Field Spectroscopy (IFS) in the optical and panchromatic broad-band imaging of the entire system. The main conclusions from this work are the following: 

\begin{enumerate}
\item Dust-attenuation profiles have been obtained using a number of methods. The mean continuum attenuation is A$_{V}$ $\sim$0.4\,mag. We find a significantly larger ionized-gas attenuation than that of the continuum in agreement with the predictions of \cite{calz01}. 

\item We have derived dust-attenuation-corrected emission-line fluxes for a total of 62 individual H~II complexes and 18 concentric annuli centered on the position of the nucleus of NGC~5668. Based on the strong-line method of Kewley \& Dopita (200X) oxygen-abundance measurements have been obtained. We find a bimodal radial distribution in metallicity with a steep negative gradient with slope -0.14$\pm$0.09(dex/kpc) within a galactocentric distance of 36\arcsec and a shallower (or even positive) metallicity gradient beyond that point, 0.002$\pm$0.002 (dex/kpc). 

\item Surface brightness and color profiles from the UV to the near-infrared have been also obtained. These profiles nicely match the predictions of the chemo-spectrophotometrical models for the evolution of galaxy disks (BP2000 models) for a circular velocity of $v_{\rm circ}=167\,km\,s^{-1}$ and a spin parameter of $\lambda$=0.053. This best-fitting model also agrees with the overall shape of the galaxy metallicity gradient and rotation curve although it cannot reproduce the steep-shallow metallicity break neither the moderate bluing in some of the color profiles ($u-g$, $g-r$) around the position of the metallicity break.

\item Out of the different mechanisms proposed in the literature to explain the change in the slope of the metallicity profile in spiral disks (and the color profile), only the presence of a bar in its formative stages agrees well with the other observational properties of NGC~5668: the position of the metallicity break in disk-scale lengths, the bluing in the colors at that position, the large HI content and star formation rate of the galaxy, the presence of HVCs and HRVCs in the galaxy velocity field and even the detection of an oval in the central region of the galaxy. 
\end{enumerate}

\acknowledgements

We would like to thank the anonymous referee for the review performed to the manuscript, the comments and suggestions helped to improved the content of the paper. We also thank Carmen Eliche-Moral, Fabi\'{a}n Rosales-Ortega, Judit Bakos and Sheila Kannappan for helpful discussions.
R.A. Marino was also funded by the spanish programme of International Campus of Excellence Moncloa (CEI).
We thank Calar Alto Observatory for allocation of
director's discretionary time to this programme.
We acknowledge support from the Spanish Programa Nacional de
Astronom\'{\i}a y Astrof\'{\i}sica under grant AYA 2009-10368.  We are also partially
funded by the Spanish MICINN under the Consolider-Ingenio 2010 Program
grant CSD2006-00070: First Science with the GTC.
JCMM receives financial support from NASA JPL/Spitzer grant RSA
1374189. He also acknowledges support from the National Radio
Astronomy Observatory, which is a facility of the National Science
Foundation operated under cooperative agreement by Associated
Universities, Inc.
This work is based in part on observations made with the {\it Spitzer} Space Telescope, which is
operated by the Jet Propulsion Laboratory, Caltech under NASA contract
1407. GALEX is a NASA Small Explorer launched in 2003 April. We
gratefully acknowledge NASA's support for construction, operation, and
scientific analysis of the GALEX mission. This research has made use
of the NASA/IPAC Extragalactic Database (NED) which is operated by the
Jet Propulsion Laboratory, California Institute of Technology, under
contract with the National Aeronautics and Space Administration.
This paper makes use of the Sloan Digital Sky Survey
data. Funding for the SDSS and SDSS-II has been provided by
the Alfred P. Sloan Foundation, the Participating Institutions,
the National Science Foundation, the U.S. Department of
Energy, the National Aeronautics and Space Administration, the
Japanese Monbukagakusho, the Max Planck Society, and the
Higher Education Funding Council for England. The SDSS Web
Site is http://www.sdss.org/. The SDSS is managed by the Astrophysical Research
Consortium for the Participating Institutions. The Participating
Institutions are the American Museum of Natural History,
Astrophysical Institute Potsdam, University of Basel, University
of Cambridge, Case Western Reserve University, University
of Chicago, Drexel University, Fermilab, the Institute for
Advanced Study, the Japan Participation Group, Johns Hopkins
University, the Joint Institute for Nuclear Astrophysics, the
Kavli Institute for Particle Astrophysics and Cosmology, the
Korean Scientist Group, the Chinese Academy of Sciences
(LAMOST), Los Alamos National Laboratory, the Max-Planck-
Institute for Astronomy (MPIA), the Max-Planck-Institute for
Astrophysics (MPA), New Mexico State University, Ohio State
University, University of Pittsburgh, University of Portsmouth,
Princeton University, the United States Naval Observatory, and
the University of Washington.

\bibliographystyle{apj}
\bibliography{referencias}

\begin{thebibliography}{140}
\expandafter\ifx\csname natexlab\endcsname\relax\def\natexlab#1{#1}\fi

\bibitem[{{Abazajian} {et~al.}(2009){Abazajian}, {Adelman-McCarthy},
  {Ag{\"u}eros}, {Allam}, {Allende Prieto}, {An}, {Anderson}, {Anderson},
  {Annis}, {Bahcall}, \& et~al.}]{aba}
{Abazajian}, K.~N., {Adelman-McCarthy}, J.~K., {Ag{\"u}eros}, M.~A., {et~al.}
  2009, \apjs, 182, 543

\bibitem[{{Acharova} {et~al.}(2005){Acharova}, {L{\'e}pine}, \&
  {Mishurov}}]{acha}
{Acharova}, I.~A., {L{\'e}pine}, J.~R.~D., \& {Mishurov}, Y.~N. 2005, \mnras,
  359, 819

\bibitem[{{Ardeberg} \& {Virdefors}(1982)}]{ardeberg}
{Ardeberg}, A., \& {Virdefors}, B. 1982, \aap, 115, 347

\bibitem[{{Athanassoula}(1994)}]{atha}
{Athanassoula}, E. 1994, 143

\bibitem[{{Azzollini} {et~al.}(2008){Azzollini}, {Trujillo}, \&
  {Beckman}}]{azzollini}
{Azzollini}, R., {Trujillo}, I., \& {Beckman}, J.~E. 2008, \apjl, 679, L69

\bibitem[{{Bakos} {et~al.}(2008){Bakos}, {Trujillo}, \& {Pohlen}}]{bakos}
{Bakos}, J., {Trujillo}, I., \& {Pohlen}, M. 2008, \apjl, 683, L103

\bibitem[{{Barbanis} \& {Woltjer}(1967)}]{barbanis}
{Barbanis}, B., \& {Woltjer}, L. 1967, \apj, 150, 461

\bibitem[{{Barden} {et~al.}(2005){Barden}, {Rix}, {Somerville}, {Bell},
  {H{\"a}u{\ss}ler}, {Peng}, {Borch}, {Beckwith}, {Caldwell}, {Heymans},
  {Jahnke}, {Jogee}, {McIntosh}, {Meisenheimer}, {S{\'a}nchez}, {Wisotzki}, \&
  {Wolf}}]{barden}
{Barden}, M., {Rix}, H., {Somerville}, R.~S., {et~al.} 2005, \apj, 635, 959

\bibitem[{{Barker} {et~al.}(2010){Barker}, {Ferguson}, {Cole}, {Ibata},
  {Irwin}, {Lewis}, {Smecker-Hane}, \& {Tanvir}}]{barker}
{Barker}, M.~K., {Ferguson}, A.~M.~N., {Cole}, A.~A., {et~al.} 2010, \mnras,
  1591

\bibitem[{{Bell}(2002)}]{bell02}
{Bell}, E.~F. 2002, \apj, 577, 150

\bibitem[{{Bell} \& {de Jong}(2000)}]{bell}
{Bell}, E.~F., \& {de Jong}, R.~S. 2000, \mnras, 312, 497

\bibitem[{{Binney}(2001)}]{binnselw}
{Binney}, J. 2001, 230, 63

\bibitem[{{Binney} \& {Tremaine}(1987)}]{bintre}
{Binney}, J., \& {Tremaine}, S. 1987

\bibitem[{{Boffi} {et~al.}(1999){Boffi}, {Sparks}, \& {Macchetto}}]{boffi}
{Boffi}, F.~R., {Sparks}, W.~B., \& {Macchetto}, F.~D. 1999, \aaps, 138, 253

\bibitem[{{Boissier} \& {Prantzos}(1999)}]{boipra}
{Boissier}, S., \& {Prantzos}, N. 1999, \mnras, 307, 857

\bibitem[{{Boissier} {et~al.}(2007){Boissier}, {Gil de Paz}, {Boselli},
  {Madore}, {Buat}, {Cortese}, {Burgarella}, {Mu{\~n}oz-Mateos}, {Barlow},
  {Forster}, {Friedman}, {Martin}, {Morrissey}, {Neff}, {Schiminovich},
  {Seibert}, {Small}, {Wyder}, {Bianchi}, {Donas}, {Heckman}, {Lee},
  {Milliard}, {Rich}, {Szalay}, {Welsh}, \& {Yi}}]{boissier}
{Boissier}, S., {Gil de Paz}, A., {Boselli}, A., {et~al.} 2007, \apjs, 173, 524

\bibitem[{{Boselli} {et~al.}(2010){Boselli}, {Eales}, {Cortese}, {Bendo},
  {Chanial}, {Buat}, {Davies}, {Auld}, {Rigby}, {Baes}, {Barlow}, {Bock},
  {Bradford}, {Castro-Rodriguez}, {Charlot}, {Clements}, {Cormier}, {Dwek},
  {Elbaz}, {Galametz}, {Galliano}, {Gear}, {Glenn}, {Gomez}, {Griffin}, {Hony},
  {Isaak}, {Levenson}, {Lu}, {Madden}, {O'Halloran}, {Okamura}, {Oliver},
  {Page}, {Panuzzo}, {Papageorgiou}, {Parkin}, {Perez-Fournon}, {Pohlen},
  {Rangwala}, {Roussel}, {Rykala}, {Sacchi}, {Sauvage}, {Schulz}, {Schirm},
  {Smith}, {Spinoglio}, {Stevens}, {Symeonidis}, {Vaccari}, {Vigroux},
  {Wilson}, {Wozniak}, {Wright}, \& {Zeilinger}}]{boselli}
{Boselli}, A., {Eales}, S., {Cortese}, L., {et~al.} 2010, \pasp, 122, 261

\bibitem[{{Bresolin} {et~al.}(2009{\natexlab{a}}){Bresolin}, {Gieren},
  {Kudritzki}, {Pietrzy{\'n}ski}, {Urbaneja}, \& {Carraro}}]{bresolin09a}
{Bresolin}, F., {Gieren}, W., {Kudritzki}, R., {et~al.} 2009{\natexlab{a}},
  \apj, 700, 309

\bibitem[{{Bresolin} {et~al.}(2012){Bresolin}, {Kennicutt}, \&
  {Ryan-Weber}}]{bresolin12}
{Bresolin}, F., {Kennicutt}, R.~C., \& {Ryan-Weber}, E. 2012, ArXiv e-prints

\bibitem[{{Bresolin} {et~al.}(2009{\natexlab{b}}){Bresolin}, {Ryan-Weber},
  {Kennicutt}, \& {Goddard}}]{bresolin09b}
{Bresolin}, F., {Ryan-Weber}, E., {Kennicutt}, R.~C., \& {Goddard}, Q.
  2009{\natexlab{b}}, \apj, 695, 580

\bibitem[{{Bruzual} \& {Charlot}(2003)}]{brucha}
{Bruzual}, G., \& {Charlot}, S. 2003, \mnras, 344, 1000

\bibitem[{{Buat}(1992)}]{buat92}
{Buat}, V. 1992, \aap, 264, 444

\bibitem[{{Buat} {et~al.}(2005){Buat}, {Iglesias-P{\'a}ramo}, {Seibert},
  {Burgarella}, {Charlot}, {Martin}, {Xu}, {Heckman}, {Boissier}, {Boselli},
  {Barlow}, {Bianchi}, {Byun}, {Donas}, {Forster}, {Friedman}, {Jelinski},
  {Lee}, {Madore}, {Malina}, {Milliard}, {Morissey}, {Neff}, {Rich},
  {Schiminovitch}, {Siegmund}, {Small}, {Szalay}, {Welsh}, \& {Wyder}}]{buat05}
{Buat}, V., {Iglesias-P{\'a}ramo}, J., {Seibert}, M., {et~al.} 2005, \apjl,
  619, L51

\bibitem[{{Calzetti}(1997)}]{calz97}
{Calzetti}, D. 1997, \aj, 113, 162

\bibitem[{{Calzetti}(2001)}]{calz01}
---. 2001, \pasp, 113, 1449

\bibitem[{{Calzetti} {et~al.}(2000){Calzetti}, {Armus}, {Bohlin}, {Kinney},
  {Koornneef}, \& {Storchi-Bergmann}}]{calz00}
{Calzetti}, D., {Armus}, L., {Bohlin}, R.~C., {et~al.} 2000, \apj, 533, 682

\bibitem[{{Calzetti} {et~al.}(1994){Calzetti}, {Kinney}, \&
  {Storchi-Bergmann}}]{calz94}
{Calzetti}, D., {Kinney}, A.~L., \& {Storchi-Bergmann}, T. 1994, \apj, 429, 582

\bibitem[{{Calzetti} {et~al.}(1996){Calzetti}, {Kinney}, \&
  {Storchi-Bergmann}}]{calz96}
---. 1996, \apj, 458, 132

\bibitem[{{Cardelli} {et~al.}(1989){Cardelli}, {Clayton}, \&
  {Mathis}}]{cardelli}
{Cardelli}, J.~A., {Clayton}, G.~C., \& {Mathis}, J.~S. 1989, \apj, 345, 245

\bibitem[{{Charlot} \& {Fall}(2000)}]{charlot}
{Charlot}, S., \& {Fall}, S.~M. 2000, \apj, 539, 718

\bibitem[{{Chiappini} {et~al.}(2001){Chiappini}, {Matteucci}, \&
  {Romano}}]{chiappini}
{Chiappini}, C., {Matteucci}, F., \& {Romano}, D. 2001, \apj, 554, 1044

\bibitem[{{Consid{\`e}re} {et~al.}(2000){Consid{\`e}re}, {Coziol}, {Contini},
  \& {Davoust}}]{considere}
{Consid{\`e}re}, S., {Coziol}, R., {Contini}, T., \& {Davoust}, E. 2000, \aap,
  356, 89

\bibitem[{{Cortese} {et~al.}(2008){Cortese}, {Boselli}, {Franzetti}, {Decarli},
  {Gavazzi}, {Boissier}, \& {Buat}}]{cortese08}
{Cortese}, L., {Boselli}, A., {Franzetti}, P., {et~al.} 2008, \mnras, 386, 1157

\bibitem[{{Cortese} {et~al.}(2006){Cortese}, {Boselli}, {Buat}, {Gavazzi},
  {Boissier}, {Gil de Paz}, {Seibert}, {Madore}, \& {Martin}}]{cortese06}
{Cortese}, L., {Boselli}, A., {Buat}, V., {et~al.} 2006, \apj, 637, 242

\bibitem[{{Dale} {et~al.}(2007){Dale}, {Gil de Paz}, {Gordon}, {Hanson},
  {Armus}, {Bendo}, {Bianchi}, {Block}, {Boissier}, {Boselli}, {Buckalew},
  {Buat}, {Burgarella}, {Calzetti}, {Cannon}, {Engelbracht}, {Helou},
  {Hollenbach}, {Jarrett}, {Kennicutt}, {Leitherer}, {Li}, {Madore}, {Martin},
  {Meyer}, {Murphy}, {Regan}, {Roussel}, {Smith}, {Sosey}, {Thilker}, \&
  {Walter}}]{dale07}
{Dale}, D.~A., {Gil de Paz}, A., {Gordon}, K.~D., {et~al.} 2007, \apj, 655, 863

\bibitem[{{de Jong} {et~al.}(2007){de Jong}, {Seth}, {Radburn-Smith}, {Bell},
  {Brown}, {Bullock}, {Courteau}, {Dalcanton}, {Ferguson}, {Goudfrooij},
  {Holfeltz}, {Holwerda}, {Purcell}, {Sick}, \& {Zucker}}]{djong}
{de Jong}, R.~S., {Seth}, A.~C., {Radburn-Smith}, D.~J., {et~al.} 2007, \apjl,
  667, L49

\bibitem[{{de Vaucouleurs} {et~al.}(1991){de Vaucouleurs}, {de Vaucouleurs},
  {Corwin}, {Buta}, {Paturel}, \& {Fouque}}]{devac}
{de Vaucouleurs}, G., {de Vaucouleurs}, A., {Corwin}, Jr., H.~G., {et~al.}
  1991, {Third Reference Catalogue of Bright Galaxies}, ed. {de Vaucouleurs,
  G., de Vaucouleurs, A., Corwin, H.~G., Jr., Buta, R.~J., Paturel, G., \&
  Fouque, P.}

\bibitem[{{Debattista} {et~al.}(2006){Debattista}, {Mayer}, {Carollo}, {Moore},
  {Wadsley}, \& {Quinn}}]{debatt}
{Debattista}, V.~P., {Mayer}, L., {Carollo}, C.~M., {et~al.} 2006, \apj, 645,
  209

\bibitem[{{D{\'{\i}}az} {et~al.}(1990){D{\'{\i}}az}, {Terlevich}, {Pagel},
  {V{\'{\i}}lchez}, \& {Edmunds}}]{diaz}
{D{\'{\i}}az}, A.~I., {Terlevich}, E., {Pagel}, B.~E.~J., {V{\'{\i}}lchez},
  J.~M., \& {Edmunds}, M.~G. 1990, \rmxaa, 21, 223

\bibitem[{{Edmunds} \& {Roy}(1993)}]{edmunds}
{Edmunds}, M.~G., \& {Roy}, J. 1993, \mnras, 261, L17

\bibitem[{{Eliche-Moral} {et~al.}(2006){Eliche-Moral}, {Balcells}, {Aguerri},
  \& {Gonz{\'a}lez-Garc{\'{\i}}a}}]{eliche06}
{Eliche-Moral}, M.~C., {Balcells}, M., {Aguerri}, J.~A.~L., \&
  {Gonz{\'a}lez-Garc{\'{\i}}a}, A.~C. 2006, \aap, 457, 91

\bibitem[{{Eliche-Moral} {et~al.}(2011){Eliche-Moral},
  {Gonz{\'a}lez-Garc{\'{\i}}a}, {Balcells}, {Aguerri}, {Gallego}, {Zamorano},
  \& {Prieto}}]{eliche11}
{Eliche-Moral}, M.~C., {Gonz{\'a}lez-Garc{\'{\i}}a}, A.~C., {Balcells}, M.,
  {et~al.} 2011, \aap, 533, A104+

\bibitem[{{Fazio} {et~al.}(2004){Fazio}, {Hora}, {Allen}, {Ashby}, {Barmby},
  {Deutsch}, {Huang}, {Kleiner}, {Marengo}, {Megeath}, {Melnick}, {Pahre},
  {Patten}, {Polizotti}, {Smith}, {Taylor}, {Wang}, {Willner}, {Hoffmann},
  {Pipher}, {Forrest}, {McMurty}, {McCreight}, {McKelvey}, {McMurray}, {Koch},
  {Moseley}, {Arendt}, {Mentzell}, {Marx}, {Losch}, {Mayman}, {Eichhorn},
  {Krebs}, {Jhabvala}, {Gezari}, {Fixsen}, {Flores}, {Shakoorzadeh}, {Jungo},
  {Hakun}, {Workman}, {Karpati}, {Kichak}, {Whitley}, {Mann}, {Tollestrup},
  {Eisenhardt}, {Stern}, {Gorjian}, {Bhattacharya}, {Carey}, {Nelson},
  {Glaccum}, {Lacy}, {Lowrance}, {Laine}, {Reach}, {Stauffer}, {Surace},
  {Wilson}, {Wright}, {Hoffman}, {Domingo}, \& {Cohen}}]{fazio}
{Fazio}, G.~G., {Hora}, J.~L., {Allen}, L.~E., {et~al.} 2004, \apjs, 154, 10

\bibitem[{{Fitzpatrick}(1999)}]{fitz}
{Fitzpatrick}, E.~L. 1999, \pasp, 111, 63

\bibitem[{{Fixsen} {et~al.}(1996){Fixsen}, {Cheng}, {Gales}, {Mather},
  {Shafer}, \& {Wright}}]{fix}
{Fixsen}, D.~J., {Cheng}, E.~S., {Gales}, J.~M., {et~al.} 1996, \apj, 473, 576

\bibitem[{{Friedli} \& {Benz}(1995)}]{friebenz}
{Friedli}, D., \& {Benz}, W. 1995, \aap, 301, 649

\bibitem[{{Friedli} {et~al.}(1994){Friedli}, {Benz}, \&
  {Kennicutt}}]{friedli94}
{Friedli}, D., {Benz}, W., \& {Kennicutt}, R. 1994, \apjl, 430, L105

\bibitem[{{Fuchs}(2001)}]{fuchs}
{Fuchs}, B. 2001, \mnras, 325, 1637

\bibitem[{{Garnett}(1992)}]{garnett}
{Garnett}, D.~R. 1992, \aj, 103, 1330

\bibitem[{{Gil de Paz} \& {Madore}(2002)}]{gilmad02}
{Gil de Paz}, A., \& {Madore}, B.~F. 2002, \aj, 123, 1864

\bibitem[{{Gil de Paz} \& {Madore}(2005)}]{gilmad}
---. 2005, \apjs, 156, 345

\bibitem[{{Gil de Paz} {et~al.}(2005){Gil de Paz}, {Madore}, {Boissier},
  {Swaters}, {Popescu}, {Tuffs}, {Sheth}, {Kennicutt}, {Bianchi}, {Thilker}, \&
  {Martin}}]{gil05}
{Gil de Paz}, A., {Madore}, B.~F., {Boissier}, S., {et~al.} 2005, \apjl, 627,
  L29

\bibitem[{{Gil de Paz} {et~al.}(2007){Gil de Paz}, {Boissier}, {Madore},
  {Seibert}, {Joe}, {Boselli}, {Wyder}, {Thilker}, {Bianchi}, {Rey}, {Rich},
  {Barlow}, {Conrow}, {Forster}, {Friedman}, {Martin}, {Morrissey}, {Neff},
  {Schiminovich}, {Small}, {Donas}, {Heckman}, {Lee}, {Milliard}, {Szalay}, \&
  {Yi}}]{gil07}
{Gil de Paz}, A., {Boissier}, S., {Madore}, B.~F., {et~al.} 2007, \apjs, 173,
  185

\bibitem[{{Gogarten} {et~al.}(2010){Gogarten}, {Dalcanton}, {Williams}, {Ro{\v
  s}kar}, {Holtzman}, {Seth}, {Dolphin}, {Weisz}, {Cole}, {Debattista},
  {Gilbert}, {Olsen}, {Skillman}, {de Jong}, {Karachentsev}, \&
  {Quinn}}]{gogarten}
{Gogarten}, S.~M., {Dalcanton}, J.~J., {Williams}, B.~F., {et~al.} 2010, \apj,
  712, 858

\bibitem[{{Gordon} {et~al.}(2000){Gordon}, {Clayton}, {Witt}, \&
  {Misselt}}]{gordon}
{Gordon}, K.~D., {Clayton}, G.~C., {Witt}, A.~N., \& {Misselt}, K.~A. 2000,
  \apj, 533, 236

\bibitem[{{Haynes} \& {Giovanelli}(1984)}]{haygio}
{Haynes}, M.~P., \& {Giovanelli}, R. 1984, \aj, 89, 758

\bibitem[{{Haywood}(2008)}]{haywood}
{Haywood}, M. 2008, \mnras, 388, 1175

\bibitem[{{Heckman} {et~al.}(1995){Heckman}, {Krolik}, {Meurer}, {Calzetti},
  {Kinney}, {Koratkar}, {Leitherer}, {Robert}, \& {Wilson}}]{heckman}
{Heckman}, T., {Krolik}, J., {Meurer}, G., {et~al.} 1995, \apj, 452, 549

\bibitem[{{Jensen} {et~al.}(1976){Jensen}, {Strom}, \& {Strom}}]{jensen}
{Jensen}, E.~B., {Strom}, K.~M., \& {Strom}, S.~E. 1976, \apj, 209, 748

\bibitem[{{Jim{\'e}nez-Vicente} \& {Battaner}(2000)}]{jimenez}
{Jim{\'e}nez-Vicente}, J., \& {Battaner}, E. 2000, \aap, 358, 812

\bibitem[{{Kalnajs}(1978)}]{kalnajs}
{Kalnajs}, A.~J. 1978, 77, 113

\bibitem[{{Kelz} {et~al.}(2006){Kelz}, {Verheijen}, {Roth}, {Bauer}, {Becker},
  {Paschke}, {Popow}, {S{\'a}nchez}, \& {Laux}}]{kelz}
{Kelz}, A., {Verheijen}, M.~A.~W., {Roth}, M.~M., {et~al.} 2006, \pasp, 118,
  129

\bibitem[{{Kewley} \& {Dopita}(2002)}]{kewdop}
{Kewley}, L.~J., \& {Dopita}, M.~A. 2002, \apjs, 142, 35

\bibitem[{{Kroupa}(2001)}]{kroupa}
{Kroupa}, P. 2001, \mnras, 322, 231

\bibitem[{{Lindblad} {et~al.}(1996){Lindblad}, {Lindblad}, \&
  {Athanassoula}}]{lindblad}
{Lindblad}, P.~A.~B., {Lindblad}, P.~O., \& {Athanassoula}, E. 1996, \aap, 313,
  65

\bibitem[{{MacArthur} {et~al.}(2004){MacArthur}, {Courteau}, {Bell}, \&
  {Holtzman}}]{mac}
{MacArthur}, L.~A., {Courteau}, S., {Bell}, E., \& {Holtzman}, J.~A. 2004,
  \apjs, 152, 175

\bibitem[{{Ma{\'i}z-Apell{\'a}niz} {et~al.}(1998){Ma{\'i}z-Apell{\'a}niz},
  {Mas-Hesse}, {Mu{\~n}oz-Tu{\~n}on}, {V{\'i}lchez}, \& {Casta{\~n}eda}}]{maiz}
{Ma{\'i}z-Apell{\'a}niz}, J., {Mas-Hesse}, J.~M., {Mu{\~n}oz-Tu{\~n}on}, C.,
  {V{\'i}lchez}, J.~M., \& {Casta{\~n}eda}, H.~O. 1998, \aap, 329, 409

\bibitem[{{Maraston} {et~al.}(2006){Maraston}, {Daddi}, {Renzini}, {Cimatti},
  {Dickinson}, {Papovich}, {Pasquali}, \& {Pirzkal}}]{mara}
{Maraston}, C., {Daddi}, E., {Renzini}, A., {et~al.} 2006, \apj, 652, 85

\bibitem[{{M{\'a}rmol-Queralt{\'o}} {et~al.}(2011){M{\'a}rmol-Queralt{\'o}},
  {S{\'a}nchez}, {Marino}, {Mast}, {Viironen}, {Gil de Paz},
  {Iglesias-P{\'a}ramo}, {Rosales-Ortega}, \& {Vilchez}}]{marmol}
{M{\'a}rmol-Queralt{\'o}}, E., {S{\'a}nchez}, S.~F., {Marino}, R.~A., {et~al.}
  2011, \aap, 534, A8

\bibitem[{{Martin} \& {Kennicutt}(2001)}]{martkenn}
{Martin}, C.~L., \& {Kennicutt}, Jr., R.~C. 2001, \apj, 555, 301

\bibitem[{{Martin} {et~al.}(2005){Martin}, {Fanson}, {Schiminovich},
  {Morrissey}, {Friedman}, {Barlow}, {Conrow}, {Grange}, {Jelinsky},
  {Milliard}, {Siegmund}, {Bianchi}, {Byun}, {Donas}, {Forster}, {Heckman},
  {Lee}, {Madore}, {Malina}, {Neff}, {Rich}, {Small}, {Surber}, {Szalay},
  {Welsh}, \& {Wyder}}]{martin05}
{Martin}, D.~C., {Fanson}, J., {Schiminovich}, D., {et~al.} 2005, \apjl, 619,
  L1

\bibitem[{{Martin} \& {Roy}(1995)}]{marroy}
{Martin}, P., \& {Roy}, J. 1995, \apj, 445, 161

\bibitem[{{Mayya} \& {Prabhu}(1996)}]{mayya}
{Mayya}, Y.~D., \& {Prabhu}, T.~P. 1996, \aj, 111, 1252

\bibitem[{{McCall}(1986)}]{mccall2}
{McCall}, M.~L. 1986, \pasp, 98, 992

\bibitem[{{McCall} {et~al.}(1985){McCall}, {Rybski}, \& {Shields}}]{mccall}
{McCall}, M.~L., {Rybski}, P.~M., \& {Shields}, G.~A. 1985, \apjs, 57, 1

\bibitem[{{McGaugh}(1991)}]{mcgaugh}
{McGaugh}, S.~S. 1991, \apj, 380, 140

\bibitem[{{Meurer} {et~al.}(1999){Meurer}, {Heckman}, \& {Calzetti}}]{meurer99}
{Meurer}, G.~R., {Heckman}, T.~M., \& {Calzetti}, D. 1999, \apj, 521, 64

\bibitem[{{Meurer} {et~al.}(1995){Meurer}, {Heckman}, {Leitherer}, {Kinney},
  {Robert}, \& {Garnett}}]{meurer95}
{Meurer}, G.~R., {Heckman}, T.~M., {Leitherer}, C., {et~al.} 1995, \aj, 110,
  2665

\bibitem[{{Miller} \& {Prendergast}(1968)}]{miller}
{Miller}, R.~H., \& {Prendergast}, K.~H. 1968, \apj, 151, 699

\bibitem[{{Minchev} {et~al.}(2011){Minchev}, {Famaey}, {Combes}, {Di Matteo},
  {Mouhcine}, \& {Wozniak}}]{minchev}
{Minchev}, I., {Famaey}, B., {Combes}, F., {et~al.} 2011, \aap, 527, A147+

\bibitem[{{Mo} {et~al.}(1998){Mo}, {Mao}, \& {White}}]{mo}
{Mo}, H.~J., {Mao}, S., \& {White}, S.~D.~M. 1998, \mnras, 295, 319

\bibitem[{{Morrissey} {et~al.}(2007){Morrissey}, {Conrow}, {Barlow}, {Small},
  {Seibert}, {Wyder}, {Budav{\'a}ri}, {Arnouts}, {Friedman}, {Forster},
  {Martin}, {Neff}, {Schiminovich}, {Bianchi}, {Donas}, {Heckman}, {Lee},
  {Madore}, {Milliard}, {Rich}, {Szalay}, {Welsh}, \& {Yi}}]{morris}
{Morrissey}, P., {Conrow}, T., {Barlow}, T.~A., {et~al.} 2007, \apjs, 173, 682

\bibitem[{{Moustakas} {et~al.}(2010){Moustakas}, {Kennicutt}, {Tremonti},
  {Dale}, {Smith}, \& {Calzetti}}]{moustakas}
{Moustakas}, J., {Kennicutt}, Jr., R.~C., {Tremonti}, C.~A., {et~al.} 2010,
  \apjs, 190, 233

\bibitem[{{Mu{\~n}oz-Mateos} {et~al.}(2011){Mu{\~n}oz-Mateos}, {Boissier}, {Gil
  de Paz}, {Zamorano}, {Kennicutt}, {Moustakas}, {Prantzos}, \&
  {Gallego}}]{mun11}
{Mu{\~n}oz-Mateos}, J.~C., {Boissier}, S., {Gil de Paz}, A., {et~al.} 2011,
  \apj, 731, 10

\bibitem[{{Mu{\~n}oz-Mateos} {et~al.}(2007){Mu{\~n}oz-Mateos}, {Gil de Paz},
  {Boissier}, {Zamorano}, {Jarrett}, {Gallego}, \& {Madore}}]{mun07}
{Mu{\~n}oz-Mateos}, J.~C., {Gil de Paz}, A., {Boissier}, S., {et~al.} 2007,
  \apj, 658, 1006

\bibitem[{{Mu{\~n}oz-Mateos} {et~al.}(2009{\natexlab{a}}){Mu{\~n}oz-Mateos},
  {Gil de Paz}, {Zamorano}, {Boissier}, {Dale}, {P{\'e}rez-Gonz{\'a}lez},
  {Gallego}, {Madore}, {Bendo}, {Boselli}, {Buat}, {Calzetti}, {Moustakas}, \&
  {Kennicutt}}]{mun09b}
{Mu{\~n}oz-Mateos}, J.~C., {Gil de Paz}, A., {Zamorano}, J., {et~al.}
  2009{\natexlab{a}}, \apj, 703, 1569

\bibitem[{{Mu{\~n}oz-Mateos} {et~al.}(2009{\natexlab{b}}){Mu{\~n}oz-Mateos},
  {Gil de Paz}, {Boissier}, {Zamorano}, {Dale}, {P{\'e}rez-Gonz{\'a}lez},
  {Gallego}, {Madore}, {Bendo}, {Thornley}, {Draine}, {Boselli}, {Buat},
  {Calzetti}, {Moustakas}, \& {Kennicutt}}]{mun09a}
{Mu{\~n}oz-Mateos}, J.~C., {Gil de Paz}, A., {Boissier}, S., {et~al.}
  2009{\natexlab{b}}, \apj, 701, 1965

\bibitem[{{Nakano} {et~al.}(2004){Nakano}, {Kushida}, {Kushida}, \&
  {Itagaki}}]{nakano}
{Nakano}, S., {Kushida}, R., {Kushida}, Y., \& {Itagaki}, K. 2004, \iaucirc,
  8272, 1

\bibitem[{{Oke}(1990)}]{oke}
{Oke}, J.~B. 1990, \aj, 99, 1621

\bibitem[{{Oort} \& {Peix{\'o}to}(1974)}]{oort}
{Oort}, A.~H., \& {Peix{\'o}to}, J.~P. 1974, \jgr, 79, 2705

\bibitem[{{Osterbrock} \& {Ferland}(2006)}]{oster}
{Osterbrock}, D.~E., \& {Ferland}, G.~J. 2006, {Astrophysics of gaseous nebulae
  and active galactic nuclei}, ed. {Osterbrock, D.~E.~\& Ferland, G.~J.}
  (University Science Books)

\bibitem[{{Pagel}(1986)}]{pagel86}
{Pagel}, B.~E.~J. 1986, \pasp, 98, 1009

\bibitem[{{Pagel} {et~al.}(1979){Pagel}, {Edmunds}, {Blackwell}, {Chun}, \&
  {Smith}}]{pagel79}
{Pagel}, B.~E.~J., {Edmunds}, M.~G., {Blackwell}, D.~E., {Chun}, M.~S., \&
  {Smith}, G. 1979, \mnras, 189, 95

\bibitem[{{P{\'e}rez-Montero} \& {D{\'{\i}}az}(2005)}]{perez}
{P{\'e}rez-Montero}, E., \& {D{\'{\i}}az}, A.~I. 2005, \mnras, 361, 1063

\bibitem[{{Poggianti} {et~al.}(1999){Poggianti}, {Smail}, {Dressler}, {Couch},
  {Barger}, {Butcher}, {Ellis}, \& {Oemler}}]{pogg99}
{Poggianti}, B.~M., {Smail}, I., {Dressler}, A., {et~al.} 1999, \apj, 518, 576

\bibitem[{{Poggianti} \& {Wu}(2000)}]{pogg00}
{Poggianti}, B.~M., \& {Wu}, H. 2000, \apj, 529, 157

\bibitem[{{Pohlen} \& {Trujillo}(2006)}]{pohlen}
{Pohlen}, M., \& {Trujillo}, I. 2006, \aap, 454, 759

\bibitem[{{Prantzos} \& {Boissier}(2000)}]{praboi}
{Prantzos}, N., \& {Boissier}, S. 2000, \mnras, 313, 338

\bibitem[{{Reach} {et~al.}(2005){Reach}, {Megeath}, {Cohen}, {Hora}, {Carey},
  {Surace}, {Willner}, {Barmby}, {Wilson}, {Glaccum}, {Lowrance}, {Marengo}, \&
  {Fazio}}]{reach}
{Reach}, W.~T., {Megeath}, S.~T., {Cohen}, M., {et~al.} 2005, \pasp, 117, 978

\bibitem[{{Roberts} \& {Hausman}(1984)}]{roberts}
{Roberts}, Jr., W.~W., \& {Hausman}, M.~A. 1984, \apj, 277, 744

\bibitem[{{Rosales-Ortega}(2009)}]{fabian}
{Rosales-Ortega}, F.~F. 2009, PhD thesis, University of Cambridge, 2010.

\bibitem[{{Ro{\v s}kar} {et~al.}(2011){Ro{\v s}kar}, {Debattista}, {Loebman},
  {Ivezi{\'c}}, \& {Quinn}}]{roskar11}
{Ro{\v s}kar}, R., {Debattista}, V.~P., {Loebman}, S.~R., {Ivezi{\'c}}, {\v
  Z}., \& {Quinn}, T.~R. 2011, ArXiv e-prints

\bibitem[{{Ro{\v s}kar} {et~al.}(2008){Ro{\v s}kar}, {Debattista}, {Quinn},
  {Stinson}, \& {Wadsley}}]{roskar08}
{Ro{\v s}kar}, R., {Debattista}, V.~P., {Quinn}, T.~R., {Stinson}, G.~S., \&
  {Wadsley}, J. 2008, \apjl, 684, L79

\bibitem[{{Roy} {et~al.}(1996){Roy}, {Belley}, {Dutil}, \& {Martin}}]{roy}
{Roy}, J., {Belley}, J., {Dutil}, Y., \& {Martin}, P. 1996, \apj, 460, 284

\bibitem[{{Roy} \& {Kunth}(1995)}]{roykun}
{Roy}, J., \& {Kunth}, D. 1995, \aap, 294, 432

\bibitem[{{Roy} \& {Walsh}(1997)}]{roywal}
{Roy}, J., \& {Walsh}, J.~R. 1997, \mnras, 288, 715

\bibitem[{{S{\'a}nchez}(2004)}]{sanchez04}
{S{\'a}nchez}, S.~F. 2004, Astronomische Nachrichten, 325, 167

\bibitem[{{S{\'a}nchez}(2006)}]{sanchez06}
---. 2006, Astronomische Nachrichten, 327, 850

\bibitem[{{S{\'a}nchez-Bl{\'a}zquez} {et~al.}(2009){S{\'a}nchez-Bl{\'a}zquez},
  {Courty}, {Gibson}, \& {Brook}}]{pat09}
{S{\'a}nchez-Bl{\'a}zquez}, P., {Courty}, S., {Gibson}, B.~K., \& {Brook},
  C.~B. 2009, \mnras, 398, 591

\bibitem[{{S{\'a}nchez-Bl{\'a}zquez} {et~al.}(2006){S{\'a}nchez-Bl{\'a}zquez},
  {Peletier}, {Jim{\'e}nez-Vicente}, {Cardiel}, {Cenarro},
  {Falc{\'o}n-Barroso}, {Gorgas}, {Selam}, \& {Vazdekis}}]{pat06}
{S{\'a}nchez-Bl{\'a}zquez}, P., {Peletier}, R.~F., {Jim{\'e}nez-Vicente}, J.,
  {et~al.} 2006, \mnras, 371, 703

\bibitem[{{Sandage} \& {Tammann}(1987)}]{sandage}
{Sandage}, A., \& {Tammann}, G.~A. 1987

\bibitem[{{Savage} \& {Mathis}(1979)}]{savage}
{Savage}, B.~D., \& {Mathis}, J.~S. 1979, \araa, 17, 73

\bibitem[{{Scarano}(2010)}]{scarano}
{Scarano}, Jr., S. 2010, Bulletin of the Astronomical Society of Brazil, 29, 65

\bibitem[{{Schulman} {et~al.}(1996){Schulman}, {Bregman}, {Brinks}, \&
  {Roberts}}]{schu96}
{Schulman}, E., {Bregman}, J.~N., {Brinks}, E., \& {Roberts}, M.~S. 1996, \aj,
  112, 960

\bibitem[{{Schulman} {et~al.}(1994){Schulman}, {Bregman}, \&
  {Roberts}}]{schu94}
{Schulman}, E., {Bregman}, J.~N., \& {Roberts}, M.~S. 1994, \apj, 423, 180

\bibitem[{{Seaton}(1979)}]{seaton}
{Seaton}, M.~J. 1979, \mnras, 187, 73P

\bibitem[{{Seibert} {et~al.}(2005){Seibert}, {Martin}, {Heckman}, {Buat},
  {Hoopes}, {Barlow}, {Bianchi}, {Byun}, {Donas}, {Forster}, {Friedman},
  {Jelinsky}, {Lee}, {Madore}, {Malina}, {Milliard}, {Morrissey}, {Neff},
  {Rich}, {Schiminovich}, {Siegmund}, {Small}, {Szalay}, {Welsh}, \&
  {Wyder}}]{seibert}
{Seibert}, M., {Martin}, D.~C., {Heckman}, T.~M., {et~al.} 2005, \apjl, 619,
  L55

\bibitem[{{Sellwood} \& {Binney}(2002)}]{sellbin}
{Sellwood}, J.~A., \& {Binney}, J.~J. 2002, \mnras, 336, 785

\bibitem[{{Sellwood} \& {Wilkinson}(1993)}]{sellwilk}
{Sellwood}, J.~A., \& {Wilkinson}, A. 1993, Reports on Progress in Physics, 56,
  173

\bibitem[{{Shaver} {et~al.}(1983){Shaver}, {McGee}, {Newton}, {Danks}, \&
  {Pottasch}}]{shaver}
{Shaver}, P.~A., {McGee}, R.~X., {Newton}, L.~M., {Danks}, A.~C., \&
  {Pottasch}, S.~R. 1983, \mnras, 204, 53

\bibitem[{{Solanes} {et~al.}(1996){Solanes}, {Giovanelli}, \&
  {Haynes}}]{solanes96}
{Solanes}, J.~M., {Giovanelli}, R., \& {Haynes}, M.~P. 1996, \apj, 461, 609

\bibitem[{{Solanes} {et~al.}(2001){Solanes}, {Manrique},
  {Garc{\'{\i}}a-G{\'o}mez}, {Gonz{\'a}lez-Casado}, {Giovanelli}, \&
  {Haynes}}]{solanes01}
{Solanes}, J.~M., {Manrique}, A., {Garc{\'{\i}}a-G{\'o}mez}, C., {et~al.} 2001,
  \apj, 548, 97

\bibitem[{{Spitzer} \& {Schwarzschild}(1953)}]{spitzer}
{Spitzer}, Jr., L., \& {Schwarzschild}, M. 1953, \apj, 118, 106

\bibitem[{{Stasi{\'n}ska} \& {Sodr{\'e}}(2001)}]{stasinska}
{Stasi{\'n}ska}, G., \& {Sodr{\'e}}, Jr., L. 2001, \aap, 374, 919

\bibitem[{{Thilker} {et~al.}(2007){Thilker}, {Bianchi}, {Meurer}, {Gil de Paz},
  {Boissier}, {Madore}, {Boselli}, {Ferguson}, {Mu{\~n}oz-Mateos}, {Madsen},
  {Hameed}, {Overzier}, {Forster}, {Friedman}, {Martin}, {Morrissey}, {Neff},
  {Schiminovich}, {Seibert}, {Small}, {Wyder}, {Donas}, {Heckman}, {Lee},
  {Milliard}, {Rich}, {Szalay}, {Welsh}, \& {Yi}}]{thilker}
{Thilker}, D.~A., {Bianchi}, L., {Meurer}, G., {et~al.} 2007, \apjs, 173, 538

\bibitem[{{Thurston} {et~al.}(1996){Thurston}, {Edmunds}, \& {Henry}}]{thur}
{Thurston}, T.~R., {Edmunds}, M.~G., \& {Henry}, R.~B.~C. 1996, \mnras, 283,
  990

\bibitem[{{Trujillo} {et~al.}(2004){Trujillo}, {Rudnick}, {Rix}, {Labb{\'e}},
  {Franx}, {Daddi}, {van Dokkum}, {F{\"o}rster Schreiber}, {Kuijken},
  {Moorwood}, {R{\"o}ttgering}, {van der Wel}, {van der Werf}, \& {van
  Starkenburg}}]{truj04}
{Trujillo}, I., {Rudnick}, G., {Rix}, H., {et~al.} 2004, \apj, 604, 521

\bibitem[{{Trujillo} {et~al.}(2006){Trujillo}, {F{\"o}rster Schreiber},
  {Rudnick}, {Barden}, {Franx}, {Rix}, {Caldwell}, {McIntosh}, {Toft},
  {H{\"a}ussler}, {Zirm}, {van Dokkum}, {Labb{\'e}}, {Moorwood},
  {R{\"o}ttgering}, {van der Wel}, {van der Werf}, \& {van
  Starkenburg}}]{truj06}
{Trujillo}, I., {F{\"o}rster Schreiber}, N.~M., {Rudnick}, G., {et~al.} 2006,
  \apj, 650, 18

\bibitem[{{Trumpler}(1930)}]{trumpler}
{Trumpler}, R.~J. 1930, \pasp, 42, 214

\bibitem[{{Vila-Costas} \& {Edmunds}(1992)}]{vila}
{Vila-Costas}, M.~B., \& {Edmunds}, M.~G. 1992, \mnras, 259, 121

\bibitem[{{V{\'i}lchez} \& {Esteban}(1996)}]{vilest}
{V{\'i}lchez}, J.~M., \& {Esteban}, C. 1996, \mnras, 280, 720

\bibitem[{{Vlaji{\'c}} {et~al.}(2009){Vlaji{\'c}}, {Bland-Hawthorn}, \&
  {Freeman}}]{vlajic}
{Vlaji{\'c}}, M., {Bland-Hawthorn}, J., \& {Freeman}, K.~C. 2009, \apj, 697,
  361

\bibitem[{{Wang} \& {Rowan-Robinson}(2009)}]{wang}
{Wang}, L., \& {Rowan-Robinson}, M. 2009, \mnras, 398, 109

\bibitem[{{Werner} {et~al.}(2004){Werner}, {Roellig}, {Low}, {Rieke}, {Rieke},
  {Hoffmann}, {Young}, {Houck}, {Brandl}, {Fazio}, {Hora}, {Gehrz}, {Helou},
  {Soifer}, {Stauffer}, {Keene}, {Eisenhardt}, {Gallagher}, {Gautier}, {Irace},
  {Lawrence}, {Simmons}, {Van Cleve}, {Jura}, {Wright}, \&
  {Cruikshank}}]{werner}
{Werner}, M.~W., {Roellig}, T.~L., {Low}, F.~J., {et~al.} 2004, \apjs, 154, 1

\bibitem[{{Witt} \& {Gordon}(2000)}]{witt}
{Witt}, A.~N., \& {Gordon}, K.~D. 2000, \apj, 528, 799

\bibitem[{{Xu} \& {Buat}(1995)}]{xu}
{Xu}, C., \& {Buat}, V. 1995, \aap, 293, L65

\bibitem[{{Yoachim} \& {Dalcanton}(2008)}]{yoadalc}
{Yoachim}, P., \& {Dalcanton}, J.~J. 2008, \apj, 682, 1004

\bibitem[{{Yoachim} {et~al.}(2010){Yoachim}, {Ro{\v s}kar}, \&
  {Debattista}}]{yoac10}
{Yoachim}, P., {Ro{\v s}kar}, R., \& {Debattista}, V.~P. 2010, \apjl, 716, L4

\bibitem[{{York} {et~al.}(2000){York}, {Adelman}, {Anderson}, {Anderson},
  {Annis}, {Bahcall}, {Bakken}, {Barkhouser}, {Bastian}, {Berman}, {Boroski},
  {Bracker}, {Briegel}, {Briggs}, {Brinkmann}, {Brunner}, {Burles}, {Carey},
  {Carr}, {Castander}, {Chen}, {Colestock}, {Connolly}, {Crocker}, {Csabai},
  {Czarapata}, {Davis}, {Doi}, {Dombeck}, {Eisenstein}, {Ellman}, {Elms},
  {Evans}, {Fan}, {Federwitz}, {Fiscelli}, {Friedman}, {Frieman}, {Fukugita},
  {Gillespie}, {Gunn}, {Gurbani}, {de Haas}, {Haldeman}, {Harris}, {Hayes},
  {Heckman}, {Hennessy}, {Hindsley}, {Holm}, {Holmgren}, {Huang}, {Hull},
  {Husby}, {Ichikawa}, {Ichikawa}, {Ivezi{\'c}}, {Kent}, {Kim}, {Kinney},
  {Klaene}, {Kleinman}, {Kleinman}, {Knapp}, {Korienek}, {Kron}, {Kunszt},
  {Lamb}, {Lee}, {Leger}, {Limmongkol}, {Lindenmeyer}, {Long}, {Loomis},
  {Loveday}, {Lucinio}, {Lupton}, {MacKinnon}, {Mannery}, {Mantsch}, {Margon},
  {McGehee}, {McKay}, {Meiksin}, {Merelli}, {Monet}, {Munn}, {Narayanan},
  {Nash}, {Neilsen}, {Neswold}, {Newberg}, {Nichol}, {Nicinski}, {Nonino},
  {Okada}, {Okamura}, {Ostriker}, {Owen}, {Pauls}, {Peoples}, {Peterson},
  {Petravick}, {Pier}, {Pope}, {Pordes}, {Prosapio}, {Rechenmacher}, {Quinn},
  {Richards}, {Richmond}, {Rivetta}, {Rockosi}, {Ruthmansdorfer}, {Sandford},
  {Schlegel}, {Schneider}, {Sekiguchi}, {Sergey}, {Shimasaku}, {Siegmund},
  {Smee}, {Smith}, {Snedden}, {Stone}, {Stoughton}, {Strauss}, {Stubbs},
  {SubbaRao}, {Szalay}, {Szapudi}, {Szokoly}, {Thakar}, {Tremonti}, {Tucker},
  {Uomoto}, {Vanden Berk}, {Vogeley}, {Waddell}, {Wang}, {Watanabe},
  {Weinberg}, {Yanny}, \& {Yasuda}}]{york}
{York}, D.~G., {Adelman}, J., {Anderson}, Jr., J.~E., {et~al.} 2000, \aj, 120,
  1579

\bibitem[{{Zaritsky} {et~al.}(1994){Zaritsky}, {Kennicutt}, \& {Huchra}}]{zari}
{Zaritsky}, D., {Kennicutt}, Jr., R.~C., \& {Huchra}, J.~P. 1994, \apj, 420, 87

\end{thebibliography}

\newpage
\onecolumn

\begin{table}[]
\caption{Global properties of NGC~5668.}\label{General}
\begin{center}
\begin{tabular}{l c c}
\hline\hline
Name & NGC~5668 & Source \\
\hline
Morphological type & SA(s)d & de Vaucouleurs et al. (1991) \\
 & Sc(s)II-III & Sandage $\&$ Tammann (1987) \\
RA(2000) & 14h 33m 24.3s & Dressel $\&$ Condon (1976) \\
Dec(2000) & +04\grado \,27$\arcmin$ \,02$\arcsec$ & Dressel $\&$ Condon (1976) \\
B & 12.13 $\pm$ 0.03 mag & Schulman (1997) \\
R & 11.28 $\pm$ 0.01 mag & Schulman (1997) \\
$L_{B}$ & 2.7 $\pm$ 0.6 × 10$^{9}$ $\Lsun^{a}$ & Schulman (1997) \\
$L_{H{\alpha}}$ & $1.0 \pm 0.3 \times 10^{8} \Lsun^{a}$ & Schulman (1997)\\
$L_{FIR}$ & $(5.8 \pm 1.2)\times 10^{9} \Lsun^{a}$ & Schulman (1997)\\
$D_{25}$ & 3.3$\arcmin$ $\pm$ 0.2$\arcmin$ & Jimenez $\&$ Battaner (2000) \\
Heliocentric systemic velocity & 1582 $\pm$ 5 km\,s$^{-1}$ & Schulman (1997) \\
Distance & 24.8 $\pm$ 1.7 Mpc & Fixen (1996) \\
Inclination & 18\grado & Schulman (1997) \\
PA& 145\grado & Schulman (1997) \\
Total dynamical mass & 5.7\,$\times$\,10$^{10}$ $M_{\odot}$ & Schulman (1997) \\
E(B-V) & 0.037 mag & NED \\
\hline\hline
\end{tabular}
\end{center}
$^{a}$ Corrected for Galactic and internal extinction. \\
$^{b}$ Uncorrected for Galactic and internal extinction $L_{\odot,B} = 5\times 10^{32}$ erg\,s$^{-1}$ \\ 
\end{table}

\begin{table}[]
\caption{NGC~5668 Observational Log.}\label{obs}
\begin{center}
\begin{tabular}{c l l c c c}
\hline\hline
Pointing$^{\dagger}$ & Obs. Date & Offsets & Exposure Time & Air-mass & Seeing\\
 & (UT) & (arcsec)& (sec) & & (arcsec) \\
(1) & (2) & (3) & (4)& (5) & (6) \\
\hline\hline
1 & 2007-06-22 & (0,0) & 3$\times$1000 & 1.21 & 1.1\\
2 & 2007-06-22 & (0,60) & 4$\times$1000 & 1.38 & 1.1 \\
3 & 2007-06-22/23 & (52,60)& 3$\times$1000 & 1.75 & 1.1\\
7 & 2007-06-23  & (52,-60) & 3$\times$1000 & 1.21 & 1.2\\
5 & 2007-06-23 & (0,-60)& 3$\times$1000 & 1.36 & 1.2 \\
4 & 2007-06-23/24 & (-52,60) & 3$\times$1000 & 1.74 & 1.0\\
\hline\hline
\end{tabular}
\end{center}
NOTE: (1): NGC~5668 pointings. (2): Observations date. (3): Pointings offsets; the original pointing of the telescope is at RA(2000) 14$^{h}$33$^{m}$24.3$^{s}$ and Dec(2000) +04$^{\circ}$33$^{\prime}$24.3$^{\prime\prime}$. (4): Exposure time in seconds and numbers of images obtained for each pointing. (5)-(6): Mean air-mass and seeing values. \\
$^{\dagger}$ Pointing number 6 in the default mapping strategy used with PPAK at the CAHA observatory has offsets (-52",-60") relative to the center of the mosaic and was not observed in the case of NGC~5668.
\end{table}

\begin{deluxetable}{lcccccrc}
\tablecolumns{8}  
\tablecaption{Spectra emission lines.}\label{supertable_ratio}
\tablehead{
\multicolumn{1}{c}{ID} & \multicolumn{1}{c}{F$_{[O\sc{II}]}$/F$_{\mathrm{H}\alpha}$} & \multicolumn{1}{c}{F$_{[O\sc{III}]}$/F$_{\mathrm{H}\alpha}$} & \multicolumn{1}{c}{F$_{[N\sc{II}]}$/F$_{\mathrm{H}\alpha}$} & \multicolumn{1}{c}{F$_{[S\sc{IIa}]}$/F$_{\mathrm{H}\alpha}$} & \multicolumn{1}{c}{F$_{[S\sc{IIb}]}$/F$_{\mathrm{H}\alpha}$} & \multicolumn{1}{c}{F$_{\mathrm{H}\alpha}$} & \multicolumn{1}{c}{A$_{V}$} \\
\colhead{(1)} & \colhead{(2)} & \colhead{(3)} & \colhead{(4)} & \colhead{(5)} & \colhead{(6)} & \colhead{(7)} & \colhead{(8)} \\ 
}
\startdata 
\multicolumn{8}{c}{Rings} \\
\cline{1-8}
1     &  \nodata       &  \nodata 	 &   \nodata 	    &   \nodata        &   \nodata 	  &   \nodata      &  \nodata \\
2     &  2.68$\pm$0.43 &  \nodata 	 &  0.32$\pm$0.05 &  0.28$\pm$0.05 &  0.27$\pm$0.05 & 18.20$\pm$2.51 & 1.57 \\
3     &  1.92$\pm$0.31 & 0.15$\pm$0.04 &  0.31$\pm$0.05 &  0.26$\pm$0.04 &  0.24$\pm$0.04 &  19.40$\pm$2.41 & 1.52 \\  
4     &  1.77$\pm$0.29 & 0.21$\pm$0.05 &  0.26$\pm$0.04 &  0.25$\pm$0.04 &  0.20$\pm$0.04 &  13.94$\pm$1.77  & 1.23 \\ 
5     &  1.13$\pm$0.21 & 0.21$\pm$0.05 &  0.29$\pm$0.06 &  0.25$\pm$0.06 &  0.23$\pm$0.05 &  4.96$\pm$0.71 & 0.35 \\  
6     &  1.70$\pm$0.32 & 0.28$\pm$0.06 &  0.26$\pm$0.05 &  0.25$\pm$0.05 &  0.19$\pm$0.05 &  9.36$\pm$1.41 & 1.00 \\  
7     &  1.83$\pm$0.24 & 0.36$\pm$0.05 &  0.21$\pm$0.03 &  0.23$\pm$0.03 &  0.18$\pm$0.03 &  10.61$\pm$1.18 & 1.33 \\  
8     &  1.30$\pm$0.18 & 0.40$\pm$0.06 &  0.17$\pm$0.03 &  0.21$\pm$0.03 &  0.17$\pm$0.03 &  7.12$\pm$0.78 & 0.88 \\ 
9     &  2.29$\pm$0.41 & 0.47$\pm$0.09 &  0.20$\pm$0.05 &  0.24$\pm$0.05 &  0.25$\pm$0.05 &  7.86$\pm$1.29 & 1.64 \\ 
10    &  2.04$\pm$0.48 & 0.61$\pm$0.14 &  0.15$\pm$0.06 &  0.23$\pm$0.07 &  0.19$\pm$0.06 &  4.36$\pm$0.93 & 1.33 \\
11    &  1.81$\pm$0.58 & 0.35$\pm$0.13 &   \nodata 	&  0.21$\pm$0.09 &  0.21$\pm$0.09 &  2.80$\pm$0.80 & 1.19 \\
12    &   \nodata      &  \nodata 	 &   \nodata 	    &   \nodata        &   \nodata 	  &   \nodata 	   & \nodata \\
13    &   \nodata      &  \nodata 	 &   \nodata 	    &   \nodata        &   \nodata 	  &   \nodata      & \nodata \\
14    &  1.99$\pm$0.87 & 0.64$\pm$0.28 &   \nodata 	&   \nodata        &   \nodata    &   1.80$\pm$0.75 & 1.10\\
15    &  1.50$\pm$0.43 & 0.66$\pm$0.18 &   \nodata      &   \nodata        &   \nodata    &   2.03$\pm$0.53 & 0.76\\  
\cline{1-8}											    
\multicolumn{8}{c}{HII Regions} \\										    
\cline{1-8}											    
1     &  1.14$\pm$0.14 & 0.64$\pm$0.04 &  0.09$\pm$0.02 &  0.28$\pm$0.03 &  0.19$\pm$0.03 &   2.23$\pm$0.09 & 0.65\\ 
2     &  1.19$\pm$0.37 & 0.48$\pm$0.11 &  0.20$\pm$0.09 &   \nodata        &   \nodata    &   0.50$\pm$0.05 & 0.12 \\ 
3     &  1.55$\pm$0.23 & 0.49$\pm$0.04 &  0.15$\pm$0.02 &  0.24$\pm$0.02 &  0.13$\pm$0.01 &  11.36$\pm$0.28 & 2.02\\ 
4     &  1.52$\pm$0.09 & 0.24$\pm$0.02 &  0.16$\pm$0.01 &  0.21$\pm$0.01 &  0.17$\pm$0.01 &  12.27$\pm$0.16 & 1.36\\
5     &  1.33$\pm$0.06 & 0.37$\pm$0.01 &  0.17$\pm$0.01 &  0.22$\pm$0.01 &  0.15$\pm$0.01 &  13.32$\pm$0.14 & 1.18\\ 
6    &   \nodata      &  \nodata        &  0.01$\pm$0.01 &  0.28$\pm$0.05 &  0.25$\pm$0.05 &  5.39$\pm$0.43 & 2.46\\ 
7    &  1.83$\pm$0.42 & 0.44$\pm$0.08 &  0.02$\pm$0.01 &  0.28$\pm$0.05 &  0.13$\pm$0.03 &  3.68$\pm$0.21 & 1.64\\
8    &  1.44$\pm$0.12 & 0.29$\pm$0.04 &  0.09$\pm$0.03 &  0.18$\pm$0.03 &  0.21$\pm$0.04 &  11.39$\pm$0.50 & 1.31\\
9     &  1.57$\pm$0.11 & 0.46$\pm$0.03 &  0.12$\pm$0.02 &  0.12$\pm$0.02 &  0.10$\pm$0.02 & 32.00$\pm$0.91 & 2.06\\
10    &  1.72$\pm$0.07 & 0.41$\pm$0.01 &  0.17$\pm$0.01 &  0.17$\pm$0.01 &  0.13$\pm$0.01 & 12.39$\pm$0.16 & 1.22\\ 
11    &  2.00$\pm$0.22 & 0.62$\pm$0.05 &  0.14$\pm$0.02 &  0.26$\pm$0.02 &  0.20$\pm$0.02 &  4.75$\pm$0.18 & 1.30\\ 
12    &  2.05$\pm$0.26 & 0.67$\pm$0.04 &  0.11$\pm$0.02 &  0.24$\pm$0.02 &  0.14$\pm$0.02 & 10.02$\pm$0.30 & 1.98\\ 
13    &  2.31$\pm$0.23 & 0.34$\pm$0.06 &  0.18$\pm$0.05 &  0.26$\pm$0.06 &  0.14$\pm$0.05 &  4.27$\pm$0.25 & 0.39\\ 
14    &  1.76$\pm$0.80 & 0.32$\pm$0.10 &  0.20$\pm$0.05 &  0.16$\pm$0.06 &   \nodata      &  23.48$\pm$1.85 & 2.60\\ 
15    &  0.48$\pm$0.08 & 0.57$\pm$0.05 &  0.10$\pm$0.04 &  0.15$\pm$0.05 &  0.12$\pm$0.04 &  6.27$\pm$0.25 & 0.73\\  
16    &  1.18$\pm$0.09 & 0.47$\pm$0.02 &  0.15$\pm$0.01 &  0.17$\pm$0.01 &  0.11$\pm$0.01 &  15.94$\pm$0.23 & 1.61\\  
17    &  3.93$\pm$0.76 & 0.96$\pm$0.16 &  0.18$\pm$0.07 &  0.22$\pm$0.06 &  0.22$\pm$0.09 &  19.32$\pm$1.93 & 2.89\\
18    &  1.26$\pm$0.37 &  \nodata      &  0.22$\pm$0.03 &  0.29$\pm$0.03 &  0.13$\pm$0.03 &  6.84$\pm$0.30 & 1.93\\
19    &  1.48$\pm$0.17 & 0.29$\pm$0.04 &  0.26$\pm$0.03 &  0.27$\pm$0.03 &  0.18$\pm$0.03 &  11.43$\pm$0.50 & 0.90\\  
20    &  0.51$\pm$0.12 & 0.41$\pm$0.04 &  0.25$\pm$0.03 &  0.28$\pm$0.04 &  0.17$\pm$0.03 &  1.375$\pm$0.05 & 0.12\\ 
21    &  1.85$\pm$0.14 & 0.29$\pm$0.03 &  0.22$\pm$0.02 &  0.15$\pm$0.02 &  0.11$\pm$0.02 &  16.73$\pm$0.59 & 0.89\\ 
22    &  2.40$\pm$0.62 & 0.26$\pm$0.10 &  0.18$\pm$0.05 &  0.35$\pm$0.05 &  0.24$\pm$0.04 &  4.30$\pm$0.34 & 1.87\\ 
23    &  0.93$\pm$0.04 & 0.97$\pm$0.01 &  0.09$\pm$0.01 &  0.12$\pm$0.01 &  0.10$\pm$0.01 &  11.94$\pm$0.12 & 0.92\\
24    &  1.69$\pm$0.15 & 0.26$\pm$0.03 &  0.23$\pm$0.02 &  0.22$\pm$0.02 &  0.16$\pm$0.02 &  7.36$\pm$0.18 & 1.31\\
25    &  1.57$\pm$0.27 & 0.71$\pm$0.07 &  0.24$\pm$0.04 &  0.14$\pm$0.03 &  0.10$\pm$0.03 &  2.41$\pm$0.12 & 1.06\\
26    &  1.41$\pm$0.51 & 0.14$\pm$0.05 &  0.31$\pm$0.07 &  0.33$\pm$0.07 &  0.21$\pm$0.05 &  9.14$\pm$1.68 & 0.88\\  
27    &   \nodata      & 2.19$\pm$0.48 &   \nodata        &  0.15$\pm$0.07 &   \nodata    & 31.30$\pm$4.64 & 4.59\\
28    &  1.74$\pm$0.47 & 0.17$\pm$0.05 &  0.30$\pm$0.05 &  0.26$\pm$0.04 &  0.19$\pm$0.04 &  20.57$\pm$2.59 & 1.67\\ 
29    &  1.28$\pm$0.20 & 0.21$\pm$0.03 &  0.25$\pm$0.03 &  0.24$\pm$0.03 &  0.19$\pm$0.03 &  14.89$\pm$0.93 & 1.08\\ 
30    &  1.34$\pm$0.24 & 0.24$\pm$0.05 &  0.25$\pm$0.04 &  0.25$\pm$0.04 &  0.16$\pm$0.03 &  11.61$\pm$0.71 & 1.15\\
31    &  1.30$\pm$0.08 & 0.38$\pm$0.02 &  0.18$\pm$0.01 &  0.18$\pm$0.01 &  0.11$\pm$0.01 &  9.23$\pm$0.12 & 1.14\\
32    &  1.13$\pm$0.09 & 0.33$\pm$0.02 &  0.18$\pm$0.01 &  0.15$\pm$0.01 &  0.11$\pm$0.01 &  29.66$\pm$0.59 & 1.18\\
33    &  1.43$\pm$0.13 & 0.20$\pm$0.02 &  0.24$\pm$0.02 &  0.19$\pm$0.02 &  0.12$\pm$0.01 &  30.84$\pm$1.21 & 1.30\\ 
34    &  2.30$\pm$0.59 & 0.41$\pm$0.10 &  0.14$\pm$0.04 &  0.24$\pm$0.03 &  0.13$\pm$0.03 &  6.59$\pm$0.41 & 2.20\\
35    &  3.42$\pm$0.38 & 0.39$\pm$0.06 &  0.35$\pm$0.04 &  0.35$\pm$0.04 &  0.21$\pm$0.04 &  50.93$\pm$3.45 & 1.50\\ 
36    &  2.32$\pm$0.15 & 0.66$\pm$0.03 &  0.14$\pm$0.01 &  0.19$\pm$0.01 &  0.12$\pm$0.01 &  17.41$\pm$0.34 & 1.96\\ 
37    &  1.15$\pm$0.05 & 0.36$\pm$0.01 &  0.18$\pm$0.01 &  0.14$\pm$0.01 &  0.11$\pm$0.01 & 57.29$\pm$1.04 & 1.00\\ 
38    &  1.55$\pm$0.23 & 0.15$\pm$0.04 &  0.26$\pm$0.03 &  0.24$\pm$0.03 &  0.15$\pm$0.03 &  13.94$\pm$0.79 & 1.25\\  
39    &  2.68$\pm$0.92 & 0.96$\pm$0.14 &   \nodata        &   \nodata        &   \nodata  &  7.80$\pm$0.59 & 2.74\\ 
40    &  3.32$\pm$1.16 & 0.51$\pm$0.11 &  0.15$\pm$0.03 &  0.26$\pm$0.03 &  0.09$\pm$0.02 &  25.61$\pm$1.71 & 3.63\\ 
41    &  1.06$\pm$0.05 & 0.32$\pm$0.01 &  0.16$\pm$0.01 &  0.19$\pm$0.01 &  0.15$\pm$0.01 &  14.14$\pm$0.16 & 1.06\\ 
42    &  2.21$\pm$0.45 & 0.27$\pm$0.07 &  0.24$\pm$0.05 &  0.23$\pm$0.04 &  0.20$\pm$0.04 &  7.85$\pm$1.30 & 1.84\\
43    &  0.76$\pm$0.16 & 0.20$\pm$0.03 &  0.22$\pm$0.03 &  0.23$\pm$0.03 &  0.18$\pm$0.03 &  18.11$\pm$1.75 & 1.22\\ 
44    &  1.78$\pm$0.19 & 0.15$\pm$0.03 &  0.32$\pm$0.03 &  0.23$\pm$0.02 &  0.24$\pm$0.02 &  28.78$\pm$2.28 & 1.47\\ 
45    &  2.99$\pm$0.41 & 0.12$\pm$0.04 &  0.29$\pm$0.04 &  0.27$\pm$0.04 &  0.21$\pm$0.03 &  25.34$\pm$3.02 & 1.75\\  
46    &  1.33$\pm$0.25 & 0.15$\pm$0.05 &  0.25$\pm$0.05 &  0.26$\pm$0.05 &  0.21$\pm$0.04 &  11.02$\pm$1.50 & 1.06\\  
47    &  1.45$\pm$0.20 & 0.16$\pm$0.04 &  0.25$\pm$0.04 &  0.26$\pm$0.04 &  0.21$\pm$0.04 &  6.23$\pm$0.71 & 0.34 \\ 
48    &  3.60$\pm$0.75 & 0.26$\pm$0.08 &  0.26$\pm$0.06 &  0.29$\pm$0.07 &  0.18$\pm$0.05 &  15.20$\pm$2.89 & 1.71\\ 
49    &  1.64$\pm$0.05 & 0.75$\pm$0.01 &  0.10$\pm$0.01 &  0.11$\pm$0.01 &  0.10$\pm$0.01 &  16.48$\pm$0.16 & 1.09\\
50    &  1.22$\pm$0.19 & 0.20$\pm$0.04 &  0.25$\pm$0.04 &  0.25$\pm$0.04 &  0.21$\pm$0.04 &  9.11$\pm$1.05 & 0.72\\ 
51    &  1.56$\pm$0.11 & 0.34$\pm$0.03 &  0.21$\pm$0.02 &  0.20$\pm$0.02 &  0.17$\pm$0.02 &  22.45$\pm$1.29 & 0.93\\ 
52    &  1.14$\pm$0.11 & 0.31$\pm$0.03 &  0.22$\pm$0.02 &  0.22$\pm$0.02 &  0.19$\pm$0.02 &  21.20$\pm$1.41 & 1.01\\ 
53    &  1.52$\pm$0.05 & 0.40$\pm$0.01 &  0.18$\pm$0.01 &  0.19$\pm$0.01 &  0.16$\pm$0.01 &  29.32$\pm$0.54 & 1.02\\  
54    &  1.65$\pm$0.06 & 0.66$\pm$0.02 &  0.12$\pm$0.01 &  0.15$\pm$0.01 &  0.12$\pm$0.01 &  10.30$\pm$0.12 & 0.94\\  
55    &  1.72$\pm$0.08 & 0.63$\pm$0.02 &  0.12$\pm$0.01 &  0.22$\pm$0.01 &  0.16$\pm$0.01 &  15.11$\pm$0.18 & 1.33\\
56    &  0.81$\pm$0.05 & 0.18$\pm$0.02 &  0.11$\pm$0.02 &  0.24$\pm$0.02 &  0.16$\pm$0.02 &  3.30$\pm$0.16 & 0.00\\ 
57    &  11.5$\pm$2.4  & 0.78$\pm$0.16   &  0.18$\pm$0.04 &  0.18$\pm$0.04 &  0.16$\pm$0.04 & 153.29$\pm$30.27 & 3.93\\
58    &  1.83$\pm$0.08 & 0.70$\pm$0.02 &  0.11$\pm$0.02 &  0.15$\pm$0.02 &  0.13$\pm$0.03 &  6.07$\pm$0.16 & 0.86\\
59    &  3.22$\pm$0.25 & 0.45$\pm$0.04 &  0.13$\pm$0.04 &  0.19$\pm$0.03 &  0.27$\pm$0.04 &  6.09$\pm$0.27 & 1.51\\ 
60    &  2.26$\pm$0.14 & 0.36$\pm$0.03 &  0.10$\pm$0.03 &  0.17$\pm$0.03 &  0.16$\pm$0.04 &  4.80$\pm$0.18 & 1.06\\  
61    &  7.2$\pm$1.3 & 0.78$\pm$0.11 &   \nodata        &  0.17$\pm$0.05 &  0.21$\pm$0.07 &  20.34$\pm$1.59 & 3.57\\  
62    &  3.24$\pm$0.61 &  \nodata        &  0.28$\pm$0.05 &  0.25$\pm$0.05 &  0.21$\pm$0.04 & 34.30$\pm$5.43 & 2.30\\ 
Total$^{\ddagger}$ & 1.31$\pm$0.17 &  0.35$\pm$0.05 & 0.16$\pm$0.04 & 0.21$\pm$0.03  & 0.15$\pm$0.03 & 3.93$\pm$0.14 & 1.03\\
\enddata \\
NOTE: (1): Region identification numbers. (2)-(7): Emission line fluxes ratios along with their errors. These ratios are measured relative to $\mathrm{H}\alpha$, where the $\mathrm{H}\alpha$ fluxes are measured in units of $10^{-16}$\,erg\,sec$^{-1}$\,\AA$^{-1}$\,cm$^{-2}$\,arcsec$^{-2}$ and are absolute-flux calibrated.  
The line fluxes listed are those of: $[\mathrm{OII}] \lambda 3727$; $[\mathrm{OIII}] \lambda 5007$; $[\mathrm{NII}]\lambda 6548$; and $[\mathrm{SII}] \lambda 6717, \lambda 6731$ (from left to right). (8): A$_{V}$ is calculated from the Balmer decrement with adopting a MW extinction law (A$_{V}$/A$_{H_{\beta}}$ =1.164,  \cite{cardelli}). 
$^{\ddagger}$ The term {\it total} refers to the integrated spectrum. In this case, the flux-lines ratios values are in units of $10^{-12}$\,erg\,sec$^{-1}$\,cm$^{-2}$.
\end{deluxetable}

\begin{table}
\begin{center}
\caption{H$\alpha$ equivalent widths in emission from the ring spectra. These values are not corrected for underlying stellar absorption.\label{ewrings}}
\begin{tabular}{c l c c l c}
\hline\hline
Ring & EW$^{em}_{\mathrm{H}\alpha}$ & Radius & Ring & EW$^{em}_{\mathrm{H}\alpha}$ & Radius  \\
 & (\AA) & ($\arcsec$) & & (\AA) & ($\arcsec$) \\
\hline\hline
1 & $\dagger$ & 1 & 9 &  38.34 & 40\\
2 &  8.75 & 5 & 10 & 37.85 & 45 \\
3 &  18.65 & 10 & 11 & 36.5 & 50 \\
4 &  29.32 & 15 & 12 & 24.04 & 55 \\
5 &  25.56 & 20 & 13 & 34.98 & 60 \\
6 &  39.18 & 25 & 14 & 50.18 & 65 \\
7 &  44.97 & 30 & 15 & 100.1 & 70 \\
8 &  44.00 & 35 & 16 & 128.5 & 75$^{\ddagger}$ \\
\hline\hline
\end{tabular}
\end{center}
$^{\dagger}$ No line emission was detected.\\
$^{\ddagger}$In this table we present the value for ring 16 that it is not shown in any of the other tables in the paper since only the H$\alpha$ line could be detected in the spectrum and its flux (and EW) measured.
\end{table}

\begin{table}
\caption{UV-continuum attenuation results.}\label{UVtable}
\begin{center}
\begin{tabular}{c c c c c c c}
\hline\hline
Radius & A$_{FUV}$ & A$_{NUV}$ &  E(B-V)$_{Cardelli}$ & A$_{V,Cardelli}$ & E(B-V)$_{Calzetti}$ & A$_{V,Calzetti}$\\
($\arcsec$) & (mag) & (mag) & (mag) & (mag) & (mag) & (mag)\\
(1) & (2) & (3) & (4) & (5) & (6) & (7) \\
\hline\hline
6 & 1.98 & 1.40 & 0.18 & 0.54 & 0.17 & 0.69  \\
12 & 1.51 & 1.13 & 0.14 & 0.44 & 0.14 & 0.56 \\
18 & 1.33 & 1.03 & 0.13 & 0.40 & 0.12 & 0.51 \\
24 & 1.20 & 0.95 & 0.12 & 0.37 & 0.12 & 0.47 \\
30 & 1.04 & 0.87 & 0.11 & 0.34 & 0.11 & 0.43 \\
36 & 0.90 & 0.80 & 0.10 & 0.31 & 0.10 & 0.39\\
42 & 1.01 & 0.86 & 0.11 & 0.33 & 0.10 & 0.42 \\
48 & 1.04 & 0.87 & 0.11 & 0.34 & 0.11 & 0.43 \\
54 & 1.23 & 0.97 & 0.12 & 0.38 & 0.12 & 0.48 \\
60 & 1.20 & 0.96 & 0.12 & 0.37 & 0.12 & 0.47\\
66 & 0.95 & 0.83 & 0.10 & 0.32 & 0.10 & 0.41 \\
72 & 0.84 & 0.77 & 0.10 & 0.30 & 0.09 & 0.38\\
78 & 0.84 & 0.77 & 0.10 & 0.30 & 0.09 & 0.38 \\
84 & 0.93 & 0.82 & 0.10 & 0.32 & 0.10 & 0.40 \\
90 & 0.75 & 0.73 & 0.09 & 0.28 & 0.09 & 0.36 \\
96 & 0.40 & 0.59 & 0.07 & 0.23 & 0.07 & 0.29\\
102 & 0.62 & 0.67 & 0.08 & 0.26 & 0.08 & 0.33 \\
108 & 0.58 & 0.66 & 0.08 & 0.26 & 0.08 & 0.32\\
114 & 0.90 & 0.80 & 0.10 & 0.31 & 0.10 & 0.40 \\
\hline\hline 
\end{tabular}
\end{center}
(1): Radius in arcsec. (2): Far-UV attenuation values in magnitudes. (3): Near-UV attenuation values in magnitudes. (4)-(5): Galactic color excess from UV data and attenuation values calculated via Cardelli law, (E(B-V) = $A_{NUV}$/8.0) $R_{V} = 3.1$, expressed in magnitudes. (6)-(7): Galactic color excess from UV data and attenuation values calculated via Calzetti law, (E(B-V) = $A_{NUV}$/8.22) $R_{V} = 4.05$, expressed in magnitudes.
\end{table}

\begin{landscape}
\begin{table}
\caption{Abundance gradient fits.}\label{tabmet}
\centering
\begin{tabular}{|l|c|c|c|c|c|c|}
\hline
\multirow{2}{*}{Fits Type} & \multirow{2}{*}{$r_{BREAK}$} & \multirow{2}{*}{Segment} & \multirow{2}{*}{$12 + \log(O/H)$ at r=0} & \multicolumn{3}{|c|}{Gradient} \\ 
& & & & (dex/$^{\prime\prime}$) & (dex/kpc) & (dex/R$_{25}$) \\ 
\hline
Single Slope & \nodata & \nodata & 8.640 $\pm$ 0.047 & -0.0042 $\pm$ 0.0009 & −0.0346 $\pm$ 0.0074 & −0.416 $\pm$ 0.089 \\
\multirow{2}{*}{Double Slope Unweighted} & \multirow{2}{*}{40$^{\prime\prime}$} & Inner & 8.876 $\pm$ 0.056 & -0.013 $\pm$ 0.002 & −0.1073 $\pm$ 0.0165 & −1.29 $\pm$ 0.20 \\
& & Outer & 8.269 $\pm$ 0.012 & 0.002 $\pm$ 0.002 & 0.002 $\pm$ 0.019 & 0.198 $\pm$ 0.198 \\
\multirow{2}{*}{Double Slope Weighted} & \multirow{2}{*}{36$^{\prime\prime}$} & Inner & 8.949 $\pm$ 0.053 & -0.017 $\pm$ 0.002 & −0.140  $\pm$ 0.016 & −1.683  $\pm$ 0.198 \\
& & Outer &  8.269 $\pm$ 0.012 & 0.002 $\pm$ 0.002 & 0.002 $\pm$ 0.019 & 0.198 $\pm$ 0.198 \\
\hline
\end{tabular}
\end{table}
\end{landscape}

\begin{deluxetable}{cccccc}
\tablecolumns{6} 
\tablecaption{Ionized-gas diagnostics in the H~II regions and concetric annuli.}\label{oxytable}
\tablewidth{0pt}
\tablehead{
\colhead{ID} & \colhead{$12 + log(O/H)$} & \colhead{$q$} & \colhead{$N_{e}$} & \colhead{$T_{e}$} & \colhead{Offsets}\\
\colhead{(1)} & \colhead{(2)} & \colhead{(3)} & \colhead{(4)} & \colhead{(5)} & \colhead{(6)}}
\startdata
\multicolumn{6}{c}{Rings} \\
\cline{1-6}
 1 & \nodata &    \nodata &    1137 &   53359 &  (-3.6,2.5) \\  
 2 & \nodata &    \nodata &     206 &   11036 &  \nodata   \\  
 3 &   8.69$\pm$0.11 &    8.9E+06 &     122 &    9822 &  \nodata   \\  
 4 &   8.60$\pm$0.16 &    9.5E+06 &     186 &    9685 &   \nodata  \\  
 5 &   8.63$\pm$0.32 &    2.2E+07 &     138 &    8362 &   \nodata  \\  
 6 &   8.61$\pm$0.18 &    1.1E+07 &     169 &    9721 &   \nodata  \\  
 7 &   8.50$\pm$0.13 &    1.5E+07 &     154 &   10184 &  \nodata   \\ 
 8 &   8.29$\pm$0.18 &    1.7E+07 &     106 &    9215 &  \nodata   \\  
 9 &   8.50$\pm$0.12 &    1.6E+07 &      84 &   11308 &  \nodata   \\  
 10 &  8.21$\pm$0.63 &    1.7E+07 &      27 &   11204 &  \nodata   \\  
 11 & \nodata &    \nodata &     256 &   10124 &  \nodata   \\  
 12 & \nodata &    \nodata &     248 &   15785 &  \nodata   \\  
 13 & \nodata &    \nodata &     287 &   16225 &  \nodata  \\  
 14 & \nodata &    \nodata &     242 &   11186 &  \nodata  \\  
 15 & \nodata &    \nodata &      56 &   10296 &   \nodata  \\  
\cutinhead{HII Regions}
   1 &   7.96$\pm$0.74 &    2.5E+07 &     123 &    9549 &  (-71.9,17.1) \\  
   2 &   7.43$\pm$1.08 &    2.0E+07 & \nodata &    9229 &  (-62.8,20.3) \\  
   3 &   8.34$\pm$0.12 &    1.7E+07 & \nodata &    9984 &  (-31.5,35.8) \\  
   4 &   8.36$\pm$0.07 &    1.1E+07 & \nodata &    9247 &  (-52.1,25.9) \\  
   5 &   8.22$\pm$0.11 &    1.6E+07 & \nodata &    9211 &  (-51.7,20.1) \\  
   6 & \nodata &    \nodata &     272 & \nodata &  (-50.7,53.6) \\  
   7 &   7.00$\pm$1.20 &    \nodata & \nodata &   10372 &  (-45.7,38.5) \\  
   8 &   8.30$\pm$0.21 &    \nodata &     405 &    9229 &  (-40.4,-0.5) \\  
   9 &   8.34$\pm$0.10 &    1.7E+07 & \nodata &    9935 &  (-39.9,-7.1) \\  
  10 & \nodata &  \nodata &      10 &   10084 &  (-38.2,18.9) \\  
  11 &   8.38$\pm$0.11 &    1.7E+07 &      68 &   11163 &  (-34.9,32.2) \\  
  12 &   8.33$\pm$0.11 &    1.8E+07 &      33 &   11368 &  (-34.5,38.2) \\  
  13 &   8.43$\pm$0.38 &    1.3E+07 & \nodata &   11047 &  (-31.8,-3.9) \\  
  14 &   8.60$\pm$0.22 &    1.5E+07 & \nodata &    9940 &  (-31.3,8.4) \\  
  15 &   8.39$\pm$0.70 &    1.1E+08 &     352 &    7959 &  (-30.9,-28.0) \\  
  16 &   8.17$\pm$0.08 &    2.0E+07 &      30 &    9175 &  (-26.2,29.5) \\  
  17 &   8.45$\pm$0.36 &    1.8E+07 & \nodata &   15032 &  (-23.6,-36.8) \\  
  18 & \nodata &    \nodata & \nodata &    8371 &  (-21.6,51.2) \\  
  19 &   8.58$\pm$0.20 &    1.3E+07 &      11 &    9295 &  (-21.1,17.6) \\  
  20 &   8.55$\pm$0.97 &    7.4E+07 &      12 &    7556 &  (-21.0,-69.4) \\  
  21 &   8.62$\pm$0.09 &    1.3E+07 &     116 &   10033 &  (-20.7,8.8) \\  
  22 &   7.80$\pm$1.00 &    1.1E+07 &     142 &   10992 &  (-18.6,36.5) \\  
  23 &   8.18$\pm$0.08 &    3.9E+07 &      78 &   10008 &  (-15.9,-42.8) \\  
  24 &   8.48$\pm$0.12 &    1.1E+07 &      12 &    9657 &  (-14.8,47.1) \\  
  25 &   8.50$\pm$0.23 &    2.8E+07 &      37 &   10573 &  (-14.2,-58.0) \\  
  26 &   8.72$\pm$0.25 &    1.0E+07 &      42 &    8739 &  (-13.6,2.8) \\  
  27 & \nodata &    \nodata & \nodata &   19479 &  (-12.5,77.3) \\  
  28 &   8.70$\pm$0.15 &    1.0E+07 &     103 &    9507 & (-8.9,-6.5) \\  
  29 &   8.61$\pm$0.26 &    1.4E+07 &      29 &    8659 &  (-9.2,21.9) \\  
  30 &   8.58$\pm$0.25 &    1.4E+07 &     194 &    8860 &  (-7.5,-15.0) \\  
  31 &   8.25$\pm$0.15 &    1.7E+07 & \nodata &    9175 &  (-5.7,-36.6) \\  
  32 &   8.50$\pm$0.23 &    \nodata & \nodata &    8699 &  (-3.7,-27.9) \\  
  33 &   8.69$\pm$0.09 &    1.2E+07 &      73 &    8941 &  (-2.3,-12.2) \\  
  34 &   8.39$\pm$0.21 &    1.5E+07 & \nodata &   11198 &  (0.8,55.9) \\  
  35 &   8.66$\pm$0.07 &    1.2E+07 & \nodata &    7534 &  (1.3,23.3) \\  
  36 &   8.41$\pm$0.04 &    1.6E+07 & \nodata &   11831 &  (1.3,77.6) \\  
  37 &   8.48$\pm$0.24 &    2.1E+07 &     116 &    8804 & (1.4,-24.9) \\  
  38 &   8.64$\pm$0.17 &    1.0E+07 &      50 &    9067 &  (3.3,-15.6) \\  
  39 & \nodata &    \nodata & \nodata &   13102 &  (3.3,-69.3) \\  
  40 &   8.47$\pm$0.18 &    1.3E+07 & \nodata &   13174 &  (3.8,46.5) \\  
  41 & \nodata &    \nodata &     168 &    8506 &  (5.3,-31.0) \\  
  42 &   8.57$\pm$0.13 &    1.1E+07 & \nodata &   10684 &  (4.8,35.9) \\  
  43 &   8.61$\pm$0.39 &    1.2E+07 &      54 &    7507 &  (6.6,20.5) \\  
  44 &   8.74$\pm$0.05 &    9.1E+06 &     242 &    9523 &  (7.9,-0.7) \\  
  45 & \nodata &   \nodata &     149 &   11738 &  (7.8,8.9) \\  
  46 &   8.57$\pm$0.26 &    9.1E+06 &     126 &    8596 &  (11.4,-18.7) \\  
  47 &   8.51$\pm$0.24 &    1.1E+07 &     241 &    8870 &  (11.5,-25.1) \\  
  48 &   8.63$\pm$0.12 &    8.5E+06 &     130 &   13086 &  (13.6,-9.6) \\  
  49 &   8.42$\pm$0.11 &    2.8E+07 &     170 &   10808 &  (14.8,68.9) \\  
  50 &   8.54$\pm$0.30 &    1.4E+07 &     201 &    8500 &  (19.0,-1.1) \\  
  51 &   8.47$\pm$0.14 &    1.7E+07 &     118 &    9599 &  (20.3,25.5) \\  
  52 &   8.53$\pm$0.27 &    1.9E+07 &     137 &    8655 &  (23.8,15.9) \\  
  53 &   8.31$\pm$0.10 &    1.5E+07 &     158 &    9683 &  (25.0,21.6) \\  
  54 &   8.43$\pm$0.08 &    2.6E+07 & \nodata &   10588 &  (25.6,59.5) \\  
  55 &   8.44$\pm$0.08 &    2.4E+07 &      54 &   10645 &  (30.7,50.5) \\  
  56 & \nodata &    \nodata &    50 &    6785 &  (33.3,8.9) \\  
  57 &   8.57$\pm$0.16 &    8.1E+06 &     259 &   23652 &  (34.3,-0.0) \\  
  58 &   8.34$\pm$0.21 &    1.9E+07 &     170 &   11032 &  (42.5,29.0) \\  
  59 &   8.34$\pm$0.37 &    1.0E+07 &    1318 &   12890 &  (45.2,13.2) \\  
  60 &   7.96$\pm$0.83 &    1.1E+07 &     335 &   10993 &  (48.3,2.9) \\  
  61 & \nodata &    \nodata &   619 &   19051 &  (69.1,14.2) \\  
  62 & \nodata &    \nodata &    60 &   12023 &  (-3.2,4.3) \\  
\enddata \\
NOTE: (1): HII regions and annuli identification. (2): Oxygen abundances along with the corresponding errors. (3): Ionization parameter {\itshape q} defined as $q = S_{H0} / n$. (4): Electron density in units of cm$^{-3}$. (5) Electron temperature in Kelvin. (6): Offset coordinates of the HII regions in units of arcsec.
\end{deluxetable}

\end{document}